\documentclass[11pt,a4paper, epsf,psfig,amsmath,amssymb]{article}

\usepackage{xspace,bm}

\usepackage{amsmath}
\usepackage{amsfonts}
\usepackage{epsfig}
\usepackage{amssymb}

\newtheorem{theorem}{Theorem}[section]
\newtheorem{proposition}[theorem]{Proposition}
\newtheorem{corollary}[theorem]{Corollary}
\newtheorem{lemma}[theorem]{Lemma}
\newtheorem{claim}[theorem]{Claim}

\newcommand{\qed}{\hfill\rule{2mm}{2mm}}
\newcommand{\remove}[1]{}

\def\s{{{\bm s}}}

\def\ms{{{\bm\sigma}}}

\def\mt{{\bm{\tau}}}

\def\tx{\textstyle}

\def\Range{{\sf Range}}

\def\myi{{\sf (i)}\xspace}
\def\myii{{\sf (ii)}\xspace}

\newtheorem{definition}{Definition}[section]
\newtheorem{observation}{Observation}[section]

\begin{document}

\title{{\bf How Many Attackers Can Selfish Defenders Catch?}\thanks{A preliminary version of this work
                                                appeared in the {\it CD-ROM Proceedings of the 41st
                                                Hawaii International Conference on System Sciences},
                                                Track on Software Technology, Minitrack on
                                                Algorithmic Challenges in Emerging Applications of Computing,
                                                January 2008.                                                                                           This work
                                                                    has been partially supported
                                                                    by the {\sf IST} Program
                                                                    of the European Union
                                                                     under contract numbers
                                                                    IST-2004-001907 ({\sf DELIS})
                                                                    and 15964 ({\sf AEOLUS}).}
      }

\author{{\sl Marios Mavronicolas}\thanks{Department of Computer Science,
                                         University of Cyprus,
                                         Nicosia CY-1678,
                                         Cyprus. Part of this work was performed while
                                         this author  was visiting the Faculty of Computer Science,
                                          Electrical Engineering
                                     and Mathematics,
                                     University of Paderborn.
                                          Email
                                         {\tt  mavronic@cs.ucy.ac.cy}
                                        }
        \and
        {\sl Burkhard Monien}\thanks{Faculty of Computer Science,
                                     Electrical Engineering
                                     and Mathematics,
                                     University of Paderborn,
                                     33102 Paderborn,
                                     Germany. Part of this work was performed while this author was visiting the Department
                                     of Computer Science, University of Cyprus.
                                     Email  {\tt bm@upb.de}
                                    }
        \and
        {\sl Vicky G.~Papadopoulou}\thanks{Department   of Computer Science and Engineering,
                                            European University Cyprus, Nicosia 1516, Cyprus. Part of this work was performed while
                                         this author  was visiting the Faculty of Computer Science,
                                          Electrical Engineering
                                     and Mathematics,
                                     University of Paderborn.
                                          Email
                                         {\tt  v.papadopoulou@euc.ac.cy}}
       }

\date{{\sc (\today)}}

\maketitle \pagenumbering{arabic} \thispagestyle{empty}

\newpage

\begin{abstract}
In a distributed system with {\it attacks} and {\it defenses,}   both {\it attackers} and {\it defenders} are
self-interested entities. We assume a {\it reward-sharing} scheme among {\it interdependent} defenders; each
defender wishes to (locally) maximize her own total {\it fair share} to the attackers extinguished due to her
involvement (and possibly due to those   of others). What is  the {\em maximum} amount of protection
achievable by a number of such defenders against a number of attackers while  the system is in a {\it Nash
equilibrium}? As a measure of system protection, we adopt the {\it Defense-Ratio}~\cite{MPPS05a}, which
provides the expected (inverse) proportion of attackers caught by the defenders. In a {\it Defense-Optimal}
Nash equilibrium, the Defense-Ratio is optimized.

We discover that the possibility of optimizing the Defense-Ratio (in a Nash equilibrium) depends in a  subtle
way on how the number of defenders compares to  two natural graph-theoretic thresholds  we identify. In this
vein, we obtain, through a  combinatorial analysis of Nash equilibria, a   collection of  trade-off results:
\begin{itemize}

\item When the number of defenders is either sufficiently small or sufficiently large,
there   are  cases where the Defense-Ratio can be optimized. The   optimization  problem is computationally
tractable for a large  number of defenders;    the problem becomes  ${\cal NP}$-complete for a small  number
of defenders and
 the intractability is inherited from     a previously unconsidered combinatorial problem in
{\em Fractional Graph Theory}.

\item Perhaps  paradoxically, there is a middle range of values for the number
of defenders where optimizing the Defense-Ratio is never   possible.

\end{itemize}
\end{abstract}

\pagebreak

\section{Introduction}
\label{introduction}

 \subsection{The Model and its Rationale}
 {\it Safety} and {\it security} are     key issues
for the design and operation of a distributed system; see, e.g., \cite{A01} or \cite[Chapter 7]{CDK05}.
Indeed, with the unprecedented advent of the Internet, there is a growing interest   to formalize, design and
analyze distributed systems prone to {\it malicious  attacks} and   ({\it non-malicious}) {\it defenses}. A
new dimension stems from the fact that   Internet {\it servers}   and {\it clients} are controlled by {\it
selfish} agents whose interest is the local maximization of their own benefits  rather than the optimization
of global performance   \cite{A01b,CGC04,GL02,GCC08,GCC08b}. So, it is a challenging task to formalize and
analyze the {\em simultaneous} impact of selfish and malicious behavior of Internet agents (cf. \cite{L81}).

 In this work, a distributed system is modeled as a graph $G = (V, E)$;   nodes represent
the {\it hosts} and edges represent the {\it  links}.
 An {\it attacker} represents a {\it virus}; it  is a malicious  client that targets a host to destroy.
A {\it defender} is a  non-malicious   server representing the {\it antivirus software}  implemented
 on a subnetwork in order to protect  all    hosts thereby connected. Here is the rationale and motivation
 for these modeling choices:
\begin{itemize}

\item
 Associating attacks with nodes makes sense since computer security  attacks are often directed to
individual hosts  such as commercial and public sector  entities.

\item Associating defenses with edges is motivated by {\it Network Edge Security}~\cite{MP01}; this is a recently
proposed, distributed  {\it firewall architecture} where   antivirus software, rather than being statically
installed and licensed at a host, is  implemented by a {\it distributed algorithm} running on a subnetwork.
Such distributed implementations are attractive since they offer  increased fault-tolerance and the benefit
of sharing the   licensing costs to the hosts.

We focus here on the simplest possible case where the subnetwork is just a  {\em single} link; a precise
understanding of the mathematical pitfalls of attacks and defenses for this simplest case is a necessary
prerequisite to  mastering   the general case.

\end{itemize}

  In reality, malicious attackers are {\it independent}; each (financially motivated) attacker
tries to maximize  on her own  the amount of harm it causes during her lifetime (cf. \cite{WP04}). Hence, it
is natural to model each attacker as a strategic {\it  player} seeking  to maximize the chance of escaping
the antivirus software; so, the  strategy of one attacker does not (directly) affect the payoff of another.
In contrast, there are  at least three approaches to  modeling the defenses:
\begin{itemize}
\item
  Defenses are {\em not} strategic; this  approach would imply   the ({\em centralized}) optimization problem of
computing  locations for the defenders that maximize the system protection   given that attackers are
strategic.
\item  Defenses are strategic  and  they   {\it cooperate}  to maximize the number of trapped  viruses. This is
modeled by assuming a {\em single} (strategic) defender, which  centrally chooses multiple links. This
approach  has been  pursued in \cite{GMPPS06}.
\item  Defenses are strategic but  {\it non-cooperative}; so, each defender still tries to maximize the number
 of trapped
viruses she catches, while    competing with the other defenders.
\end{itemize}

\noindent We have chosen to adopt the third approach. Our   choice of approach is motivated as follows:

\begin{itemize}
\item  In a large network, the defense policies are {\it independent} and {\it decentralized}. Hence,
it may    be not    so realistic to assume that a {\it centralized} (even selfish) entity   coordinates all
defenses.
\item
 There are   financial incentives offered by hosts to   defense mechanisms on
the basis of   the   number of sustained attacks; consider, for example, the following scenaria:
\begin{itemize}
\item  Prices for antivirus software are determined through {\it recommendation systems}, which
collect data   from networks where scrutinized  hosts were witnessed. Such price incentives induce a
competition  among   defenders, resulting to non-cooperation.
\item
 Think of a {\it network owner}    interested in maximizing the network protection.
Towards  that end, the owner has subcontracted the protection task to a set of independent, deployable
agents. Clearly, each such agent   tries  to optimize the protection she offers in order to increase her
reward; again, this manifests   non-cooperation.
\end{itemize}
\end{itemize}

We   materialize the assumption that defenses are independent and non-cooperative on the basis of  an
intuitive {\it reward-sharing} scheme:  Whenever more than one colocated defenders are extinguishing the
attacker(s)
 targeting   a host, each defender will be rewarded with the {\it fair share} of the number of attackers
extinguished.  So,   each defender is   modeled as a strategic player seeking to maximize her {\em total}
fair share to the number of extinguished attackers.

 We assume    two selfish species with  $\alpha$ attackers and
$\delta$ defenders;  both species   may  use mixed strategies. Note  that $\delta$ is proportional to the
real cost of purchasing and installing several units of (licensed) antivirus software. The very special but
yet highly non-trivial case   with a single defender was originally introduced in~\cite{MPPS05a} and further
studied in~\cite{GMPPS06,MMPPS06,MPPPS06,MPPS05}.
In a    {\it Nash equilibrium}~\cite{N50,N51}, no
 player can unilaterally increase her (expected) {\it utility}.

 To evaluate Nash equilibria, we employ  the
{\it Defense-Ratio};  this  is the ratio of the optimum number  $\alpha$ over   the expected number of
attackers extinguished  by the defenders (cf. \cite{MMPPS06,MPPPS06}). Motivated by {\it best-case} Nash
equilibria and the {\it Price of Stability}~\cite{ADKTWR04}, we introduce  {\it Defense-Optimal} Nash
equilibria where the Defense-Ratio  attains the      value    $\max\Big\{1, \frac{\tx |V|}{\tx 2\delta}
\Big\}$ (Definition \ref{defense optimal nash equilibrium}); we choose  this value since we observe that it
is a (tight) lower bound on Defense-Ratio (Corollary \ref{lower bound on defense ratio}).
 (Contrast  Defense-Optimal Nash equilibria and the smallest possible Defense-Ratio to
{\em worst-case}  Nash equilibria and the {\it Price of Anarchy} from the seminal work of Koutsoupias and
Papadimitriou~\cite{KP99}.) A {\it Defense-Optimal} graph (for a given  $\delta$) is one that  admits  a
Defense-Optimal Nash equilibrium.

\subsection{Contribution}
We are interested in the possibility of achieving,  and the complexity of computing, a Defense-Optimal Nash
equilibrium    for a {\em given} number of defenders $\delta$. We discover that this possibility and the
associated complexity  depend on $\delta$ in a quantitatively subtle way: They are determined by   two
graph-theoretic thresholds for $\delta$, namely $\frac{\textstyle |V|} {\textstyle 2}$ and $\beta'(G)$ (the
size of a {\it Minimum Edge Cover}).  (Recall that $\frac{\textstyle |V|} {\textstyle 2} \leq \beta'(G)$.)

Our chief tool is a combinatorial characterization of the associated Nash equilibria we obtain (Proposition
\ref{characterization of nash equilibria}). For Pure Nash equilibria where both species use pure strategies,
this characterization yields some interesting necessary graph-theoretic conditions for Nash equilibria
(Proposition \ref{necessary condition for pure nash equilibria}). Furthermore, this characterization yields
some sufficient conditions for Defense-Optimal Nash equilibria (Theorems \ref{lemma one} and \ref{defender
pure is defense optimal}).

Our end findings are as follows:
\begin{itemize}

\item When either $\delta \leq             \frac{\textstyle |V|}                  {\textstyle 2}$ or $\delta \geq
 \beta'(G)$, there {\em are} cases allowing for a Defense-Optimal Nash equilibrium.
\begin{itemize}

\item \underline{The case  of {\em few} defenders $\left(\textrm{with }  \delta \leq \frac{\textstyle |V|}
           {\textstyle 2}\right)$:}
We provide a combinatorial characterization of Defense-Optimal  graphs    (Theorem~\ref{characterization of
defense optimal graphs}), which points out an interesting connection to {\em Fractional} ({\em Perfect}) {\em
Matchings}~\cite[Chapter 2]{SU97}. Roughly speaking, these graphs make a strict subset of  the class of
graphs with a  Fractional Perfect Matching: for a Defense-Optimal graph,  and assuming that $\delta\leq
\frac{\tx |V| }{\tx 2}$, it is possible to partition {\em some} Fractional Perfect Matching of it into
$\delta$ smaller, {\it vertex-disjoint} Fractional Perfect Matchings so that the total  weight (inherited
from the original Fractional Perfect Matching) in each partite is  equal   to  $  \frac{\textstyle |V|}
{\textstyle 2\, \delta}  $ (Theorem~\ref{characterization of defense optimal graphs}). Call such a Fractional
Perfect Matching a {\it$\delta$-Partitionable Fractional Perfect Matching}; this is a previously
unconsidered, combinatorial concept in {\it Fractional Graph Theory}~\cite{SU97}.

 We  prove that  the recognition problem for the   class of graphs with a $\delta$-Partitionable Fractional Perfect Matching
  is ${\cal
NP}$-complete  (Corollary \ref{Fractional Perfect Matching theo NPcompleteness1});  this intractability
result holds for an {\em arbitrary} value  of $\delta$.  Hence, so is the decision problem for
  a Defense-Optimal Nash equilibrium  $ \left({\rm for \, }\delta \leq \frac{\textstyle |V|}
{\textstyle 2} \right)$ (Corollary~\ref{Fractional Perfect Matching theo NPcompleteness}).
To establish the ${\cal NP}$-completeness of the recognition problem, we develop some techniques for the reduction of Fractional
(Perfect) Matchings (Section \ref{Fractional Perfect Matchings Section}); these may be of independent interest.

We note that the recognition problem for the class of graphs  with  a $\delta$-Partitionable Fractional
Perfect Matching simultaneously generalizes a tractable and an intractable recognition problem: the first one
concerns the class of graphs with a Perfect Matching \cite{E65}, while the second concerns   the class of
graphs whose vertex set can be partitioned into triangles \cite[GT11]{GJ79}.

 A further interesting number-theoretic consequence of the
combinatorial  characterization we have derived for Defense-Optimal graphs $\left( {\rm for \, } \delta \leq \frac{\textstyle |V|}
{\textstyle 2} \right)$ is that $\delta$ divides $|V|$ in a Defense-Optimal graph
(Corollary \ref{defense optimal graphs corollary}).

On the positive side, we identify  another  restriction of  the class of graphs with a Fractional Perfect
Matching  that are Defense-Optimal   in certain, well-characterized   cases (Theorem~\ref{perfect matching
few defenders characterization}); these are the  graphs   with a {\it Perfect Matching}.

\item \underline{The case of {\em too many} defenders   (with $\delta \geq \beta'(G)$):} We identify two  cases where there are
Defense-Optimal  Nash equilibria with a special structure, namely the {\em vertex-balanced} Nash equilibria
(Definition \ref{Defender-pure Vertex-balanced Profile definition});
their
  structure enables  their  polynomial time computation (Theorems~\ref{defender pure balanced
equilibria} and~\ref{pure balanced equilibria}). The two corresponding algorithms rely on the efficient
computation of {\it  Minimum Edge Cover}; the second algorithm requires some relation between $\delta$ and
$\alpha$ (namely, that $2\delta$ divides $\alpha$).

\end{itemize}

\item \underline{The case of {\em many} defenders  (with $\frac{\textstyle |V|}      {\textstyle 2} < \delta < \beta'(G)$):}
We provide a combinatorial proof that there is {\em no} Defense-Optimal  graph for $\frac{\textstyle |V|}
{\textstyle 2} < \delta < \beta'(G)$ (Theorem~\ref{impossibility result for many defenders}). This is somehow
paradoxical  since with   fewer  defenders $\left( \delta \leq \frac{\textstyle |V|}      {\textstyle
2}\right)$, we already identified cases with a Defense-Optimal Nash equilibrium. However, since    the
Defense-Ratio in a Defense-Optimal Nash equilibrium has a transition   around the value $\delta
=\frac{\textstyle |V|}{\textstyle 2}$, this paradox may not be wholly surprising.

\end{itemize}

\begin{figure}[p]
\epsfclipon
  \centerline{\hbox{
  \leavevmode
   \epsffile{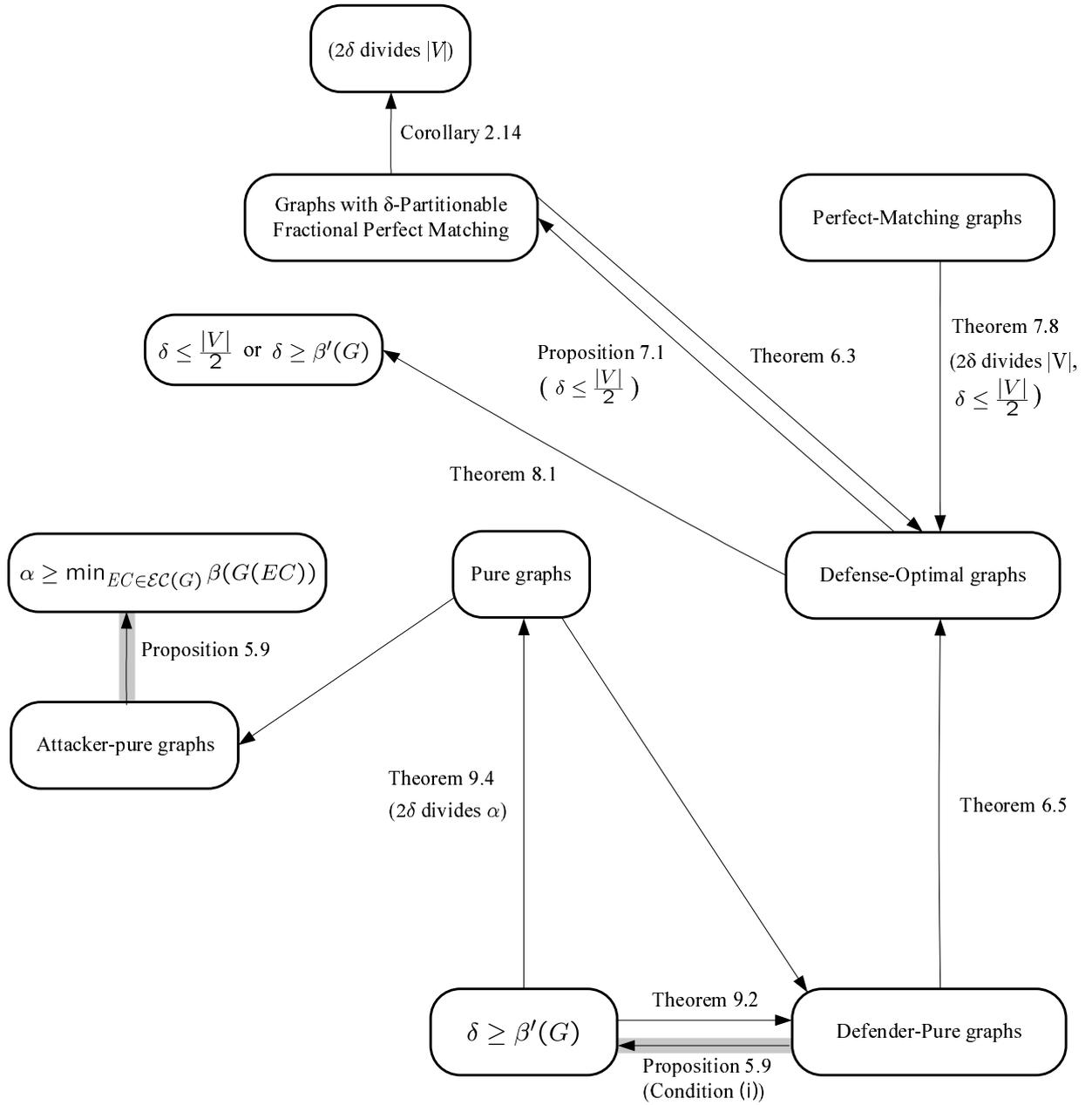}
  }
} \epsfclipoff \caption{Some inclusion relationships among the graph classes associated with Nash equilibria
we have introduced. A directed edge from class ${\cal C}_1$ to class ${\cal C}_2$ indicates that ${\cal
C}_1\subseteq  {\cal C}_2$; a condition on the edge indicates the condition under which the inclusion holds.
Clouded directed edges indicate inclusions that have been demonstrated to be non-strict.
 } \label{statusfigure}
\end{figure}

Our techniques have identified several new classes of graphs for any arbitrary pair of values of $\delta$ and
$\alpha$, such as graphs with $\delta$-Partitionable Fractional Perfect Matching, Defense-Optimal graphs and
{\em Pure graphs} (which admit Pure Nash equilibria); each such class was defined to support the existence of
some Nash equilibria with a particular structure (for example, Defense-Optimal Nash equilibrium or Pure Nash
equilibrium). Our results have revealed a fine structure among these classes, which is summarized in Figure
\ref{statusfigure}.

\subsection{Related Work and Comparison}
We emphasize that the assumption of $\delta > 1$ defenders  has required a far more challenging combinatorial
and graph-theoretic analysis than those used for   the    case of a single defender in~\cite{GMPPS06,
MMPPS06,MPPPS06,MPPS05,MPPS05a}. Hence, we view this work as a   major  generalization of the work
in~\cite{GMPPS06,MMPPS06,MPPPS06,MPPS05,MPPS05a} towards the more realistic case of $\delta > 1$ defenders.

The notion of Defense-Ratio generalizes a corresponding definition from~\cite[Section 3.4]{MMPPS06} to the
case of $\delta
> 1$ defenders. The special case where $\delta = 1$ of Theorem~\ref{characterization of defense optimal graphs}
was considered  in~\cite[Corollary 2]{MPPPS06}; this case   allowed for a polynomial time algorithm to decide
the existence of (and compute)  a Defense-Optimal Nash equilibrium by reduction to the recognition problem
for a graph with a Fractional Perfect Matching. In contrast, the decision problem for a Defense-Optimal Nash
equilibrium  $\Big($for an arbitrary $\delta \leq \frac{\tx |V|}{\tx 2}\Big)$ is ${\cal NP}$-complete
(Corollary~\ref{Fractional Perfect Matching theo NPcompleteness}).

Schechter and Smith \cite{SS03}  considered the complementary question of determining the minimum number of
defenders to catch a {\em single} attacker in a  related model of economic threats.

\subsection{Road Map}
The rest of this paper is organized as follows. Section \ref{background and preliminaries} collects together
some background and preliminaries from Graph Theory. A preliminary combinatorial lemma is formulated and
proved in Section \ref{Combinatorics Section}. Section \ref{framework} presents the game-theoretic framework.
The combinatorial structure of the associated Nash equilibria is treated in Section \ref{structure of nash
equilibria}. Section \ref{Defense Optimal Graphs Section} considers  Defense-Optimal Nash equilibria.
Sections \ref{few defenders}, \ref{moderate defenders} and    \ref{too many defenders} treat the cases of
few, many and too many defenders, respectively.   We conclude, in Section \ref{conclusion}, with a discussion of the results and
some open problems.

Throughout, for an integer $n \geq 1$, denote
 $[n] = \{ 1, \ldots, n \}$; for a number $x\neq 0$, ${\tt sgn}(x) $ denotes the {\em sign} of $x$
(which is $+1$ or $-1$).

\section{Background and Preliminaries from Graph Theory}
\label{background and preliminaries}

Some basic   definitions are articulated in Section \ref{Graph Theory Section}. Fractional
  Matchings  are recalled in Section \ref{Fractional Matchings Section intro}. Some reduction
techniques for Fractional (Perfect) Matchings are developed in Section \ref{Fractional Perfect Matchings
Section}. Section \ref{m-Partitionable Fractional Perfect Matchings Section} treats $\delta$-Partitionable
Fractional Perfect Matchings.

\subsection{Basics}\label{Graph Theory Section}
We consider a simple undirected graph $G = \left\langle V, E     \right\rangle$ (with no isolated vertices).
The {\it trivial} graph consists of a single edge. We will sometimes model an edge as the set of its two
vertices.  Denote as ${\sf d}_G(v)$ the {\it degree} of vertex $v$ in $G$. An edge $(u,v)\in E$ is {\it
pendant} if ${\sf d}_G(u)=1$ but ${\sf d}_G(v)
>1$. A {\em path}     is a
sequence of vertices $v_1, v_2, \cdots, v_{n+1} $ from $V$ such that for each index $k\in [n]$, $(v_k,
v_{k+1})\in E$; in a {\it cycle} ${\cal C}$, $v_{n+1} = v_1$.   The cycle ${\cal C}$ has length $n$, and
${\cal C}$ is {\it even} (resp., {\it odd}) if $n$ is even (resp.,  odd). A {\em triangle} is a cycle of
length three. We shall sometimes   treat ${\cal C}$ as a set of vertices;   $E({\cal C})$ denotes the edge
set induced by ${\cal C}$ in the natural way.

Vertex sets and edge sets induce subgraphs in the natural way.  For a vertex set $U \subseteq V$, denote  as
$G(U)$ the subgraph of $G$ induced by $U$; denote   ${\sf Edges}_{G}(U) = \{ (u,v) \in E \mid u,v \in U
  \}$. For an edge set $F \subseteq E$, denote as $G(F)$ the subgraph of $G$ induced
by $F$; denote  ${\sf Vertices}_G(F) = \bigcup_{(u,v)\in F} \{ u,v\} $. (We shall sometimes omit the index
$G$ when it is clear from context.) A {\it component} is a maximal connected subgraph. A cycle is  {\it
isolated} (as a subgraph) if it is a component; else, it is {\it non-isolated}. A component is {\it cyclic}
if it contains a cycle; else, it is {\it acyclic}.

For an undirected graph, both an odd and an even cycle are  computable in polynomial time. A linear time
algorithm to compute an odd cycle is based on incorporating {\em breadth-first search}  into the constructive
proof for the characterization of bipartite graphs due to K\"{o}nig   \cite{K16} (cf. \cite[Proposition
2.27]{KV00}). Polynomial time algorithms to compute an even cycle have appeared in \cite{LP84,M83,YZ97}.

 A {\it Vertex Cover} is a vertex set $VC \subseteq V$ such that for each edge $(u,v)  \in E$ either $u\in
VC$ or $v \in VC$; a {\it Minimum Vertex Cover} is one that has minimum size, which is denoted as $\beta(G)$.
An {\it Edge Cover} is an edge set $EC \subseteq E$ such that for each vertex $v \in V$, there is an edge
$(u, v) \in EC$; a {\it Minimum Edge Cover} is one that has minimum size, which is denoted as
$\beta^{\prime}(G)$. Clearly,   $\frac{\tx|V| }{\tx 2}\leq \beta'(G)$.  Denote as ${\cal EC}(G)$ the set of
all Edge Covers of $G$.

 A {\it Matching} is a set $M \subseteq E$ of non-incident edges; a {\it Maximum Matching} is one that has
maximum size. The first polynomial time algorithm to compute a Maximum Matching is due to Edmonds~\cite{E65}.
It is known that computing a Minimum Edge Cover is polynomial time reducible  to computing a Maximum
Matching---see, e.g.,~\cite[Theorem 3.1.22]{W01} or \cite{L76}.

A {\it Perfect Matching} is a Matching that is also an Edge Cover; so, a Perfect Matching has size
$\frac{\textstyle |V|} {\textstyle 2}$. A {\it Perfect-Matching} graph is one that has a Perfect  Matching;
note that in a Perfect-Matching graph, $   \beta'(G) = \frac{\textstyle |V|} {\textstyle 2} $. Since a
Perfect Matching is a Maximum Matching, any polynomial time algorithm to compute a Maximum Matching yields a
polynomial time algorithm to recognize Perfect-Matching graphs and compute a Perfect Matching.

\subsection{Fractional (Perfect) Matchings}\label{Fractional Matchings Section intro}
  A {\it Fractional Matching} is a function $f: E \rightarrow [0, 1]$ where  for each vertex $v \in V$,
$\sum_{e \in E \mid v \in e}   f(e) \leq 1$. (Matching is the special case where $f(e) \in \{ 0, 1 \}$ for
each edge $e \in E$.)  For a Fractional  Matching $f$, induced is the set   $E(f) = \{ e\in E \mid f(e)> 0
\}$; $|E(f)|$ is the {\it size} of $f$. The {\it range} of a Fractional Matching $f$ is the set $\Range (f) =
\{ f(e) \mid e\in E \}$; so, $\Range(f) \subseteq [0,1]$.

 Given two Fractional Matchings $f$ and $f'$, write  $f'\subseteq f$ (resp., $f'\subset f$) if
 $E(f')\subseteq E(f)$ (resp., $E(f')\subset  E(f)$). Say that two functions $f:E\rightarrow [0,1]$
and $f': E\rightarrow [0,1]$ are  {\it equivalent} if for each vertex $v\in V$, $\sum_{e\in E \mid v\in e } f
(e)= \sum_{e \in E \mid v\in e } f '(e)$. Clearly, a function  $f :E\rightarrow [0,1]$ that is equivalent to
a Fractional Matching    is also  a Fractional Matching.

A {\it Fractional Perfect Matching} is a Fractional Matching $f$ such that for each vertex $v\in V$, $\sum_{e
\in E \mid v \in e} f(e) = 1$.  (Perfect Matching is the special case where $f(e) \in \{ 0, 1 \}$ for each
edge $e \in E$. Note that in this special  case, $E(f)$ is a Perfect Matching; for  an arbitrary Fractional
Perfect Matching, $E(f)$ need not be a Perfect Matching.) In this case,  for each vertex $v \in V$, there is
at least one edge $e\in E$ with $v\in e$ such that   $f(e)>0$, so that $e\in E(f)$;  hence,  for a Fractional
Perfect Matching $f$, $E(f)$ is an Edge Cover. Note  that a function $f :E\rightarrow [0,1]$ which  is
equivalent to a Fractional Perfect Matching is also a Fractional Perfect Matching.

A {\em Fractional Maximum Matching} is a Fractional Matching
 $f$  that   maximizes   $\sum_{e\in E}f(e)$ among  all Fractional Matchings. A Fractional Perfect Matching is
 a Fractional Maximum Matching (but not vice versa).
 We observe  a simple property of Fractional Perfect Matchings:
\begin{lemma}
\label{Fractional Perfect Matchings have no pendant edges} For a Fractional Perfect Matching $f$, the graph
$G(E(f))$ has no pendant edge.
\end{lemma}
\begin{proof}
 Assume, by way of contradiction, that $G(E(f))$ has a pendant edge $ (u,v)$
with ${\sf d}_{G(E(f))}(u) =1$ and ${\sf d}_{G(E(f))}(v) >1$. Since $f$ is a Fractional Perfect Matching,
$\sum_{e \in E \mid u \in e } f(e) =1$ and $\sum_{e \in E \mid v \in e } f(e) =1$. By assumption on $ u $,
the first equality implies   that $f((u,v)) = 1$. By  assumption on $v$, the second equality  implies that
$f((u,v)) <1$. A contradiction. \qed\end{proof}

\noindent  Lemma~\ref{Fractional Perfect Matchings have no pendant edges} implies that  for a  Fractional
Perfect Matching  $f$, each component of   $G(E(f))$
 is either a single edge or a (non-trivial) subgraph of $G$
with no pendant edges;  in particular, each  acyclic component of $G(E(f))$ is a single edge. The proof for
\cite[Theorem 2.1.5]{SU97} 
 establishes as a by-product that a Fractional Maximum Matching $f$ with smallest size  has no pendant edge; so,
Lemma \ref{Fractional Perfect Matchings have no pendant edges} provides a complementary  property  for  the
special case of Fractional Perfect Matchings.

 The class of graphs with  a Fractional Perfect Matching is recognizable in polynomial time
via a Linear Programming formulation. (See~\cite{BP89} for an efficient combinatorial algorithm.) The same
holds for the corresponding search problem.

\subsection{Reductions of Fractional (Perfect) Matchings}\label{Fractional Perfect Matchings Section}
Our starting point is a combinatorial property of a special case of   a Fractional Maximum Matching; this
property is reported in   \cite[Theorem 2.1.5]{SU97}.

\begin{proposition}\label{theorem2.1.5}
Consider a  Fractional Maximum Matching $f$ with smallest size. Then,  $f$ has  only  single edges and odd
cycles.
\end{proposition}

\noindent  Proposition \ref{theorem2.1.5}, outlaws, in particular, the induction of even cycles and
non-isolated odd cycles in a Fractional Maximum Matching with smallest size.
 In the spirit of Proposition
\ref{theorem2.1.5}, we shall present two new reduction techniques for a Fractional (Perfect) Matching.   The
first reduction will eliminate all induced even cycles from an {\em arbitrary} Fractional Matching. The
second reduction is applicable only to Fractional Perfect Matchings; it will eliminate all induced
non-isolated odd cycles when run on a Fractional Perfect Matching with no induced even cycles.

The corresponding elimination algorithms ({\sf EliminateEvenCycles} and {\sf IsolateOddCycles} in Figures
\ref{EliminateEvenCycles_alg}  and \ref{IsolateOddCycles_alg}, respectively) are inspired from the
corresponding inexistence proof for Proposition \ref{theorem2.1.5}. In more detail, that  proof assumes the
existence of an even or a non-isolated odd cycle and derives a contradiction by relying on the property that
the Fractional Matching is  a Maximum one of smallest size; the contradiction is derived by eliminating edges
to get a Fractional (Maximum) Matching with less size. In contrast, our elimination algorithms compute in
polynomial time an even or a non-isolated odd cycle (as long as there are such), respectively; they keep
eliminating  edges (as long as possible) till there are no more even or  non-isolated  odd cycles,
respectively.

\subsubsection{Elimination of Even Cycles}
 We prove:

\begin{proposition}\label{Fractional Matchings with no even cycles}
Consider a Fractional   Matching $f$. Then, there is a polynomial time algorithm to transform $f$ into an
equivalent Fractional Matching $f'\subseteq f$   with  no even cycle.
 \end{proposition}

\begin{figure}[h]
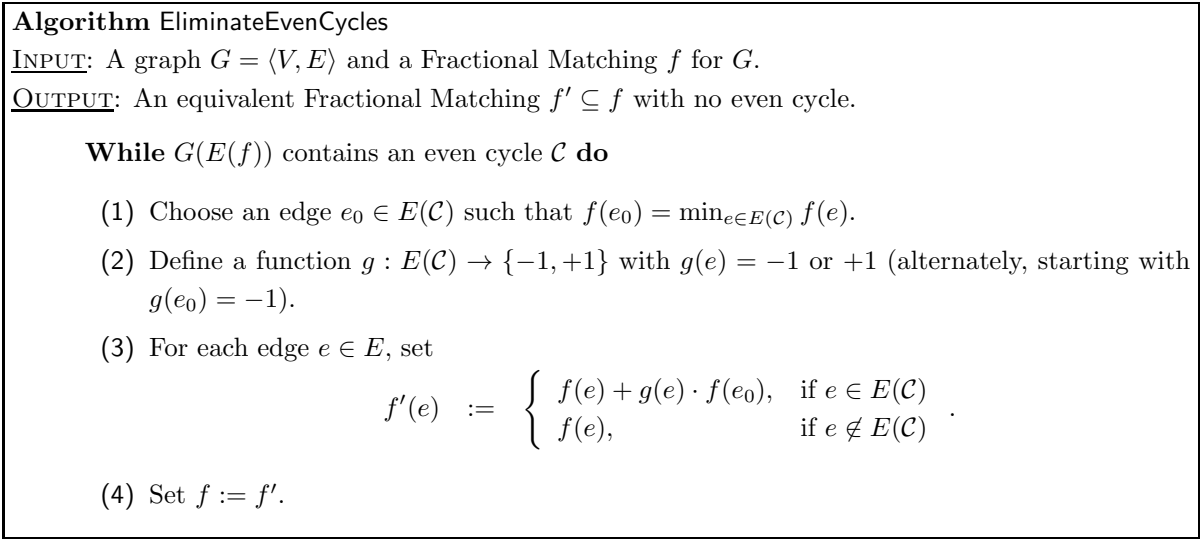

\fbox{\parbox{15.5cm}
 { \small
 { {\bf Algorithm} {\sf EliminateEvenCycles}

{\underline{\sc Input}}:  A graph $G=\langle  V,E\rangle $              and a Fractional Matching $f$ for
$G$.

{\underline{\sc Output}}: An equivalent  Fractional  Matching $f' \subseteq f$   with  no even cycle.
\begin{itemize}
\item[] {\bf  While} $G(E(f))$  contains  an even cycle ${\cal C}$ {\bf do}
\begin{itemize}
\item[{\sf(1)}] Choose an edge $e_0\in  E({\cal C})$ such that   $ f (e_0)  = \min_{e\in E({\cal C})} f (e)$.
\item[{\sf(2)}] Define a function $g:E ({\cal C})\rightarrow \left\{-1, +1\right\}$ with
 $g(e ) =
    - 1 $   or  $ +1$    (alternately,     starting   with  $ g(e_0)=-1$).
\item[{\sf(3)}] For each edge $e\in E$, set \vspace{-0.3cm}
\begin{eqnarray*}
f'(e) &:= &\left\{
\begin{array}{ll}
f (e) +    g(e) \cdot  f(e_0), & \mbox{if $e\in E({\cal C})$} \\
f (e) , & \mbox{if $e\not\in E({\cal C})$}
\end{array} \right. .\vspace{-0.25cm}
\end{eqnarray*}
\item[{\sf(4)}] Set   $f  := f' $.
\end{itemize}
\end{itemize}
}}}
 \caption{The algorithm {\sf EliminateEvenCycles}, which consists of a single loop. The
precondition for the loop is the existence of an even cycle ${\cal C}$; so, upon termination, there will be
no even cycle for the output $f'$. (Note that if there are no loop iterations, then $f'=f$.) Step {\sf (1)}
chooses an edge $e_0$ on the cycle ${\cal C}$ on which $f$ is minimized, while Step {\sf (2)} assigns a sign
to each edge $e$ on ${\cal C}$. (Since ${\cal C}$ is an even cycle, alternating signs are possible.) The new
values for $f'$ are assigned in Step {\sf (3)}; note that $f'(e_0) = 0$.  Step {\sf (4)} prepares the input
($f$) for the next loop iteration. An example execution of the algorithm {\sf EliminateEvenCycles} is
illustrated in Figure~\ref{algorithm1_example}.}\label{EliminateEvenCycles_alg}
\end{figure}

\noindent To prove the claim,  we present
    the algorithm {\sf EliminateEvenCycles}  in Figure \ref{EliminateEvenCycles_alg}.

 \begin{proof}
We start with a first  invariant of the algorithm {\sf EliminateEvenCycles}:

\begin{figure}[p]
\epsfclipon
  \centerline{\hbox{
  \epsfxsize=13.6cm
  \leavevmode
   \epsffile{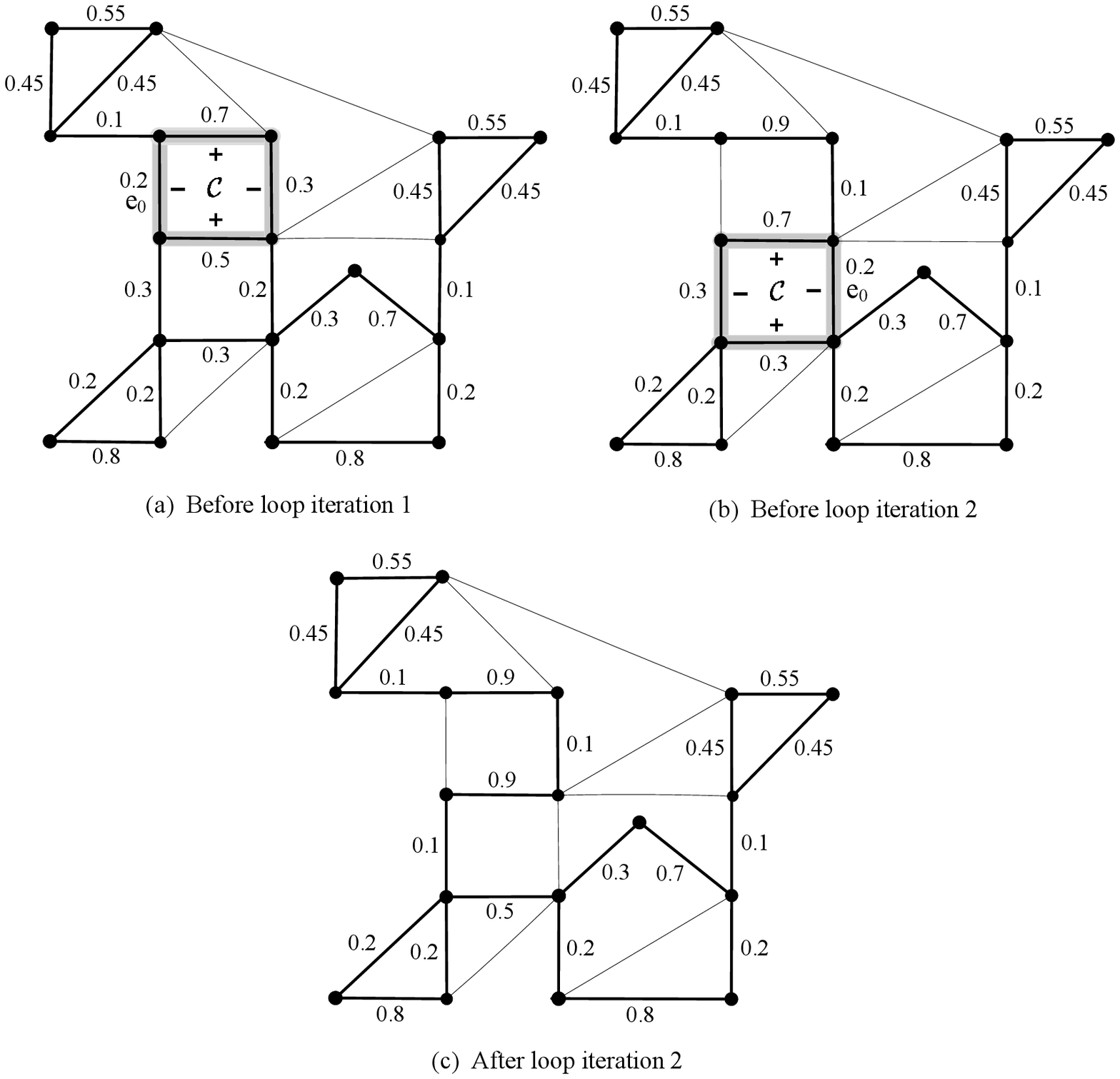}
  }
} \epsfclipoff \caption{An example execution of the   algorithm {\sf EliminateEvenCycles} on a graph   with a
Fractional Perfect Matching $f$; the execution terminates after two loop iterations. For each loop iteration,
edges in $E(f)$ are drawn thick;  edges on the  cycle ${\cal C}$ are drawn clouded. A number next to each
(thick) edge $e\in E(f)$ indicates the value $f(e)$; the sign of $g(e)$ is also indicated  for each (clouded)
edge  $e$ on  the cycle ${\cal C}$.}    \label{algorithm1_example}
\end{figure}

\begin{lemma}\label{Fractional Matchings with no even cycles_lemma1}
For each loop iteration of {\sf EliminateEvenCycles}, upon completion of {\it Step} {\sf (3)}, $f'$ is a
Fractional Matching equivalent to $f$.
\end{lemma}

\noindent Note that the input Fractional   Matching $f$ is already modified in the first (if any) loop
iteration of {\sf EliminateEvenCycles} (in Step {\sf (4)}), while the statement of  Lemma \ref{Fractional
Matchings with no even cycles_lemma1} refers to the input Fractional   Matching $f$. The proof of Lemma
\ref{Fractional Matchings with no even cycles_lemma1}  will use  the current Fractional Matching $f$;
 reference to the input $f$ will be restored in an inductive way upon completing   the proof.

\begin{proof}
Fix any loop iteration of {\sf EliminateEvenCycles}, upon completion of Step {\sf (3)}.  Consider  any vertex
$v\in V$. Then, by Step {\sf (3)},
 {\small
\begin{eqnarray*}
 \sum_{e\in E \mid v\in e } f'(e) &
  =  &      \sum_{  e  \in E({\cal C} ) \mid v\in e }  f' (e) +  \sum_{  e   \in E\backslash E({\cal C}) \mid v\in e }  f' (e)   \\
  &=   &   \sum_{  e  \in E({\cal C}) \mid v\in e }  f' (e) +  \sum_{  e   \in E\backslash E({\cal C})  \mid v\in e }  f  (e)
\end{eqnarray*}
}
If there is no edge  $e\in E({\cal C})$ such that $v\in e$, then $ \sum_{  e  \in E({\cal C} ) \mid v\in e }
\! f' (e)   \!  = \!      \sum_{  e \in E({\cal C} ) \mid v\in e } \! f  (e)  $ $=0$, and we are done. So,
assume otherwise. Since ${\cal C}$ is a cycle, there are (exactly) two   edges $e_1$, $e_2 \in E({\cal C})$
such that $\, v\in e_1$ and $v\in $ $ e_2$. Note that by Step {\sf (2)}, $g(e_1)+g(e_2)=0$. Hence, by Step
{\sf (3)}, {\small
  \begin{eqnarray*}
 \sum_{e\in E \mid v\in e } f'(e)
&=  &
        f' (e_1) + f'(e_2) + \sum_{  e  E\backslash E({\cal C}) \mid v\in e }  f  (e)   \\
 &  =   & f  (e_1) +g(e_1)\cdot f(e_0)  + f  (e_2) +g(e_2)\cdot f(e_0)     +
\sum_{  e  \in E\backslash E({\cal C}) \mid v\in e }  f (e)     \\
 &  =   & f  (e_1)  + f  (e_2) + \left( g(e_1) + g(e_2)\right) \cdot f(e_0)     +
\sum_{  e  \in E\backslash E({\cal C}) \mid v\in e }  f  (e)     \\
 &  = & f  (e_1)  + f  (e_2)     + \sum_{  e  \in E\backslash E({\cal C}) \mid v\in e }  f  (e)
    \\
 &  = &    \sum_{  e  \in E({\cal C}) \mid v\in e }  f  (e) + \sum_{  e  \in E\backslash E({\cal C}) \mid v\in e }  f(e) \\
 &  = &     \sum_{  e \in E  \mid v\in e }  f  (e) ,
\end{eqnarray*}
}which implies that $f'$ is equivalent to $f$. By Step {\sf (4)}, it follows inductively that $f'$ is
equivalent to the input Fractional Matching $f$.  Since $f$ is a Fractional Matching,  this implies that $f'$
is a Fractional Matching, and the claim follows. \qed\end{proof}

 \noindent We continue with a second   invariant of the algorithm {\sf EliminateEvenCycles}:

  \begin{lemma}\label{Fractional Matchings with no even cycles_lemma2}
For each loop iteration of {\sf EliminateEvenCycles}, upon completion of Step {\sf (3)}, {\sf (i)} $f'\subset
f$ and {\sf (ii)} the even cycle ${\cal C}$ is eliminated from $G(E(f'))$.
\end{lemma}
Similarly to Lemma \ref{Fractional Matchings with no even cycles_lemma1}, the statement of Lemma
\ref{Fractional Matchings with no even cycles_lemma2} (Condition {\sf (i)}) refers to the input Fractional
Matching $f$. The  proof of Lemma \ref{Fractional Matchings with no even cycles_lemma2} will use the current
Fractional   Matching $f$; reference to the input $f$ will be restored in an inductive way upon completing
the proof.

\begin{proof}
Fix any  loop iteration of {\sf EliminateEvenCycles}, upon completion of  Step {\sf (3)}. Consider any edge
$e\in E$. We proceed by case analysis.
\begin{itemize}
\item
Assume that    $e\not\in  E({\cal C})$. Then,   Step {\sf (3)}, implies that $e\in E(f')$ if and only if $e\in E(f)$.
\item
Assume that  $e\in E({\cal C})$. Then,  $e\in E(f)$; so,  it holds vacuously that if $e\in E(f')$, then $e\in E(f)$.
\end{itemize}
The case analysis implies that $f'\subseteq f$. Since $f'(e_0)=0$  while $f(e_0)> 0$, this implies that
$f'\subset f $. By Step {\sf (4)}, Condition {\sf (i)} follows  now inductively.
 Since $f'(e_0) =0 $,  edge $e_0$ is eliminated from $G(E(f'))$, so that  the even cycle
 ${\cal C}$ is eliminated from $G(E(f'))$ and Condition {\sf (ii)}
follows. \qed
\end{proof}

   Lemma~\ref{Fractional Matchings with no even cycles_lemma1}  and Lemma \ref{Fractional Matchings with no even cycles_lemma2}
(Condition {\sf (i)}) together imply  that the output $f'$ of algorithm {\sf EliminateEvenCycles}, which
contains no even cycle due to the loop precondition, is a Fractional Matching which is equivalent to and
contained in $f$. (By Lemma \ref{Fractional Matchings with no even cycles_lemma2} (Condition {\sf (ii)}),
containment is strict exactly when there is at least one loop iteration.)

Lemma~\ref{Fractional Matchings with no even cycles_lemma2} (Condition {\sf (i)} or {\sf (ii)}) implies that
at least one edge is eliminated from $f$ in each loop iteration and no edge is added. Hence,  there are at
most $|E|$ loop iterations. Note that each loop iteration takes $O(|E|)$ time. Since an even cycle is
computable in polynomial time, it follows that the algorithm {\sf EliminateEvenCycles} is polynomial time, and
we are done. \qed\end{proof}

\subsubsection{Elimination of Non-Isolated Odd Cycles}
 We prove:

\begin{proposition}\label{Fractional Matchings with odd cycles entire components}
Consider  a Fractional Perfect Matching $f$  with   no  even cycle.  Then, there  is a polynomial time
algorithm to transform  $f$ into an equivalent Fractional Perfect Matching $f'\subseteq f$
 with  no  non-isolated  odd cycle.
\end{proposition}

\noindent To prove the claim,   we present  the algorithm {\sf IsolateOddCycles} in  Figure
\ref{IsolateOddCycles_alg}.

\begin{figure}[p]\vspace{-0.5cm}
\vspace{-1.2cm}  \fbox{\parbox[t]{15.7cm}{ \small { {\bf Algorithm} {\sf IsolateOddCycles}\\
{\underline{\sc Input}}:  A graph $G=\langle V,E\rangle$              and  a Fractional    Perfect Matching
$f$  for $G$ with   no even cycle.\\
{\underline{\sc Output}}: An equivalent  Fractional Perfect Matching $f'\subseteq f$ with no non-isolated odd
cycle.
\begin{itemize}
\item[]$\,\,\, \hspace{-1.2cm}$ {\bf   While}   $G(E(f))$ contains  a non-isolated  odd cycle ${\cal C}$     {\bf
do}   \hspace{-0.3cm}
\begin{itemize}
\item[]$\,\,\, \hspace{-1.90cm}${\sf (1)} $\;$ Choose a  vertex $v_0\in  {\cal C} $ with
${\sf d}_{G(E(f))}(v_0)\geq 3$ and an
 edge  $ (v_0,v_1) \in E(f)$ with   $v_1 \not \in   {\cal C }  $.
\item[]$\,\,\, \hspace{-1.90cm}${\sf (2)} $\,$ {\bf While} $ E(f) $
includes  all edges from   $ E({\cal C}) \cup \{ (v_0,v_1) \}$ {\bf do}
\begin{itemize}
\item[]$\,\,\, \hspace{-1.9cm}${\sf (2/a)} $\,$ Choose  a   path $v_1, v_2, \cdots, v_r$ with $v_r = v_l$
for some $l \in 0 \cup  [r-2]$.
\item[]$\,\,\, \hspace{-1.9cm}${\sf (2/b)} $\,$ Define a  function $g:E({\cal C})\cup \{(v_{k}, v_{k+1})
\mid 0 \leq k \leq r-1 \} \rightarrow \left\{+1, -1, +\frac{\textstyle 1}{\textstyle 2}, -
  \frac{\textstyle 1}{\textstyle 2} \right\}$ with

  $\hspace{-2.5cm}  g(e )  =  \left\{
\begin{array}{ll}
 \!\!\! +\frac{\textstyle 1}{\textstyle 2} \mbox{ or}-\!\!\frac{\textstyle 1}{\textstyle 2} ,  &
 \mbox{if } e \in E({\cal C})
\mbox{ (alternately, starting with  $+\frac{\textstyle 1}{\textstyle 2}$   for an edge incident to
 $v_0$)}\\
 \!\!\! +1 \mbox{ or}   -\!1, \, \,  &
  \mbox{if } e=(v_k, v_{k+1}) \, \mbox{ for } \, 0\leq k \leq   l-1 \mbox{ with } l>0
\mbox{ (alternately, starting with } -\!1\mbox{)}\\
  \!\!\! +\frac{\textstyle 1}{\textstyle 2} \mbox{ or}-\!\!\frac{\textstyle 1}{\textstyle 2},
 &   \mbox{if } e=(v_k, v_{k+1}) \mbox{ for }  l\leq k \leq r-1
\mbox{ (alternately, starting with  a sign opposite} \\
& \mbox{to the sign of the last   value assigned by $g$)}
\end{array} \right. \hspace{-0.2cm} . $      \\
\item[]$\,\,\, \hspace{-1.8cm}${\sf (2/c)} $\,$
Choose an edge $e_0\in E({\cal C})\cup \{(v_k, v_{k+1})\mid 0 \leq k \leq r-1 \}$ that realizes the quantity
\begin{eqnarray*}
f_0 &: =&
\min\left\{  \min_{e\in E({\cal C})} \frac{\tx f(e)}{\tx |g(e)|},\,  \min_{\substack{l>0 \\
0\leq k \leq l-1} } \frac{\tx f((v_k,v_{k+1}))}{\tx |g((v_k,v_{k+1}))| }, \,   \min_{l\leq k  \leq r-1}
\frac{\tx f((v_k,v_{k+1}))}{\tx |g((v_k,v_{k+1}))|} \right\} ;
\end{eqnarray*}
\item[]$\,\,\,  \hspace{-1.8cm}${\sf (2/d)} $\,$    If $g(e_0)>0$, then set $g:=-g$.
 \item[]$\,\,\, \hspace{-1.8cm}${\sf (2/e)} $\,$   For each edge $e\in E$, set

  $
 f'(e )  :=\left\{
\begin{array}{ll}
f(e) +g(e) \cdot  \frac{\tx f(e_0)}{\tx |g(e_0)|},
&   \mbox{ if } e\in E({\cal C})\cup \{(v_k, v_{k+1})\mid 0 \leq k \leq r-1 \} \\
   f(e), &  \mbox{otherwise}    \nonumber
\end{array} \right. \!\!  . $
 \item[]$\,\,\, \hspace{-1.80cm}${\sf (2/f)} $\;$   Set $f:=f'$.
  \end{itemize}
 \end{itemize}
\end{itemize}
}}} \caption{The algorithm {\sf IsolateOddCycles}, which consists of an outer loop; the outer loop includes
an inner loop (Step {\sf (2)}). The precondition for the outer loop is the existence of a non-isolated odd
cycle; so, upon termination, there will be no non-isolated odd cycle for $f'$. (Note that if there are no
(outer) loop iterations, then $f'=f$.)  For Step {\sf (1)}, note that a vertex $v_0\in {\cal C}$ with ${\sf
d}_{G(E(f))}(v_0)\geq 3$ exists since ${\cal C}$ is non-isolated; $v_0$ has two incident edges from $\cal C$
and   at least one incident edge $(v_0, v_1)$ outside  $\cal C$. The precondition for the inner loop is the
inclusion of all edges from $E({\cal C})\cup \{(v_0,v_1)\}$ in $E(f)$; note that ${\cal C}$ remains a
(non-isolated) cycle (and the inner loop continues) as long as no such  edge  has been eliminated from $f$
(by   Step {\sf (2/e)}).
     For Step {\sf (2/a)}, note that  a path $v_1, \cdots, v_r$
with $v_r = v_l$ for some $l\in 0 \cup [r-2]$ exists  since $G(E(f))$ has  no pendant edges (by
Lemma~\ref{Fractional Perfect Matchings have no pendant edges});   this path together with ${\cal C}$ make a
{\it bicycle graph}.  For Step {\sf (2/b)}, note that Lemma~\ref{Fractional Matchings with odd cycles entire
components lemma1} implies that for any vertex $v_k$ with $0< k \leq r $, it holds that $v_k \not \in {\cal
C}\backslash \{v_0
\}$. 
So, Step {\sf (2/b)} assigns a signed coefficient to each edge  $e   \in E({\cal C}) \, \cup \{ (v_k,
v_{k+1}) \, \mid \, 0\leq k \leq r-1\} $. Step {\sf (2/c)}  chooses an edge $e_0$ on either the cycle ${\cal
C}$ or the outgoing path $v_1, v_2,\cdots, v_r$ that minimizes a certain quantity $f_0$ determined from $f$
and $g$; so, $f_0 = \frac{\tx f(e_0)}{\tx |g(e_0)|}$. Step {\sf (2/d)} adjusts $g$ so that $g(e_0)<0$. The
new values for $f'$ are  assigned in Step {\sf (2/e)}; note that $f'(e_0) =0$ (by Step {\sf (2/d)}). Step
{\sf (2/f)} prepares the input ($f$) for the next (inner) loop iteration. An example execution of the
algorithm {\sf IsolateOddCycles} is illustrated in Figure~\ref{algorithm2_example}. }
\label{IsolateOddCycles_alg}
\end{figure}


\begin{figure}[p]
\epsfclipon
  \centerline{\hbox{
 \epsfxsize=14.5cm
  \leavevmode
   \epsffile{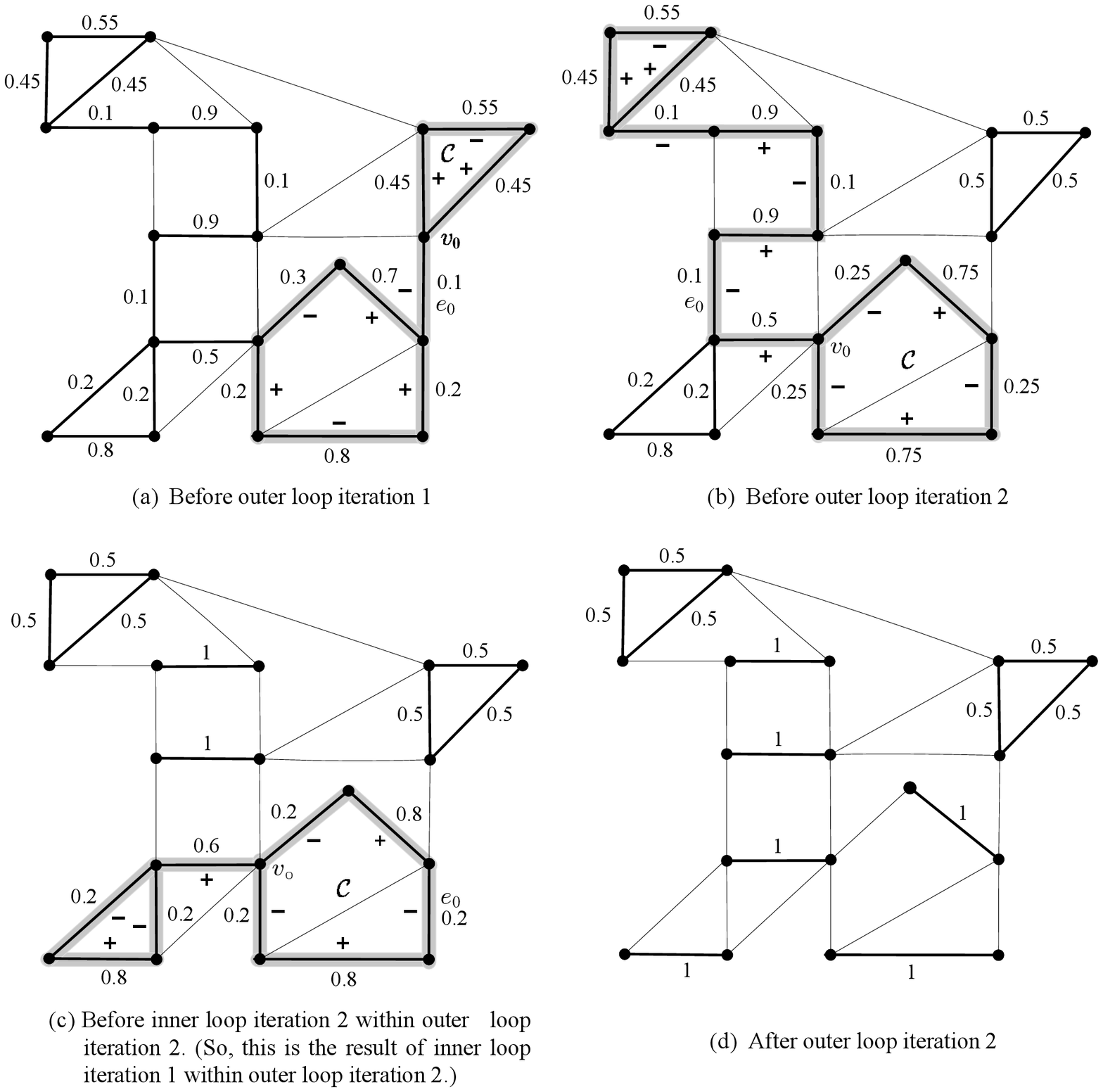}
  }
} \epsfclipoff \caption{An example execution of the   algorithm {\sf IsolateOddCycles} on the graph  with a
Fractional Perfect Matching $f$ (with no induced even cycle) from Figure \ref{algorithm1_example}(c). The
execution terminates after two outer loop iterations;  the first outer loop iteration incurs   one inner loop
iteration, while the second outer loop iteration incurs   two inner loop iterations. For each (inner or
outer) loop iteration, edges in $E(f)$ are drawn thick;    edges on the cycle ${\cal C}$ are drawn clouded. A
number next to each (thick) edge $e\in E(f)$ indicates $f(e)$; the sign of $g(e)$ (for each edge $e$ on the
cycle ${\cal C}$), the
  vertex $v_0$ and the edge
 $e_0$  are also indicated  for each iteration. } \label{algorithm2_example}
\end{figure}

 \begin{proof}
Since $G(E(f))$ has  no even cycle,   the cycle $v_l, \cdots, v_r =v_l$ determined in Step {\sf (2/a)} is
odd. We now prove a  preliminary property of the algorithm {\sf IsolateOddCycles}:

\begin{lemma}\label{Fractional Matchings with odd cycles entire components lemma1}
The     path $v_1, v_2, \cdots, v_r$ is disjoint from     ${\cal C}\backslash \{ v_0\}$.
\end{lemma}
\begin{proof}
By way of contradiction, assume that   there is a vertex $v_k$ with  $  k\in [r] $  such that $v_k\in {\cal
C}\backslash \{ v_0\} $. Since ${\cal C}$ has odd length, the vertices $v_0$ and $v_k$ partition ${\cal C}$
into two paths ${\cal C}_1$ and ${\cal C}_2$ of odd and even length, respectively. Consider the two
concatenations of the path $v_1, \cdots, v_k$ with ${\cal C}_1$ and ${\cal C}_2$, respectively; each of them
is a cycle in $G(E(f))$ and one of them  has even length. A contradiction. \qed\end{proof}

\noindent
   We start with a first invariant of the algorithm {\sf IsolateOddCycles}.

\begin{lemma}\label{Fractional Matchings with odd cycles entire components lemma2}
For each inner loop iteration in an outer loop iteration of {\sf IsolateOddCycles}, upon completion of {\em
Step} {\sf (2/e)},
 $f'$ is a Fractional Perfect Matching equivalent to
$f$.
\end{lemma}

\noindent Note that the input Fractional  Perfect Matching $f$ is already modified in the first inner  loop
iteration in the  first outer loop iteration of   {\sf IsolateOddCycles} (in Step {\sf (2/f)}). Reminiscent
of Lemma \ref{Fractional Matchings with no even cycles_lemma1}, the statement of Lemma \ref{Fractional
Matchings with odd cycles entire components lemma2} refers to the input Fractional Perfect Matching $f$. The
proof of Lemma \ref{Fractional Matchings with odd cycles entire components lemma2} will use the current
Fractional Perfect Matching $f$;
 reference to the input $f$ will be restored in an inductive way upon completing   the proof.

\begin{proof}
 The proof   consists of two technical claims. The first claim  determines the range of $f'$.
Fix any inner loop iteration in an outer loop iteration of   {\sf IsolateOddCycles}, upon    completion of
  Step {\sf (2/e)}. We prove:

\begin{claim}\label{f' is a Fractional Matching}
  $\Range(f')\subseteq [0,1]$.
\end{claim}
\begin{proof}
By Step {\sf (2/e)}, it suffices  to consider inductively an edge   $e $ from $ E({\cal C})\cup \{(v_k,
v_{k+1})\mid 0 \leq k \leq r-1 \}$. By Step    {\sf (2/f)}, it follows inductively that $f$ is a Fractional
Perfect Matching.

We first prove that  $f'(e)\geq 0$. By Step {\sf (2/e)},   it suffices to consider the case where $g(e)<0$,
so that $g(e) = -|g(e)|$. Then, by Step {\sf (2/e)} and the choice  of the edge $e_0$,
\begin{eqnarray*}
 f'(e ) &  = &      f(e )  -   | g(e ) |  \cdot
 \frac{\tx f(e_0 ) } {\tx |g(e_0)|}   \\
& \geq  &  0  ,
\end{eqnarray*}
as needed.

We now prove   that $f'(e)\leq 1$. By  Step {\sf (2/e)}, it suffices to  consider the case where $g(e)>0$, so
that $g(e) = |g(e)|$. We proceed by case analysis on whether there is an  edge $e'$ adjacent to $e$ such that
 $e$ and $e'$ are  either  both   on the cycle ${\cal C}$  or both on the path $ v_0,\cdots,  v_l $  (with  $ l>
0$) or both on the cycle $  v_l, \cdots, v_{r} =  v_l$.

Assume first that there is such an edge $e'$; clearly,  $|g(e')| = |g(e)|$. Then,  by Step {\sf (2/e)},
\begin{eqnarray*}
\lefteqn{f'(e )}\\
   = & f(e) +  |g(e )| \cdot   \frac{\tx  f(e_0)    } {\tx |g(e_0)|}   &  \\
 \leq  & 1 - f(e') + \frac{\tx  g(e )       } {\tx |g(e_0)|} \cdot f(e_0) & \mbox{(since $f$ is a Fractional Matching)}\\
 \leq  & 1  - |g(e ') |   \cdot \frac{\tx f(e_0)}{\tx |g(e_0)|} +  g(e )  \cdot \frac{\tx   f(e_0)}{\tx |g(e_0)|}
& \mbox{(by the choice of the edge $e_0$)}\\
 =  & 1  - |g(e )|\cdot \frac{\tx  f(e_0) }{\tx |g(e_0)|}  + g(e )  \cdot \frac{\tx   f(e_0)     } {\tx |g(e_0)|}
 & \\
= & 1,&
\end{eqnarray*}
as needed.

Assume now that there is no edge $e'$ adjacent to $e$ such that   $e$ and $e'$ are   either   both on the
cycle ${\cal C}$  or both on the path $ v_0, \cdots,  v_l $  (with  $ l> 0$)  or  on the cycle $  v_l,
\cdots, v_{r} = v_l$. Since both the cycle ${\cal C}$ and the cycle $  v_l, \cdots, v_{r} =  v_l$ are odd,
each of them includes at least three edges. It follows that edge $e$ lies neither on the cycle ${\cal C}$ nor
on the cycle $ v_l, \cdots, v_{r} =  v_l$. Hence, edge $e$ lies on the path $v_0, \cdots,  v_l $ (with  $ l>
0$). Since there is no edge $e'$ adjacent to $e $ on this path, it follows that $l=1$, so that $e=(v_0,v_1)$.
So, consider the edges $e_1$ and $e_2$ on the cycle ${\cal C}$ that are adjacent to $e$. By the choice of
$g$, it follows that  $|g(e_1)|+|g(e_2)| = |g(e)| $. Hence,

{\small \begin{eqnarray*}
\lefteqn{f'(e)}\\
=& f(e) + |g(e)|\cdot \frac{\tx f(e_0)}{\tx |g(e_0)| } &\mbox{(by Step {\sf (2/e)})}\\
\leq & 1- f(e_1) -f(e_2) +   |g(e)|\cdot \frac{\tx f(e_0)}{\tx |g(e_0)| } &\mbox{(since $f$ is a Fractional Matching)}  \\
\leq & 1- |g(e_1)|\cdot \frac{\tx f(e_0)}{\tx |g(e_0)| } -
 |g(e_2)|\cdot \frac{\tx f(e_0)}{\tx |g(e_0)| }  +   |g(e)|\cdot \frac{\tx f(e_0)}{\tx |g(e_0)| } &
\mbox{(by definition of $e_0$)}  \\
 =  & 1- \Big( |g(e_1)| +
 |g(e_2)| - |g(e)| \Big)    \cdot \frac{\tx f(e_0)}{\tx |g(e_0)| }   \\
 =& 1,
\end{eqnarray*}
} as needed. The proof is now complete. \qed
  \end{proof}

We continue with the second technical claim:
 \begin{claim}\label{f' is equivalent to f}
   $f'$ is equivalent to $f$.
\end{claim}
\begin{proof}
Consider any vertex $v\in V$. Then, by Step    {\sf (2/e)}), {\small
\begin{eqnarray*}
 \sum_{e\in E \mid v\in e } f'(e)
    &  =&  \sum_{  e   \in E({\cal C}) \, \cup \{ (v_k, v_{k+1}) \, \mid \,
0\leq k \leq r-1\} \,  \mid \, v\in e }  f' (e)          +
         \sum_{  e   \in E  \backslash ( E({\cal C}) \, \cup \{ (v_k, v_{k+1}) \, \mid \, 0\leq k \leq r-1\} ) \, \mid \,  v\in e }  f' (e)   \\
   &      =& \sum_{  e   \in E({\cal C}) \, \cup \{ (v_k, v_{k+1}) \, \mid \, 0\leq k \leq r-1\} \,  \mid \, v\in e }  f'
(e) + \sum_{  e   \in E  \backslash ( E({\cal C}) \, \cup \{ (v_k, v_{k+1}) \, \mid \, 0\leq k \leq r-1\} )
\, \mid \,  v\in e }  f (e)   .
\end{eqnarray*}
}
 If there is no edge $e\in E({\cal C}) \cup \{ (v_k, v_{k+1}) \, \mid \, 0\leq k \leq r-1\}$
such that $v\in e$, then {\small
\begin{eqnarray*}
 \sum_{  e   \in E({\cal C})\,  \cup \{ (v_k, v_{k+1}) \, \mid \, 0\leq k \leq r-1\}\, \mid \, v\in e }  f'
(e) \, &=& \, \sum_{  e    \in E({\cal C})\,  \cup \{ (v_k, v_{k+1}) \, \mid \, 0\leq k \leq r-1\}\,   \mid
\, v\in e }  f (e) \\
 & =& 0,
\end{eqnarray*}
}
 and we are done. So, assume otherwise. Note that by Step    {\sf (2/b)},
{\small
\begin{eqnarray*}
\sum_{ e \in E({\cal C})\, \cup \{ (v_k, v_{k+1}) \, \mid \, 0\leq k \leq r-1\} \, \mid \, v\in e }  g  (e)
&=& 0.
\end{eqnarray*}
}
 Hence, by Step {\sf (2/e)},
{\small
\begin{eqnarray*}
   \sum_{e\in E \mid v\in e } f'(e)
 & = &  \sum_{  e   \in E({\cal C}) \, \cup \{ (v_k, v_{k+1}) \, \mid \,
0\leq k \leq r-1\} \,  \mid \, v\in e }  f' (e)          +
         \sum_{  e   \in E  \backslash ( E({\cal C}) \, \cup \{ (v_k, v_{k+1}) \, \mid \, 0\leq k \leq r-1\} ) \, \mid \,  v\in e }   f  (e)    \\
     &     =&  \sum_{  e   \in E({\cal C}) \, \cup \{ (v_k, v_{k+1}) \, \mid \, 0\leq k \leq r-1\} \,  \mid \, v\in e }  \left( f(e) +g(e)\cdot  f_0
\right)
          + \hspace{-1.5cm} \sum_{  e   \in E  \backslash ( E({\cal C}) \, \cup \{ (v_k, v_{k+1}) \, \mid \, 0\leq k \leq r-1\} ) \, \mid \,  v\in e }  \hspace{-0.4cm} f  (e)    \\
       &   =&  \sum_{  e   \in E({\cal C}) \, \cup \{ (v_k, v_{k+1}) \, \mid \, 0\leq k \leq r-1\} \,  \mid \, v\in e }    f(e)
 +
f_0 \cdot \sum_{  e   \in E({\cal C}) \, \cup \{ (v_k, v_{k+1}) \, \mid \, 0\leq k \leq r-1\} \,  \mid \,
v\in
e }   g(e)    \\
& & +   \sum_{  e   \in E  \backslash ( E({\cal C}) \, \cup \{ (v_k, v_{k+1}) \, \mid \, 0\leq k \leq r-1\} )
\, \mid \,  v\in e }\hspace{-1cm}  f(e)
  \\
  &    =&      \hspace{2cm}    \sum_{  e   \in E  \, \mid \,  v\in e }  f  (e)    ,
\end{eqnarray*}
}
 which implies that $f'$  is equivalent to $f$.
By Step {\sf (2/f)}, it follows inductively that $f'$ is equivalent to the input Fractional Perfect Matching
$f$. \qed
\end{proof}

 Since $f$ is a Fractional Perfect Matching, Claims \ref{f'
is a Fractional Matching} and  \ref{f' is equivalent to f}  imply   together    that $f'$ is  a Fractional
Perfect Matching, and the claim follows.
 \qed\end{proof}

 We continue with a second   invariant of the algorithm {\sf IsolateOddCycles}:
\begin{lemma}\label{Fractional Matchings with odd cycles entire components lemma3}
For each outer   loop iteration of {\sf IsolateOddCycles}, {\sf (a)} for each inner loop iteration, upon
completion of Step {\sf (2/e)}, \myi $f' \subset f$, and   \myii  some edge from $E({\cal C}) \cup \{(v_k,
v_{k+1}) \mid 0\leq k \leq r-1\}$ is eliminated from $E(f')$,  and {\sf (b)} for the last inner loop
iteration, upon completion of Step {\sf (2/e)}, the non-isolated odd cycle   ${\cal C}$   is eliminated from
$G(E(f'))$.
\end{lemma}
Similarly to Lemma \ref{Fractional Matchings with odd cycles entire components lemma2}, the statement of
Lemma \ref{Fractional Matchings with odd cycles entire components lemma3} (Condition {\sf (a/i)}) refers to
the input Fractional Perfect Matching $f$. The proof of Lemma \ref{Fractional Matchings with odd cycles
entire components lemma3} will use the current Fractional Perfect Matching $f$; reference to the input $f$
will be restored in an inductive way upon completing the proof.

\begin{proof}
Consider  any  outer loop iteration. For Condition {\sf (a)}, consider    any  inner loop iteration within
this outer loop iteration, upon completion of Step {\sf (2/e)}. Consider any edge $e\in E$.
\begin{itemize}
\item
Assume that  $e\not\in E({\cal C})\cup \{ (v_k, v_{k+1}) \, \mid \, 0\leq k \leq r-1\}$. Then,   Step {\sf (2/e)} implies
that $e\in E(f')$ if and only if $e\in E(f)$.
\item
Assume that  $e\in E({\cal C})\cup \{ (v_k, v_{k+1}) \, \mid \, 0\leq k \leq r-1\}$. Then,  $e \in E(f)$; so,  it holds
vacuously that
 if $e\in E(f')$ then   $e\in E(f)$.
\end{itemize}
 The case analysis implies that $f'\subseteq f$. Since $f'(e_0)=0$  while $f(e_0)> 0$, this implies that
$f'\subset f $. By Step {\sf (2/f)}, Condition {\sf (a/i)} follows inductively.

Since $f'(e_0) =0 $,    $e_0$ is   eliminated from $E(f')$, so that some edge from   $E({\cal C}) \cup \{
(v_k, v_{k+1}) \, \mid \, 0\leq k \leq r-1\}$ is eliminated from $E(f')$, and Condition {\sf (a/ii)} follows.

To prove Condition {\sf (b)}, note that Condition {\sf (a/i)} implies that there is a last inner loop
iteration (and the outer loop terminates). So, consider the last inner loop iteration.  The precondition for
the inner loop implies that some edge from  $ E({\cal C}) \cup \{ (v_0,v_1) \}$ has been eliminated from
$E(f')$. Hence, the non-isolated odd cycle ${\cal C}$ is eliminated from $G(E(f'))$, and  Condition {\sf (b)}
follows.
\qed\end{proof}

  Lemma~\ref{Fractional Matchings with odd cycles entire components lemma2} and  Lemma
\ref{Fractional Matchings with odd cycles entire components lemma3} (Condition {\sf (a/i)}) together  imply
that the output $f'$ of the algorithm {\sf IsolateOddCycles}, which contains no non-isolated odd cycle due to
the outer loop precondition,   is a Fractional Perfect Matching which
  is equivalent to and   contained in $f$. (By Lemma \ref{Fractional Matchings with odd cycles entire components lemma3}
(Condition {\sf (b)}), containment is strict exactly when there is at least one outer loop iteration.)

Lemma~\ref{Fractional Matchings with odd cycles entire components lemma3} implies that at least one edge is
eliminated from $f$ in each inner  loop iteration and no edge is added. Hence, there are at most $|E|$ inner
loop iterations  in all  outer loop iterations. Note that each iteration of the inner loop takes $O(|E|)$
time. Since an odd cycle   is computable  in polynomial time, it follows that the algorithm {\sf
IsolateOddCycles} is polynomial time, and we are done. \qed\end{proof}

\remove{ We note that the arguments used in the proof of Propositions \ref{Fractional Matchings with no even
cycles} and \ref{Fractional Matchings with odd cycles entire components} are similar to the following result:
Such a Fractional Matching is shown  there to satisfy the properties of the Fractional (Perfect) Matching
$f'$ of Propositions \ref{Fractional Matchings with no even cycles} and \ref{Fractional Matchings with odd
cycles entire components}.  However, we remark that that proof does not show  how to compute such a
Fractional Matching. }

 \subsubsection{Recap}

We are now ready to prove:

\begin{proposition}\label{min Fractional Perfect Matching prop}
Consider   a  Fractional Perfect Matching $f$. Then, there is a polynomial time algorithm to transform $f$
into  an equivalent    Fractional Perfect Matching $f'  \subseteq f$ with
  only       single edges and odd cycles.
\end{proposition}

\begin{figure}[h]
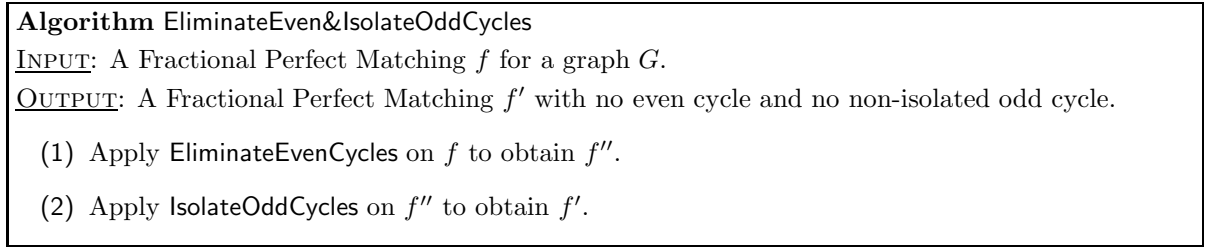

\fbox{\parbox{15.5cm}
 { \small
 { {\bf Algorithm} {\sf EliminateEven\&IsolateOddCycles}

{\underline{\sc Input}}:  A    Fractional Perfect Matching $f$ for a graph $G$.

{\underline{\sc Output}}:  A  Fractional Perfect Matching $f'$
                           with no even cycle and no non-isolated odd  cycle.
\begin{itemize}

\item[{\sf (1)}]  Apply   {\sf EliminateEvenCycles} on $f$ to obtain $f''$.
\item[{\sf (2)}]  Apply  {\sf IsolateOddCycles}   on $f''$ to obtain $f'$.
 \end{itemize}
}}}
 \caption{The algorithm {\sf EliminateEven\&IsolateOddCycles}, incorporating the algorithms
 {\sf EliminateEvenCycles}  and   {\sf IsolateOddCycles} from Figures \ref{EliminateEvenCycles_alg} and
  \ref{IsolateOddCycles_alg}, respectively.}\label{EliminateEvenOddCycles_alg}
\end{figure}

\noindent To prove the claim,  we present
    the algorithm {\sf EliminateEven\&IsolateOddCycles} in Figure
    \ref{EliminateEvenOddCycles_alg}. The algorithm is the sequential cascade of the algorithms
{\sf EliminateEvenCycles}  and   {\sf IsolateOddCycles} from Figures \ref{EliminateEvenCycles_alg} and
  \ref{IsolateOddCycles_alg}, respectively.

%
\begin{proof}
By Proposition~\ref{Fractional Matchings with no even cycles}, $f''\subseteq f $ is a  Fractional Perfect
Matching with no even cycle, which is equivalent to $f$. By Proposition~\ref{Fractional Matchings with odd
cycles entire components}, $f'\subseteq f''  $ is a Fractional Perfect Matching with no non-isolated odd
cycle, which is equivalent to $f''$. It follows that
 {\sf (1)} $f'$ is equivalent to $f$ and   $f'\subseteq f $, and  {\sf (2)}
  $f'$ has no even cycle and no  non-isolated odd cycle. Since $f$ is a Fractional Perfect Matching,
Condition  {\sf (1)}  implies that $f'$ is a Fractional Perfect Matching; by Lemma \ref{Fractional Perfect
Matchings have no pendant edges}, this implies that each acyclic component of $G(E(f'))$ is a single  edge.
 By Condition {\sf (2)},  it follows  that each cyclic component of $G(E(f'))$ is an (isolated) odd cycle.
It follows that
   $f'$ consists of single edges and odd cycles,
and the proof is complete.
 \qed\end{proof}


\subsection{$\delta$-Partitionable Fractional Perfect Matchings}
\label{m-Partitionable Fractional Perfect Matchings Section}

\subsubsection{Definition and Preliminaries}
 \noindent  We   introduce   a
special class of Fractional Perfect Matchings:
 \begin{definition}
\label{well partitioned FPM} Fix an integer $\delta \geq 1$. A Fractional Perfect Matching  $f: E \rightarrow
\mathbb{R}$ is {\em \emph{\textbf{$\delta$-Partitionable}}} if the edge set $E(f)$ can be partitioned into
$\delta$ (non-empty,)  vertex-disjoint partites $E_{1}, \cdots, E_{\delta}$ so that for each partite  $E_{j}$
with $j\in[\delta]$,  $\sum_{e \in E_{j}}   f(e) = \frac{\textstyle |V|}      {\textstyle 2\, \delta}$.
\end{definition}

\noindent Note that  a   $1$-Partitionable Fractional
 Perfect Matching is a  Fractional Perfect Matching. Hence, the  decision problem for a  $1$-Partitionable Fractional
 Perfect Matching   is  solved
in polynomial time. Note also that the restriction of a $\delta$-Partitionable Fractional Perfect Matching to
each partite $E_j$ with $j\in [ \delta ]$ is a Fractional Perfect Matching; so, for each partite $E_j$ with
$j\in[\delta]$, for each vertex $v\in V(E_j)$, $\sum_{e\in E \mid v\in e} f(e) =1$. Since the partites are
vertex-disjoint, this implies that for each partite $E_j$ with $j\in [\delta]$, for each vertex $v\in
V(E_j)$, $\sum_{e\in E_j  \mid v\in e }f(e)  =1 $.
 We now prove  a necessary condition for a $\delta$-Partitionable Fractional
Perfect Matching:

\begin{proposition}\label{m-PFPMs size of Vj}
Consider a   $\delta$-Partitionable Fractional Perfect Matching  $f $. Then,  for each partite $E_j$ with
 $j\in[\delta]$, $
|V(E_j)  | =
    \frac{\textstyle |V|}      {\textstyle  \delta} \, $.
\end{proposition}
\begin{proof}
Fix a partite $E_j$ with $j\in[\delta]$.  Then,
{\small
\begin{eqnarray*}
 \sum_{e\in E_j}f(e)
& =& \frac{\tx 1}{\tx 2} \sum_{v\in V(E_j) } \Big(
\sum_{e\in E_j \mid v\in e  } f(e) \Big)  \\
& =& \frac{\tx 1}{\tx 2} \sum_{v\in V(E_j) } 1   \\
& = &\frac{\textstyle |V(E_j)  |}{\textstyle 2}  .
\end{eqnarray*}
}
 \noindent Since $f$ is $\delta$-Partitionable, it follows that
\begin{eqnarray*}
 |V(E_j)  | &= &   2\cdot \sum_{e\in E_j} f(e)  \\
             & = &  2 \cdot  \frac{\textstyle |V|}      {\textstyle 2\, \delta}\\
            &= &    \frac{\textstyle |V|}      {\textstyle  \delta},
\end{eqnarray*}
 as needed.
  \qed
\end{proof}

\noindent     Proposition \ref{m-PFPMs size of Vj} immediately implies:
\begin{corollary}
 \label{Partitionable Fractional Perfect Matchings divides lemma} If $G$ has a $\delta$-Partitionable
  Fractional  Perfect Matching,   then  $\delta$ divides $|V|$, so that
$\delta\leq \frac{\tx |V|}{\tx 2}$.
 \end{corollary}

\noindent We observe that the equality in the necessary condition $\delta \leq \frac{\tx |V|}{\tx 2}$ in Corollary
\ref{Partitionable Fractional Perfect Matchings divides lemma} is not always necessary:

\begin{proposition}\label{counterexample3}
There is  a graph $G$ and an integer $\delta$ such that   $G$ has a $\delta$-Partitionable Fractional Perfect
Matching while $\delta< \frac{\tx |V|}{\tx 2}$.
\end{proposition}
\begin{proof}
Consider the cycle   graph ${\cal C}_3$ and fix $\delta=1$. Clearly,   $\beta'({\cal C}_3)=2$ so that $\delta
< \frac{\tx |V|}{\tx 2}$. Consider the function $f: E({\cal C}_3)\rightarrow [0,1]$ with  $f(e) =\frac{\tx
1}{ \tx 2 }$ for each edge $e\in E({\cal C}_3)$. Clearly,  $f$ is an  $1$-Partitionable Fractional Perfect
Matching, and the claim follows.
 \qed\end{proof}

We  finally  prove that the equivalence relation on Fractional Perfect Matchings preserves
$\delta$-Partitionability under a certain   containment assumption:
\begin{proposition}\label{m-PFPMs equivalence}
Consider a   $\delta$-Partitionable Fractional Perfect Matching  $f $ and  an equivalent  Fractional Perfect
Matching   $f'\subseteq f$. Then,  $f'$ is   $\delta$-Partitionable.
 \end{proposition}
\begin{proof}
Consider the $\delta$ (non-empty,) vertex-disjoint partites $E_{1}, \cdots, E_{\delta}$. Define edge sets
$E'_1, \cdots , E'_{\delta}$ so  that for each $j\in[\delta]$,  $E'_j  = \{ e\in E_j \mid f'(e) >0 \}$. Since
$f'\subseteq f$, it follows that for each $j\in [\delta] $, $E'_j\subseteq E_j$. This implies that the
collection $E'_{1}, \cdots, E'_{\delta}$ partitions $E(f')$. Since the partites $E_{1}, \cdots, E_{\delta}$
are vertex-disjoint, this also implies that the edge sets $E'_{1}, \cdots, E'_{\delta}$ are vertex-disjoint;
so call them partites. Fix any partite $E'_j$ with $j\in[\delta]$. Then,

{\small
\begin{eqnarray*}
\lefteqn{\sum_{e\in E'_j} f'(e) }\\
=&  \sum_{e\in E_j} f'(e)  &\mbox{(since $f'(e) = 0$  for each $e\in E_j \backslash E'_j$)}\\
 =&  \frac{\tx 1 } {\tx 2} \cdot \sum_{v \in V(E_j) }  \sum_{e\in E_j  \mid v\in e}   f'(e)  & \\
 =&  \frac{\tx 1 } {\tx 2} \cdot \sum_{v \in V(E_j) }  \sum_{e\in E   \mid v\in e}   f' (e)   &\\
 =&  \frac{\tx 1 } {\tx 2} \cdot \sum_{v \in V(E_j) }  \sum_{e\in E   \mid v\in e}   f  (e)   &
\mbox{(since $f $ and $f'$ are equivalent)}\\
=&  \frac{\tx 1 } {\tx 2} \cdot  \sum_{v \in V(E_j) } 1 & \mbox{(since $f$ is Perfect)}\\
=&  \frac{\tx 1 } {\tx 2} \cdot  |V(E_j)|    & \\
=&    \frac{\tx |V| } {\tx 2 \delta}& \mbox{(by Proposition \ref{m-PFPMs size of Vj})}.
\end{eqnarray*}
}
Hence,   $f'$ is $\delta$-Partitionable, as needed. \qed
\end{proof}

\subsubsection{Characterization}


  We show:

 \begin{proposition}\label{min Fractional Perfect Matching equivalence} A graph $G$ has a $\delta$-Partitionable
 Fractional Perfect Matching if and only if $E$ contains    a collection of  $\delta$ (non-empty) vertex-disjoint edge sets
$E_1, \cdots, E_{\delta}$
   such that {\sf (1)}    $\bigcup _{j\in [\delta]} E_j $ is an Edge Cover,  and  {\sf (2)}
for each edge set   $E_j$ with  $j\in [\delta] $,
  {\sf (i)}    $E_j$ consists   of
single edges and  odd     cycles, and {\sf (ii)} $|V(E_j) | = \frac{\textstyle |V|}{\textstyle \delta}$.
\end{proposition}

\noindent Note that the edge sets $E_1, \cdots,E_{\delta} $ need not form a partition of $E$; in contrast, by
Condition  {\sf (1)},  the induced vertex sets $V(E_1), \cdots, V(E_{\delta})$ are required to form a
partition of $V$.

\begin{proof}
Assume first that $G$ has  a $\delta$-Partitionable Fractional Perfect Matching  $f$. By Proposition \ref{min
Fractional Perfect Matching prop}, there is an equivalent Fractional Perfect Matching  $f'\subseteq f$ with
only single edges and odd cycles. Since $f$ is $\delta$-Partitionable and $ f'\subseteq f $, Proposition
\ref{m-PFPMs equivalence}
  implies that $f'$ is $\delta$-Partitionable. So, the edge set $E(f')$ can be partitioned into $\delta$
(non-empty), vertex-disjoint partites  $E_1, \cdots, E_{\delta}$ so that for each partite $E_j$ with
$j\in[\delta]$,
 $\sum_{e \in E_{j}}   f(e) = \frac{\textstyle |V|}      {\textstyle 2\, \delta}$.

Consider now the (vertex-disjoint) edge sets $E_1, \cdots, E_{\delta}$.   Since $f'$ is a Fractional Perfect
Matching, $E(f')$ is an Edge Cover; so,   $ \bigcup_{j\in [\delta]}
   E_j      $ is an Edge Cover. Consider now any edge set $E_j$ with $j \in [\delta]$.
Since $f'$  consists of single edges and odd cycles, Condition {\sf (2/i)} follows;  since $f'$ is a Fractional
Perfect Matching, Condition {\sf (2/ii)} follows from Proposition \ref{m-PFPMs size of Vj}.

Assume now that $E$ contains a collection  of $\delta$ (non-empty,) vertex-disjoint edge sets $E_1, \cdots,
E_{\delta}$ such that {\sf (1)} $\bigcup_{j\in [\delta]} E_j $ is an Edge Cover, and {\sf (2)}  for each edge
set $E_j$ with $j\in [\delta] $, {\sf (i)} $E_j$ is a collection of single edges and    odd cycles, and {\sf
(ii)} $|V(E_j) | = \frac{\textstyle |V|}{\textstyle \delta}$. We shall prove  that $G$ has a
$\delta$-Partitionable Fractional Perfect Matching   $f $. The proof is constructive. Define the function
$f:E\rightarrow [0,1]$ with
\begin{eqnarray*}
f(e) & = & \left\{
\begin{array}{ll}
 1, & \mbox{if  $e\in E_j$   with  $j\in[\delta]$ and    $E_j$ is a single edge} \\
\frac{\textstyle 1}{\textstyle 2},  & \mbox{if $e\in E_j$   with  $j\in[\delta]$  and   $E_j$  is an odd
cycle}\\
0, & \mbox{if $e\in E \backslash \bigcup_{j\in [\delta]} E_j$}
\end{array} \right. .
\end{eqnarray*}

To prove that $f$ is a Fractional Perfect Matching, consider any vertex $v\in V$. Since     $\bigcup_{j\in
[\delta]}  E_j
 $ is an Edge Cover,  this implies that      $v\in V(E_{j})$ for some     $j\in [\delta]$.
There are two   cases:
\begin{itemize}
\item Assume that $E_j$ is a single edge $e_j$. Then, by construction, $\sum_{e\in E \mid v\in e} f(e)= f(e_j) = 1$.

\item Assume that $E_j$ is an (isolated) odd cycle, so that $v = e_j\cap e'_j$ for a pair of consecutive edges $e_j,e'_j$ on the cycle.
Then, by construction,
 $\sum_{e\in E \mid v\in e} f(e)= f(e_j) +f(e'_j)= \frac{\tx 1}{\tx 2} +\frac{\tx 1}{\tx 2} = 1$.
\end{itemize}
The case analysis implies that $f$ is a Fractional Perfect Matching.  To prove  that $f $ is
$\delta$-Partitionable,  consider the partites $E_1, \cdots, E_{\delta}$.
  Fix any partite $E_j$ with $j \in [\delta]$. Then,
{\small \begin{eqnarray*}
\lefteqn{\sum_{e\in E_j}f (e)}\\
 =& \frac{\tx 1}{\tx 2} \sum_{v\in V(E_j) } \sum_{e\in E_j \mid v\in e } f (e) &   \\
 =& \frac{\tx 1}{\tx 2} \sum_{v\in V(E_j) } 1 & \mbox{(since $f$ is     Perfect)} \\
 = &  \frac{\tx 1}{\tx 2}   |V(E_j)  |  &\\
  = &\frac{\textstyle |V   |}{\textstyle 2 \, \delta}  & \mbox{(by Condition  {\sf (2/ii)})}.
\end{eqnarray*}
}
\noindent    It follows that $f$ is        $\delta$-Partitionable, and the proof is now complete.
  \qed\end{proof}

  We observe  an interesting special case  of Proposition \ref{min Fractional Perfect Matching
equivalence}:
\begin{proposition} \label{m-fractional perfect matching for perfect graphs}
  A graph  $G$ has a $\frac{\textstyle |V|}      {\textstyle 2}$-Partitionable Fractional Perfect
Matching  if and only if  $G$ is   Perfect-Matching.
\end{proposition}
\begin{proof}
Assume first that $G$ has a $\frac{\textstyle |V|} {\textstyle 2}$-Partitionable Fractional Perfect Matching.
    Proposition  \ref{min Fractional Perfect Matching equivalence} implies that    $E$ contains
 a collection
 of  $\frac{\textstyle |V|}      {\textstyle 2}$  (non-empty,) vertex-disjoint edge sets
$E_1, \cdots, E_{\frac{\textstyle |V|}      {\textstyle 2}}$    such that {\sf (1)} $ \bigcup
_{j\in\Big[\frac{\tx |V|}{\tx 2}\Big]}E_j$ is an Edge Cover,  and  {\sf (2)} for each edge set $E_j$ with
$j\in\Big[\frac{\tx |V|}{\tx 2} \Big] $, {\sf (i)}   $E_j$ consists of single edges and odd cycles, and {\sf
(ii)} $|V(E_j) | =2$. By Conditions {\sf (2/i)} and {\sf (2/ii)}, it follows that each edge set  $E_j$ with
$j\in [\delta]$ is a single edge. Hence, the collection of the (vertex-disjoint) edge sets  is a Matching. By
Condition {\sf (1)}, this implies  that the collection of the  edge sets  is   a Perfect Matching.

  \noindent $\,$Assume    that  $G$  is  Perfect-Matching   with a Perfect Matching   $M$. Consider the {\it indicator function}
 $f:E\rightarrow  \{ 0,1\}$  for $M$,  where $f (e)  = 1 $ if and only if $e\in M$; so,  
   $f$ is a Fractional Perfect
Matching, and   it remains to show that  $f$ is      $\frac{\textstyle |V|}      {\textstyle 2} \,
$-Partitionable.
 For  each edge $e_j\in M$ with
$j \in\Big[ \frac{\textstyle |V|}      {\textstyle 2} \Big]$, define the partite  $E_j := \{e_j\}$. Since $M$
is a Perfect Matching, the partites  are vertex-disjoint. So,    for each   $E_j$  with
$j\in[\delta]$, $\sum_{e\in E_j}f(e)
  =     1  = \frac{\textstyle |V|} {\textstyle 2 \, \frac{\textstyle |V|}{\tx 2}}
 $, and this   completes  the proof.
\qed
\end{proof}

\subsubsection{Complexity}
\noindent We  define a natural decision problem  about  $\delta$-Partitionable
  Fractional  Perfect Matchings:
 \vspace{0.1cm} \hrule
    \noindent{\sf  $\delta$-PARTITIONABLE FPM}  \\
\noindent{\sc Instance}: A graph $G = \left\langle V, E\right\rangle$ and  an integer  $\delta$ which
  divides $|V|$.  \\
\noindent{\sc Question}: Is there a $\delta$-Partitionable Fractional Perfect Matching for $G$?
\vspace{0.1cm} $\;$  \hrule   \vspace{0.2cm}

\noindent Note that the restriction to instances for which   $\delta$ divides $|V|$ is inherited from
Corollary~\ref{Partitionable Fractional Perfect Matchings divides lemma} in order to exclude the
non-interesting   instances.

\noindent Proposition \ref{m-fractional perfect matching for perfect graphs} identifies a tractable special
case of {\sf $\delta$-PARTITIONABLE FPM} (namely, {\sf $\frac{\textstyle |V|}{\textstyle 2}$-PARTITIONABLE
FPM}). We shall use Proposition \ref{min Fractional Perfect Matching equivalence} to show that in the general
case where $\delta$ is arbitrary,  {\sf $\delta$-PARTITIONABLE FPM} is ${\cal NP}$-complete. To do so, we
shall observe an interesting relation of some other  (intractable) special case of the problem  to a well
known graph-theoretic problem: \vspace{0.3cm} \hrule
    \noindent{\sf  PARTITION INTO TRIANGLES}  \\
\noindent{\sc Instance}: A graph $G = \left\langle V, E\right\rangle$
with $|V|=3\delta$ for some integer $\delta$.  \\
\noindent{\sc Question}: Can  $V$ be partitioned into $\delta$ disjoint vertex sets $V_1, \cdots,
V_{\delta}$, each containing exactly three  vertices, such that for each $j\in[\delta]$,   $E(V _j  )$   is a
triangle? \vspace{0.3cm} $\;$ \hrule  \vspace{0.3cm}

\noindent This problem is known to be ${\cal NP}$-complete~\cite[GT11, attribution to (personal communication
with) Schaefer]{GJ79}. (This restriction to graphs $G=\langle V, E \rangle $ with $|V| = 3\delta$ is made in
order to exclude the non-interesting instances.) To prove that {\sf $\delta$-PARTITIONABLE FPM} is ${\cal
NP}$-complete (for an arbitrary $\delta$), we consider the  special case of it with $\delta = \frac{\tx
|V|}{\tx 3}$: \vspace{0.3cm} \hrule
    \noindent{\sf
$\frac{\tx |V|}{\tx 3}$-PARTITIONABLE FPM}  \\
\noindent{\sc Instance}: A graph $G = \left\langle V, E\right\rangle$ with $|V| = 3\delta$ for some integer  $\delta$.  \\
\noindent{\sc Question}: Is there a $\frac{\textstyle |V|}{\textstyle 3}$-Partitionable Fractional Perfect
Matching for $G$?   \vspace{0.3cm}    $\;$ \hrule \vspace{0.3cm}

 \noindent
(The restriction to graphs $G=\langle  V, E\rangle$ with $|V|=3\delta$ is necessary since $\delta = \frac{\tx
|V|}{\tx 3}$ is an integer.)
 To prove that this special case is intractable,
 we prove that it coincides with    {\sf  PARTITION INTO TRIANGLES}: it incurs  an identical set  of   positive instances. We prove:
\begin{proposition}\label{Fractional Perfect Matching theo NPcompleteness1}
 {\sf
$\frac{\tx |V|}{\tx 3}$-PARTITIONABLE FPM}   $=$ {\sf  PARTITION INTO TRIANGLES}
\end{proposition}
\begin{proof}
Consider a graph $G=\langle V, E\rangle $ with $|V| =  3\delta$ for some integer $\delta$. Assume first that
$G$ is a positive instance for {\sf $\frac{\tx |V|}{\tx 3}$-PARTITIONABLE FPM}. By Proposition~\ref{min
Fractional Perfect Matching equivalence}, $E$ contains a   collection  of $\frac{\tx |V|}{\tx 3}$
 (non-empty,) vertex-disjoint edge sets $E_1,\cdots , E_{\frac{ \tx |V|}{ \tx 3}}$ such  that {\sf (1)}  $\bigcup _{j\in
\left[\frac{\tx |V|}{\tx 3}\right]} E_j $ is an Edge Cover, and {\sf (2)}
 each edge set $E_j$   consists of single edges and odd
cycles with    $|V(E_j) |   = 3$.  It follows that each edge set  $E_j$ with $j\in\left[ \frac{\tx  |V|}{\tx 3} \right]$
is a
  triangle. This implies that   $G$ is a positive instance for    {\sf PARTITION INTO TRIANGLES} (with vertex sets
 $ V(E_1), \cdots, V(E_{\frac{\tx |V|}{\tx 3} }) $).

Assume now that   $G$ is a positive instance for  {\sf PARTITION INTO TRIANGLES}. Consider the corresponding
partition of $V$
  into $\delta = \frac{\tx |V|}{\tx 3} $ disjoint vertex sets $V_1 , \cdots, V_{\frac{\tx |V|}{\tx 3} } $.  This partition induces a
corresponding partition of $E$ into a collection  of $\frac{ \textstyle |V|}{ \textstyle 3} $ vertex-disjoint
partites $E_1,\cdots , E_{\frac{ \tx |V|}{ \tx 3}}$, where each partite $E_j$ is a single triangle.
Proposition~\ref{min Fractional Perfect Matching equivalence} implies that $G$ has a  $\frac{ \textstyle
|V|}{  \textstyle 3} $ -Partitionable Fractional Perfect Matching. Hence, $G$ is a positive  instance for
{\sf $\frac{\tx |V|}{\tx 3}$-}{\sf PARTITIONABLE FPM}, and we are done. \qed
\end{proof}

\noindent By Proposition \ref{Fractional Perfect Matching theo NPcompleteness1},     it follows that  {\sf
$\frac{\tx |V|}{\tx 3}$-PARTITIONABLE FPM} is ${\cal NP}$-complete. Since  {\sf $\frac{\tx |V|}{\tx
3}$-PARTITIONABLE FPM} is a special case of {\sf $\delta$-PARTITIONABLE FPM}, this implies:

\begin{corollary}
\label{m-partitionable FPM is NP complete}
  {\sf $\delta$-PARTITIONABLE FPM} is ${\cal NP}$-complete.
\end{corollary}
\section{A Combinatorial Lemma}\label{Combinatorics Section}
 In this section, we   prove a combinatorial lemma that will be useful  later.

  For a
{\it probability} $x$, we define two {\it probability literals,} or {\it  literals} for short: the {\it
positive} literal $x$ and the {\it negative} literal $\bar{x} = 1-x$. A {\it probability product}, or {\it
product} for short, is a product of probability literals $x_{1} \cdots x_{n}$ for any $n \geq 1$; we  adopt
the convention that $x_{\ell_1} \cdots x_{\ell_2}=1$ whenever $\ell_2< \ell_1$. A {\it constant} probability
product is the trivial one which equals to $1$ and has no literals. The {\em expansion} of a probability
product is obtained when substituting each negative literal $\overline{x}$ with $1-x$. So, an expansion
contains positive literals and no negative literals.

The probability product $x_1  \cdots  x_n$ is {\it positive} if all its probability literals are positive.
More generally,  for any integer $\ell \leq n$, the probability product $x_{1} \cdots x_{n}$ is {\it
$\ell$-positive} if exactly $\ell$ of its probability literals are positive; so, an $n$-positive probability
product is a positive probability product.  For each $\ell \in[ n]$, denote as ${\sf Pos}_{\ell}(x_{1},
\ldots, x_{n})$ the collection of all $\ell$-positive probability products with literals defined from the
probabilities $x_{1}, \ldots, x_{n}$. We prove a combinatorial identity for sums of   probability products:

\begin{lemma}
\label{combinatorial lemma} For each integer $n\geq 2$, { \small
\begin{eqnarray*}
      \sum_{\ell \in [n]}        \frac{1}             {\ell}\,    \cdot     \sum_{x_{2} \ldots   x_{n}
       \in              {\sf Pos}_{\ell-1}(x_{2}, \ldots, x_{n})}          x_{2} \cdots   x_{n}
& = & \sum_{\ell \in [n]}         (-1)^{\ell-1}\,         \cdot         \frac{1}              {\ell}\, \cdot
\sum_{x_{2}
          \ldots               x_{\ell}               \in               {\sf Pos}_{\ell-1}(x_{2}, \ldots, x_{n}) }
         x_{2} \cdots x_{\ell}\, .
\end{eqnarray*}
}
\end{lemma}
Note that the right-hand side ({\sf RHS}) is a weighted sum of positive probability products, with   weights
of alternating signs.  In   contrast, the left-hand side ({\sf LHS}) is a weighted sum of arbitrary (not
necessarily positive) probability products, with positive weights; an $(\ell -1)$-positive product in the
{\sf LHS} is multiplied by $\frac{\tx 1}{\tx \ell}$.

\begin{proof}
It suffices to establish that for each $\ell\in[n]$, each (positive) probability product   $x_{2} \cdots
x_{\ell} \in {\sf Pos}_{\ell-1}(x_{2}, \ldots, x_{n})$ from the  {\sf RHS}  appears in the  expansion of the
{\sf LHS} with the same coefficient. We proceed by case analysis on $\ell$.
\begin{itemize}
\item Assume first that $\ell=1$, and fix  any product  $x_2\cdots x_\ell\in {\sf Pos}_{\ell-1}(x_{2},
\ldots, x_{n})$ with $\ell=1$ in the {\sf RHS}.  By convention, there is only one such product and it is
constant.
  The coefficient of this product is   $(-1)^{1-1} \cdot \frac{1}      {1} = 1$.

In the {\sf LHS}, the only constant term is the constant term in the sum
\begin{eqnarray*}
\left.  \sum_{x_{2} \cdots x_{n} \in {\sf Pos}_{\ell-1}(x_{2}, \ldots, x_{n})}     x_{2} \cdots  x_{n}
\right|_{\ell=1} &=& \overline{x}_{2} \cdots \overline{x}_{n} .
\end{eqnarray*}

 Clearly, this constant term is $1$ and its coefficient is
$\frac{1}      {1} = 1$. The claim follows for $\ell=1$.
\item Assume now that $\ell \geq 2$, and fix     any    product
$x_{2} \cdots x_{\ell} \in {\sf Pos}_{\ell-1}(x_{2}, \ldots, x_{n})$ from the sum $\sum_{x_{2} \cdots
x_{\ell}       \in       {\sf Pos}_{\ell-1}(x_{2}, \ldots, x_{n})} $ $ x_{2} \cdots x_{\ell}$ in the {\sf
RHS}. Note that all   products  in ${\sf Pos}_{\ell-1}(x_{2}, \ldots, x_{n}) $ (in the {\sf RHS}) have the
same coefficient, which is $(-1)^{\ell-1} \cdot \frac{\textstyle 1}
           {\textstyle \ell}$.  We calculate the coefficient of this particular product in the expansion of the  {\sf LHS}.

Clearly, a $k$-positive product with $k \geq \ell$ in the {\sf LHS} cannot include $x_{2} \cdots x_{\ell}$ in
its expansion. So, we only need to consider contributions from the expansions of $k$-positive products with $0
\leq k \leq \ell-1$ (in the {\sf LHS}) to the coefficient of the product $x_{2} \cdots x_{\ell}$ in the
expansion of the  {\sf LHS}.
\begin{itemize}
\item Note that there are ${\ell-1 \choose k}$ ways to choose $k$ positive   literals
(or $\ell-1-k$ negative   literals) out of the $(\ell-1)$ literals $x_{2}, \ldots, x_{\ell}$ in order  to
form a $k$-positive product that includes   $x_{2} \cdots x_{\ell}$ (multiplied with a coefficient) in its
expansion. (All literals $x_{\ell+1}, \ldots, x_{n}$   have to be negative since they do not appear in the
product $x_{2} \cdots x_{\ell}$.)
\item The sign of the resulting $k$-positive product is $(-1)^{(\ell-1)-k}$, since each of the $(\ell-1)-k$ negative
    literals
in it contributes one minus  sign. (The negative literals $x_{\ell+1}, \ldots, x_{n}$ do not contribute to
the sign.).
\item The absolute value of the coefficient of the resulting $k$-positive product is $\frac{\tx 1}         {\tx k+1}$.
\end{itemize}
So,  the coefficient of $x_{2} \cdots x_{\ell}$ in   the  expansion of the {\sf LHS} is

{ \small
\begin{eqnarray*}
 \sum_{0 \leq k \leq \ell-1}        {\ell-1 \choose k}\,        (-1)^{(\ell-1)-k}\,
 \frac{\textstyle 1}             {\textstyle k+1}
&  = & \sum_{0 \leq k \leq \ell-1}        {\ell-1 \choose (\ell-1)-k} \,        (-1)^{(\ell-1)-k}\,
\frac{\textstyle 1}
             {\textstyle \ell-((\ell-1)-k) }                            \\
&  = & \sum_{0 \leq k \leq \ell-1}        {\ell-1 \choose k}\,        (-1)^{k}\,        \frac{\textstyle 1}
             {\textstyle \ell-k}                            \\
&  = & \frac{\textstyle 1}           {\textstyle \ell}\,      \sum_{0 \leq k \leq \ell-1}        {\ell
\choose k}
        (-1)^{k}
                                                                        \\
 & = & \frac{\textstyle 1}           {\textstyle \ell}\,      \left( \sum_{0 \leq k \leq \ell}               {\ell \choose k}
(-1)^{k} - {\ell \choose \ell}\,             (-1)^{\ell}
      \right)                                                                          \\
 &  = & \frac{\textstyle 1}           {\textstyle \ell}\,
      (0 + (-1)^{\ell-1})                                                                 \\
&  = & \frac{\textstyle 1}           {\textstyle \ell}\,      (-1)^{\ell-1}\, ,
\end{eqnarray*}
}
 and the claim follows   for $\ell\geq 2$.
 \end{itemize}

The proof is now complete.
\qed
\end{proof}

\section{Game-Theoretic Framework}
\label{framework}

Section \ref{The Strategic Game section} introduces the strategic game ${\sf AD}_{\alpha, \delta}(G)$.
 The associated pure Nash equilibria are defined in Section \ref{Pure Nash Equilibria section}. Section
\ref{Mixed Profiles Section} considers mixed profiles; their associated Expected Utilities are treated
in Section \ref{Expected Utilities Section}. (Mixed) Nash equilibria are introduced in Section \ref{Nash
Equilibria section}. Some special profiles and corresponding special classes of Nash equilibria are treated
in Section \ref{Some Special Profiles section}. Some notation is articulated in Section \ref{Notation
section}.

\subsection{The Strategic Game ${\sf AD}_{\alpha, \delta}(G)$}\label{The Strategic Game section}
Fix integers $\alpha \geq 1$ and $\delta \geq 1$. Associated with a graph $G$ is a
(\emph{\textbf{strategic}}) \emph{\textbf{game}} ${\sf AD}_{\alpha, \delta} (G)$:

  \begin{center}
\fbox{
\begin{minipage}{15.5cm}
\begin{itemize}

\item The set of \emph{\textbf{players}} is $    {\cal  A} \cup {\cal  D}$;
 ${\cal  A}$ contains $\alpha$ \emph{\textbf{attackers}} ${\sf a}_{i}$ with  $i\in [ \alpha]$, and ${\cal D}$
contains $\delta$ \emph{\textbf{defenders}} ${\sf d}_{j}$  with $  j \in [ \delta]$.

  \item The \emph{\textbf{strategy set}} $S_{{\sf a} }$ of each attacker ${\sf a} $ is $V$;   the
\emph{\textbf{strategy set}} $S_{{\sf d} }$ of each defender ${\sf d} $ is $E$. So, the
\emph{\textbf{strategy space}} $S$   is $S = \left(  \times \normalfont_{{\sf a}  \in {\cal  A}} S_{{\sf a} }
\right) \times \left( \times_{{\sf d}  \in {\cal  D}} S_{{\sf d} } \right)  = V^{\alpha} \times E^{\delta} $.

A \emph{\textbf{profile}} (or \emph{\textbf{pure profile}}) is an $(\alpha + \delta)$-tuple ${\bf s} =
\langle s_{{\sf a}_{1}},         \ldots, s_{{\sf a}_{\alpha}},         s_{{\sf d}_{1}},         \ldots,
s_{{\sf d}_{\delta}} \rangle \in S$.  The  profile  ${\bf s}_{-{\sf b}} \diamond t_{\sf b} $ is obtained from
the profile ${\bf s}$ and a strategy $t_{\sf b}$ for player ${\sf b} \in {\cal A}\cup {\cal D}$ by
substituting $t_{\sf b}$ for $s_{\sf b}$ in the profile ${\bf s}$.

For each vertex $v \in V$, ${\sf A}_{{\bf s}}(v) = \{ {\sf a}  \in {\cal  A}\mid   s_{{\sf a} } = v \} $ and
${\sf D}_{{\bf s}}(v) = \{ {\sf d} \in {\cal  D} \mid v \in s_{{\sf d} } \}$. Assume that $v \in s_{{\sf d}
}$. Then, the \emph{\textbf{proportion}} ${\sf Prop}_{{\bf s}}({\sf d} , v)$  of defender ${\sf d} $ on
vertex $v$ in the profile ${\bf s}$ is given by  ${\sf Prop}_{{\bf s}}({\sf d} , v) = \frac{\textstyle 1}
{\textstyle |{\sf D}_{{\bf s}}(v)|}$.

  \item
\begin{itemize}
\item The \emph{\textbf{Utility}} of attacker ${\sf a}$ is a function ${\sf U}_{{\sf a}}: S
                 \rightarrow \{ 0,1 \}$ with
\begin{eqnarray*}
      {\sf U}_{{\sf a}}({\bf s}) & = & \left\{
\begin{array}{ll}
0\, , &       \mbox{ if  $s_{{\sf a} }  \in  s_{{\sf d} }$  for some defender ${\sf d} \in {\cal  D}$} \\
1\, , &       \mbox{ if $ s_{{\sf a} } \not\in  s_{ {\sf d} }  $ for every defender ${\sf d} \in {\cal  D}$ }
\end{array}
      \right. .
\end{eqnarray*}
Intuitively, when the attacker ${\sf a} $ chooses vertex $v$, she receives $0$ if it is caught by a defender;
otherwise, she receives 1.

\item The \emph{\textbf{Utility}} of defender ${\sf d} $ is a function
${\sf U}_{{\sf d} }: S                    \rightarrow \mathbb{Q}$ with
\begin{eqnarray*}
      {\sf U}_{{\sf d} }({\bf s}) & = & \frac{|{\sf A}_{{\bf s}}(u)| } {|{\sf D}_{{\bf s}}(u)|}
      + \frac{|{\sf A}_{\bf s}(v)|} {|{\sf D}_{\bf s}(v)|}\, ,
\end{eqnarray*}
where $ s_{\sf d} =(u,v)  $.  Intuitively, the defender ${\sf d} $ receives the \emph{\textbf{fair share}} of
the total number of attackers choosing each of the two  vertices of the edge it chooses.

\end{itemize}

\end{itemize}
\end{minipage}
}
\end{center}

 \subsection{Pure Nash Equilibria}\label{Pure Nash Equilibria section}
The profile ${\bf s}$ is a \emph{\textbf{Pure Nash equilibrium}}~\cite{N50,N51}  if for each player ${\sf b}
\in
  {\cal  A} \cup {\cal  D}$, for each strategy $t_{\sf b}\in S_{\sf b}$,
${\sf U}_{\sf b}({\bf s})\geq {\sf U}_{\sf b}( {\bf s}_{-{\sf b}} \diamond t_{\sf b})$;
  so, a Pure Nash equilibrium is a
local maximizer for  the Utility of each player. Say that $G$ \emph{\textbf{admits a Pure Nash equilibrium}},
or   $G$ is \emph{\textbf{Pure}}, if there is a Pure Nash equilibrium for the strategic game ${\sf
AD}_{\alpha,\delta} (G)$.

 \subsection{Mixed Profiles}\label{Mixed Profiles Section}
 A \emph{\textbf{mixed  strategy}} for a player    is  a
 probability distribution over her strategy set; so, a mixed strategy for an attacker (resp., a defender)
is a probability distribution over vertices (resp.,  edges). A \emph{\textbf{mixed profile}} (or
\emph{\textbf{profile}} for short) $ \ms  = \langle \sigma_{{\sf a}_{1}},
             \ldots,                   \sigma_{{\sf a}_{\alpha}}, \sigma_{{\sf d}_{1}}, \ldots,
             \sigma_{{\sf d}_{\delta}}           \rangle$ is a collection of mixed strategies, one
for each player; $\sigma_{{\sf a} }(v)$ is the probability that attacker ${\sf a} $ chooses vertex $v$, and
$\sigma_{{\sf d} }(e)$ is the probability that defender ${\sf d} $ chooses edge $e$.

\subsubsection{Supports}

  Fix now a mixed profile $\ms$. The \emph{\textbf{support}} of player
${\sf b}    \in {\cal A}\cup {\cal D}$ in the profile ${\ms}$, denoted as ${\sf Support}_{{\ms}}({\sf b})$,
is the set of pure strategies in $S_{{\sf b}}$ to which ${\sf b}$ assigns strictly positive probability.
Denote ${\sf Supports}_{{\ms}}({\cal A} )           =
   \bigcup_{{\sf a}  \in {\cal  A}}
   {\sf Support}_{{\ms}}({\sf a} )$ and ${\sf Supports}_{{\ms}}({\cal D} )
        =           \bigcup_{{\sf d}  \in {\cal  D}}
         {\sf Support}_{{\ms}}({\sf d} )$.  A mixed profile ${\ms}$
induces a probability measure $\mathbb{P}_{{\ms}}$ (on pure profiles) in the natural way.  Note that in a
pure profile  ${\bf s}$, ${\sf Supports}_{{\bf s}}({\cal A}) \leq \alpha$ and ${\sf Supports}_{{\bf s}}({\cal
D}) \leq \delta$.

\subsubsection{Expectations about Attackers}

For each vertex $v\in V$, denote as $|{\sf A}|_{{\ms}}(v)$ the expected number of attackers choosing vertex
$v$ in $\ms$; so,
\begin{eqnarray*}
 |{\sf A}|_{{\ms}}(v)& =&  \sum_{{\sf a} \in {\cal A}} \sigma_{{\sf a}
}(v)\,  .
\end{eqnarray*}

 Clearly, $|{\sf A}|_{{\ms}}( v)  >0 $ if and only if $v\in {\sf Supports}_{\ms}({\cal A}) $.
For an edge $ (u,v)\in E$, denote
\begin{eqnarray*}
 |{\sf A}|_{{\ms}}((u,v)) & = &   |{\sf A}|_{{\ms}}(u)  +  |{\sf A}|_{{\ms}}(v).
\end{eqnarray*}
We observe:
\begin{observation}\label{sum of VPs}
For a mixed profile $\ms$,
\begin{eqnarray*}
\sum_{v\in {\sf Supports}_{\ms}({\cal A})} |{\sf A}|_{\ms}(v) &=&\alpha.
\end{eqnarray*}
\end{observation}
\begin{proof}
Clearly,
\begin{eqnarray*}
 \sum_{v\in {\sf Supports}_{\ms}({\cal A})} |{\sf A}|_{\ms}(v) &=&
\sum_{v\in {\sf Supports}_{\ms}({\cal A})} \sum_{{\sf a} \in {\cal A}}        \sigma_{{\sf a} }(v) \\
& =& \sum_{{\sf a} \in {\cal A}}    \sum_{v\in {\sf Supports}_{\ms}({\cal A})}     \sigma_{{\sf a} }(v) \\
&=& \sum_{{\sf a} \in {\cal A}}    1 \\
&=&   \alpha \, ,
\end{eqnarray*}
as needed.
\qed
\end{proof}

\subsubsection{Hitting Events and Vertices}

   Fix a vertex $v \in V$. For a defender ${\sf d} $,
denote as ${\sf Hit}({\sf d} , v)$ the event that defender ${\sf d} $ chooses an edge incident to vertex $v$;
clearly, for the mixed profile $\ms$,
\begin{eqnarray*}
\mathbb{P}_{{\ms}}({\sf  Hit} ({\sf d}, v )) &=& \sum_{e\in {\sf Support}_{\ms}({\sf d}) \mid v\in e}
\sigma_{{\sf d }}(e).
\end{eqnarray*}
 Denote as ${\sf Hit}(v)$ the event that some defender chooses an edge
 incident to vertex $v$. Clearly,
  \begin{eqnarray*}
{\sf Hit}(v) & =& \bigcup_{{\sf d}  \in {\cal  D}}
  {\sf Hit}({\sf d} , v).
 \end{eqnarray*}
Finally, denote as $ {\sf   D}_{\ms}(v)$ the set\\
\begin{eqnarray*}
{\sf   D}_{\ms}(v) & =&  \Big\{ {\sf d}  \in {\cal  D}
   \mid\      \mbox{there is an edge  }  e \in {\sf Support}_{{\ms}}({\sf d} )
\mbox{ such that }  v\in e    \Big\} ;
\end{eqnarray*}
so, ${\sf   D}_{\ms}(v)$ is the set of defenders ``hitting'' vertex $v$.

 A vertex $v\in V$ is \emph{\textbf{multidefender}}
 in the profile ${\ms}$ if
 $|{\sf   D}_{\ms}(v)|\geq 2$;
that is, a multidefender vertex is ``hit'' by more than one defenders. A vertex $v\in V$  is
\emph{\textbf{unidefender}} in $\ms$
if $|{\sf   D}_{\ms}(v)|\leq 1$;
  $v$ is
\emph{\textbf{monodefender}} in $\ms$ if
 $|{\sf   D}_{\ms}(v)| =  1$.
 So,   for each unidefender (resp., monodefender) vertex $v$, there is at most (resp., exactly)
one defender ${\sf d} $ with an edge $e\in {\sf Support}_{\ms}({\sf d})$  such that $v\in e$; if there is
such a defender, denote it   as ${\sf d}_{\ms}(v)$, else, set, by convention, $\mathbb{P}_{{\ms}}{\sf
Hit}(({\sf d}_{\ms}(v), v )) =0$.

A profile ${\ms}$ is \emph{\textbf{unidefender}} (resp., \emph{\textbf{monodefender}}) if every vertex $v \in
V$ is unidefender (resp., monodefender) in ${\ms}$; else the profile $\ms$ is \emph{\textbf{multidefender}}.
Note that for a unidefender (resp., monodefender) profile $\ms$, for each edge $e\in E$, there is at most
(resp., exactly) one defender ${\sf d}$ such that $\sigma_{{\sf d}}(e)
> 0$; if there is such a defender  ${\sf d}$, denote it as ${\sf d}_{\ms}(e)$, else set, by convention,
$\mathbb{P}_{{\ms}}({\sf d}_{\ms}(e), e ) = 0$.

\subsubsection{Hitting Probabilities}

Since the  events ${\sf Hit}({\sf d}_j,v)$   and ${\sf Hit}({\sf d}_{j'},v)$ with $j\neq j'$ are independent
and  not mutually exclusive (for a fixed vertex $v$), we immediately obtain a strengthening of the Union
Bound:

\begin{observation}
\label{multidefender vertex} For each vertex   $v\in V$,
 \begin{eqnarray*}
\mathbb{P}_{{\ms}} ({\sf Hit}(v))  &
  \left\{
\begin{array}{l}
<  \\
=\\
\end{array}
\right. &
\begin{array}{ll}
  \sum_{{\sf d}  \in {\cal  D}}  \mathbb{P}_{{\ms}} ({\sf Hit}({\sf d} , v)), & \mbox{if $v$ is multidefender in $\ms$}\\
    \sum_{{\sf d}  \in {\cal  D}}  \mathbb{P}_{{\ms}} ({\sf Hit}({\sf d}, v)) , &\mbox{if $v$  is
unidefender in $\ms$}
\end{array}.
\end{eqnarray*}
\end{observation}

\noindent By the {\it Principle of Inclusion-Exclusion,} we immediately observe:
\begin{lemma}\label{Prob Hit is equal}
For a  vertex $v\in V$,
  \begin{eqnarray*}
    \mathbb{P}_{{\ms}}({\sf Hit}(v))
& =&   \sum_{l \in [\delta]}        (-1)^{l-1}        \sum_{{\cal D}' \subseteq {\cal  D}
       \mid              |{\cal D}'| = l} \prod_{{\sf d}  \in {\cal D}'}
   \mathbb{P}_{{\ms}}\left( {\sf Hit}({\sf d} , v)
      \right).
 \end{eqnarray*}
\end{lemma}

  \noindent We continue to prove:

\begin{lemma}
\label{upper bound on sum of hit probabilities} For a mixed profile $\ms$,
\begin{eqnarray*}
\sum_{v \in V}
 \mathbb{P}_{{\ms}}({\sf Hit}(v))
\left\{
\begin{array}{ll}
 < 2\, \delta,  & \mbox{if   $\ms$ is   multidefender}\\
 = 2\, \delta,  & \mbox{if   $\ms$ is   unidefender}\\
\end{array}
\right. .
\end{eqnarray*}
 \end{lemma}
\begin{proof}
  Clearly,
\begin{eqnarray*}
 \sum_{v \in V}   \sum_{{\sf d}  \in {\cal D}}
  \mathbb{P}_{{\ms}}({\sf Hit}({\sf d}, v))  &=&
\sum_{v \in V}       \sum_{{\sf d} \in {\cal D}} \sum_{e \in {\sf Support}_{{\ms}}({\sf d})\mid v\in e}
          \sigma_{\sf d} (e)                                                                \\                                                                                           \\
& =& 2\,            \sum_{e \in E}   \sum_{{\it {\sf d} } \in {\cal  D}}         \sigma_{{\sf d}}(e) \\
& =& 2\,      \sum_{{\it {\sf d} } \in {\cal  D}}        \sum_{e \in E}          \sigma_{{\sf d}}(e) \\
&= & 2\, \delta\, .
\end{eqnarray*}
Hence, by Observation \ref{multidefender vertex},
\begin{eqnarray*}
\sum_{v \in V}    \mathbb{P}_{{\ms}}({\sf Hit}(v)) & &\left\{
\begin{array}{ll}
< \,\, \sum_{v\in V} \sum_{{\sf d}  \in {\cal  D}}  \mathbb{P}_{{\ms}} ({\sf Hit}({\sf d} , v)), & \mbox{if
$\ms$
is multidefender}\\
= \,\, \sum_{v\in V} \sum_{{\sf d}  \in {\cal  D}}  \mathbb{P}_{{\ms}} ({\sf Hit}({\sf d} , v)), &  \mbox{if
$\ms$ is unidefender}
\end{array} \right.\\
& &\left\{
\begin{array}{ll}
< \,\, 2\, \delta , & \mbox{if $\ms$
is multidefender}\\
=   \,\, 2\, \delta, &  \mbox{if $\ms$ is unidefender}
\end{array} \right. ,
\end{eqnarray*}
as needed.
\qed\end{proof}

\subsubsection{Minimum Hitting Probability, Maxhit Vertices and  Maxhitters}

  \noindent Denote as
\begin{eqnarray*} {\sf MinHit}_{\ms} &=& \min_{v \in V}
  \mathbb{P}_{{\ms}}({\sf Hit}(v))
\end{eqnarray*}
 the \emph{\textbf{Minimum Hitting Probability}}
 associated with the mixed profile $\ms$.  We observe:

\begin{lemma}
\label{upper bound on minimum hitting probability} For a mixed profile ${\ms}$,
\begin{eqnarray*}
{\sf MinHit}_{{\ms}} &\leq& \frac{\textstyle 2 \delta}      {\textstyle |V|}.
\end{eqnarray*}
\end{lemma}
\begin{proof}
Assume, by way of contradiction, that ${\sf MinHit}_{{\ms}}
 >
 \frac{\textstyle 2 \delta}      {\textstyle |V|}$.
 Then,
\begin{eqnarray*}
\sum_{v \in V}   \mathbb{P}_{{\ms}}           ({\sf Hit}(v)) &\geq & |V| \cdot {\sf MinHit}_{{\ms}} \\
 &> &  2 \delta,
\end{eqnarray*}
 a contradiction to Lemma~\ref{upper bound on sum of hit probabilities}.
\qed\end{proof}

A vertex $v \in V$ is \emph{\textbf{maxhit}} in the profile ${\ms}$ if $\mathbb{P}_{\ms}({\sf Hit}(v))
  =    1$; say that a defender ${\sf d}  \in {\cal  D}$ is a \emph{\textbf{maxhitter}} in ${\ms}$ if there is a vertex
$v \in {\sf Vertices}({\sf Support}_{{\ms}}({\sf d} ))$ such that $\mathbb{P}_{{\ms}}({\sf Hit}({\sf d} , v))
= 1$. We observe:

\begin{lemma}\label{PsHit equals one case}
Consider a maxhit vertex $v$ in a profile $\ms$. Then, there is a (maxhitter) defender ${\sf d}$ (in $\ms$)
with   $\mathbb{P}_{{\ms}}({\sf Hit}({\sf d}, v))=1 $.
\end{lemma}
\begin{proof}
Assume, by way of contradiction, that for each defender ${\sf d}\in {\cal D}$, $\mathbb{P}_{{\ms}}({\sf
Hit}({\sf d}, v))<1$. Since the set $\{ {\sf Hit}({\sf d}, v)\mid {\sf d}\in {\cal D} \}$ is a family of
independent events with none of them being certain, this implies that the event $ {\sf Hit} (v) =
\bigcup_{{\sf d} \in {\cal D}} {\sf Hit}({\sf d}, v) $ is not certain. So, $\mathbb{P}_{{\ms}}({\sf
Hit}(v))<1$. A contradiction.
 \qed
\end{proof}

\subsection{Expected Utilities}\label{Expected Utilities Section}

\noindent The mixed profile ${\ms}$ induces   an \emph{\textbf{Expected Utility}} ${\sf U}_{{\sf b}}({\ms})$
for each player ${\sf b} \in {\cal A}\cup {\cal D}$, which is the expectation (according to $\ms$) of the
Utility of player ${\sf b}$. We shall derive some formulas for Expected Utilities. To do so, we first define
and derive formulas for some auxiliary quantities. In more detail, we define the Conditional Expected
Proportion associated with the   defenders; we then use it to derive an expression for the Conditional
Expected Utility for each attacker. The Expected Utility of each attacker is then derived as a weighted sum
of Conditional Expected Utilities. Similarly, the Expected Utility of each defender is derived as a weighted
sum of Conditional Expected Utilities defined for the defenders in the natural way.

\subsubsection{Conditional Expected Proportion}

  Induced by ${\ms}$ is   the \emph{\textbf{Conditional Expected Proportion}}
${\sf Prop}_{{\sf d}}({\ms}_{-{\sf d}} \diamond v)$ of
 defender ${\sf d}                 \in {\cal  D}$ on vertex $v$, which  is the  expectation (induced by $\ms$) of the
proportion of   defender ${\sf d}$ on vertex $v$ had she chosen an edge incident to vertex $v$.  Clearly, {
\small
\begin{eqnarray*}
{\sf Prop}_{{\sf d}}({\ms}_{-{\sf d} } \diamond v) & = & \sum_{\ell \in [\delta]} \frac{\textstyle 1}
     {\textstyle \ell}\,        \sum_{{\cal D}' \subseteq {\cal  D}
       \setminus                                 \{ {\sf d} \}              \mid   |{\cal D}'| = \ell-1}
         \prod_{{\sf d}_{k} \in {\cal D}'}             \mathbb{P}_{{\ms}}
        ({\sf Hit}({\sf d}_{k}, v)) \prod_{{\sf d}_{k} \not\in {\cal D}' \cup \{ {\sf d} \}}
        (1 - \mathbb{P}_{{\ms}}
                          ({\sf Hit}({\sf d}_{k}, v)))
\end{eqnarray*}
}

 \noindent  Lemma \ref{combinatorial lemma} implies now an alternative expression for Conditional Expected Proportion.
\begin{lemma}\label{Proportion of Defender}
For each pair of a defender ${\sf d}\in {\cal D}$ and a vertex $v\in V$,
\begin{eqnarray*}
\hspace{-3.3cm}{\sf Prop}_{{\sf d} }({\ms}_{-{\sf d} } \diamond v) & = & \sum_{\ell \in [\delta]}
\frac{\textstyle 1} {\textstyle \ell}\,\, (-1)^{\ell-1}\, \sum_{{\cal D}' \subseteq {\cal D} \setminus \{
{\sf d} \} \mid |{\cal D}'| = \ell-1} \prod_{{\sf d}_{k} \in {\cal D}'} \mathbb{P}_{{\ms}} ({\sf Hit}({\sf
d}_{k}, v))\,.
\end{eqnarray*}
\end{lemma}

 \subsubsection{Attackers}
  Induced by ${\ms}$ is   the \emph{\textbf{Conditional Expected Utility}}
${\sf U}_{{\sf a} }({\ms_{-{\sf a}}} \diamond v)$ of attacker ${\sf a}  \in {\cal  A}$ on vertex $v$, which
is the conditional expectation (induced by   ${\ms}$) of the Utility of attacker ${\sf a} $  had she chosen
vertex $v$. Clearly,
\begin{eqnarray*}
{\sf U}_{{\sf a} }({\ms_{-{\sf a}  }}\diamond v)& = & 1 - \mathbb{P}_{{\ms}}({\sf Hit}(v))\,.
\end{eqnarray*}

\noindent By the Law of Conditional Alternatives, we immediately obtain:

\begin{lemma}\label{Expected Utility of attacker}
Fix a mixed profile $\ms$. Then, the Expected Utility ${\sf U}_{{\sf a}}(\ms)$ of an attacker ${\sf a}\in
{\cal A}$ is
\begin{eqnarray*}
{\sf U}_{{\sf a} }({\ms}) &  =& \sum_{v \in V} \sigma_{{\sf a}}(v)        \cdot \left( 1 -
\mathbb{P}_{{\ms}}({\sf Hit}(v)) \right)\, .
\end{eqnarray*}
\end{lemma}

We continue with a preliminary observation:
\begin{lemma}
\label{tiny} Fix a mixed profile $\ms$. Then, for each vertex    $v \in V$,
\begin{eqnarray*}
 \mathbb{P}_{{\ms}}({\sf
Hit}(v)) &= & \sum_{{\sf d} \in {\cal  D}}        \mathbb{P}_{{\ms}} ({\sf Hit}({\sf d}, v)) \cdot {\sf
Prop}_{{\sf d}}({\ms}_{-{\sf d}} \diamond v)\, .
\end{eqnarray*}
\end{lemma}
\begin{proof}
By Lemma \ref{Proportion of Defender}, { \small
\begin{eqnarray*}
  & & \sum_{{\sf d} \in {\cal  D}}        \mathbb{P}_{{\ms}}                ({\sf Hit}({\sf d}, v)) \cdot       {\sf Prop}_{{\sf d}}
             (\ms_{-{\sf d}} \diamond   v)\\
 & = & \sum_{{\sf d} \in {\cal  D}}        \mathbb{P}_{{\ms}}                ({\sf Hit}({\sf d}, v)) \cdot      \left( \sum_{\ell \in [\delta]}               \frac{\textstyle 1} {\textstyle \ell}\, \,
(-1)^{\ell-1}               \sum_{\substack{ {\cal D} '\subseteq {\cal  D} \setminus \{ {\sf d} \} \\ | {\cal
D}'| = \ell-1}}                  \prod_{{\sf d}_{k} \in {\cal D}'} \mathbb{P}_{{\ms}}({\sf Hit}({\sf d}_{k},
v))
      \right)                                                                                  \\
 & = & \sum_{{\sf d} \in {\cal  D}}        \sum_{\ell \in [\delta]}          \frac{\textstyle 1}
{\textstyle \ell}\, \, (-1)^{\ell-1} \sum_{\substack{ {\cal D}' \subseteq {\cal  D} \setminus \{ {\sf d} \}
\\     |{\cal D}'| = \ell-1}} \hspace{0.4cm} \prod_{{\sf d}_{k} \in {\cal D}' \cup \{{\sf d}\}}
               \mathbb{P}_{{\ms}}({\sf Hit}({\sf d}_{k}, v))                              \\
&= &    \sum_{\ell \in [\delta]}          (-1)^{\ell-1} \cdot \frac{\tx 1} {\tx \ell}\sum_{{\sf d} \in {\cal
D}}   \sum_{\substack{ {\cal D}' \subseteq {\cal  D} \setminus \{ {\sf d} \}\\     |{\cal D}'| = \ell-1}}
\hspace{0.4cm} \prod_{{\sf d}_{k} \in {\cal D}' \cup \{{\sf d}\}}
               \mathbb{P}_{{\ms}}({\sf Hit}({\sf d}_{k}, v))                 .
\end{eqnarray*}
}

\noindent  Note that for each integer $\ell \in [\delta]$, for each set ${\cal D}''\subseteq  {\cal D}$ with
$|{\cal D}''|=\ell$, there are $\ell$ pairs of a defender ${\sf d} \in {\cal D}$ such that ${\sf d} \in {\cal
D}''$ and a set ${\cal D}' \subseteq {\cal D}''$ such that ${\cal D}' \subseteq {\cal D} \setminus \{ {\sf
d}\}$ and $|{\cal D}'| = \ell -1$. Hence,

{\small
\begin{eqnarray*}
   \sum_{\substack{ {\cal D}'' \subseteq {\cal  D} \\
|{\cal D}''| = \ell}} \hspace{0.4cm} \prod_{{\sf d}_{k} \in {\cal D}''}
               \mathbb{P}_{{\ms}}({\sf Hit}({\sf d}_{k}, v))
& = & \frac{\tx 1} {\tx \ell}
\sum_{{\sf d} \in {\cal D}}   \sum_{\substack{ {\cal D}' \subseteq {\cal  D} \setminus \{ {\sf d} \}\\
|{\cal D}'| = \ell-1}} \hspace{0.4cm}  \prod_{{\sf d}_{k} \in {\cal D}' }
               \mathbb{P}_{{\ms}}({\sf Hit}({\sf d}_{k}, v)) .
\end{eqnarray*}
}

\noindent It follows that
{\small
\begin{eqnarray*} \sum_{{\sf d} \in {\cal  D}}        \mathbb{P}_{{\ms}}
({\sf Hit}({\sf d}, v)) \cdot {\sf Prop}_{{\sf d}}
             (\ms_{-{\sf d}} \diamond   v)
 & = &          \sum_{\ell \in [\delta]}          (-1)^{\ell-1} \sum_{\substack{{\cal D}'
\subseteq {\cal  D}    \\  |{\cal D}'| = \ell}}\hspace{0.2cm} \prod_{{\sf d}_{k} \in {\cal D}'}
\mathbb{P}_{{\ms}}({\sf Hit}({\sf d}_{k},
v))\,\\
 & = & \mathbb{P}_{{\ms}}({\sf Hit}(v)) ,
\end{eqnarray*}
} as needed. \qed\end{proof}

\subsubsection{Defenders}

  Induced by $\ms$ is also the  \emph{\textbf{Conditional Expected Utility}}
${\sf U}_{{\sf d} }({\ms}_{-{\sf d} } \diamond  (u,v))$ of defender ${\sf d} $ on edge $  (u, v) \in E$,
which is    the conditional expectation (induced by  ${\ms}$) of the Utility of defender ${\sf d} $ had she
chosen edge $(u,v)$. Clearly,
\begin{eqnarray*}
{\sf U}_{{\sf d} }(( {\ms}_{-{\sf d} }  \diamond  (u,v))) &=& {\sf Prop}_{{\sf d} }({\ms}_{-{\sf d} }
\diamond u) \cdot |{\sf A}|_{{\ms}}(u)      +      {\sf Prop}_{{\sf d} }({\ms}_{-{\sf d} } \diamond v) \cdot
|{\sf A}|_{{\ms}}(v)\, .
\end{eqnarray*}
We prove:
\begin{lemma}\label{expected utility of defender is}
Fix a mixed profile $\ms$. Then, the Expected Utility of a defender ${\sf d} \in {\cal D}$ is
\begin{eqnarray*}
      {\sf U}_{{\sf d} }({\ms})
& = & \sum_{v \in V}
  \mathbb{P}_{{\ms}}  ({\sf Hit}({\sf d} , v)) \cdot {\sf Prop}_{{\sf d} }({\ms}_{-{\sf d} } \diamond v) \cdot        |{\sf A}|_{{\ms}}(v)\, .
\end{eqnarray*}
\end{lemma}
\begin{proof}
By the Law of Conditional Alternatives,
 { \small
\begin{eqnarray*}
      {\sf U}_{{\sf d} }({\ms})
& = &  \sum_{ (u,v) \in E}        \sigma_{{\sf d} }((u,v)) \cdot    {\sf U}_{{\sf d} }(( {\ms}_{-{\sf d} }
\diamond
(u,v)))\\
& = & \sum_{ (u,v) \in E}    \sigma_{{\sf d} }((u,v)) \cdot \Big({\sf Prop}_{{\sf d} }({\ms}_{-{\sf d} }
\diamond u) \cdot |{\sf A}|_{{\ms}}(u)      +      {\sf Prop}_{{\sf d} }({\ms}_{-{\sf d} } \diamond v) \cdot
|{\sf A}|_{{\ms}}(v)\Big)\, \\
& = & \sum_{v \in V} \left(\sum_{e\mid v\in e}   \sigma_{{\sf d} }(e) \right) \cdot {\sf Prop}_{{\sf d}
}({\ms}_{-{\sf d}
} \diamond  v) \cdot        |{\sf A}|_{{\ms}}(v)\, \\
& = & \sum_{v \in V}
  \mathbb{P}_{{\ms}}  ({\sf Hit}({\sf d} , v)) \cdot {\sf Prop}_{{\sf d} }({\ms}_{-{\sf d} } \diamond v) \cdot        |{\sf A}|_{{\ms}}(v)\, ,
\end{eqnarray*}
}
as needed.
\qed
\end{proof}

\subsection{Nash Equilibria}\label{Nash Equilibria section}
A mixed profile ${\ms}$ is a \emph{\textbf{Nash equilibrium}}~\cite{N50,N51} if for each player ${\sf b} \in
{\cal A}\cup {\cal D}$, for each mixed strategy $\tau_{\sf b}$ of player ${\sf b}$, ${\sf U}_{{\sf
b}}(\ms)\geq {\sf U}_{{\sf b}}(\ms_{-{\sf b}}\diamond \tau_{{\sf b}})$;
 so, a Nash equilibrium is a local maximizer of the Expected Utility of each player. A (necessary and) sufficient
condition for a Nash equilibrium $\ms$ is that for each player ${\sf b} \in {\cal A}\cup {\cal D}$, for each
pure strategy $t_{{\sf b}}$ of player $ {\sf b}$, ${\sf U}_{{\sf b}}(\ms)\geq {\sf U}_{{\sf b}}(\ms_{-{\sf
b}}\diamond t_{{\sf b}})$.  By the celebrated Theorem of Nash~\cite{N50,N51}, ${\sf AD}_{\alpha,
\delta}(G)$ has at least one Nash equilibrium. Say that $G$ \emph{\textbf{admits}} a Nash equilibrium with a
particular property if the game ${\sf AD}_{\alpha, \delta}$ has a Nash equilibrium with this particular
property.

Clearly, in a Nash equilibrium ${\ms}$, for each attacker ${\sf a} \in {\cal A}$, ${\sf U}_{{\sf a}
}({\ms_{-{\sf a} }} \diamond v)$ is {\em constant} over all vertices $v \in {\sf Support}_{{\ms}}({\sf a} )$;
for each defender ${\sf d}\in {\cal D} $, ${\sf U}_{{\sf d}}({\ms_{-{\sf d} }} \diamond e)$ is {\em constant}
over all edges $e \in {\sf Support}_{{\ms}}({\sf d})$. It follows that in a Nash equilibrium ${\ms}$, for
each attacker ${\sf a} \in {\cal A}$,
\begin{eqnarray*}
{\sf U}_{{\sf a} }({\ms}) &=& 1 - \mathbb{P}_{{\ms}}({\sf Hit}(v)),
\end{eqnarray*}
 for any vertex $v \in {\sf Support}_{{\ms}}({\sf a} )$. So, for each attacker ${\sf a} \in {\cal A}$, the quantity $\mathbb{P}_{{\ms}}({\sf Hit}(v))$ is
constant  over all vertices $v \in {\sf Support}_{{\ms}}({\sf a} )$. In the same way,  for each defender
${\sf d} \in {\cal D}$,
\begin{eqnarray*}
       {\sf U}_{{\sf d}} (\ms)   & = &
       {\sf Prop}_{{\sf d}}({\ms}_{-{\sf d}} \diamond u)
\cdot      |{\sf A}|_{{\ms}}(u)      +      {\sf Prop}_{{\sf d}}({\ms}_{-{\sf d}} \diamond v) \cdot
|{\sf A}|_{{\ms}}(v) ,
\end{eqnarray*}
for any edge $(u, v) \in {\sf Support}_{{\ms}}({\sf d})$. So, for each defender ${\sf d}\in {\cal D} $, the
quantity ${\sf Prop}_{{\sf d}}({\ms}_{-{\sf d}} \diamond u) \cdot      |{\sf A}|_{{\ms}}(u) + {\sf
Prop}_{{\sf d}}({\ms}_{-{\sf d}} \diamond  v) \cdot      |{\sf A}|_{{\ms}}(v)$ is constant  over all edges
$(u, v) \in {\sf Support}_{{\ms}}({\sf d})$. Note that in a Nash equilibrium $\ms$, for each defender ${\sf
d}\in {\cal D}$, ${\sf U}_{{\sf d}}(\ms) > 0 $; in contrast, it is possible that
$ {\sf U}_{{\sf a}}(\ms) = 0$ for some attacker ${\sf a}\in {\cal A}$. (See, for an example, the proof of Theorem
\ref{defender pure balanced equilibria}.)

\subsection{Some Special Profiles}\label{Some Special Profiles section}
 A profile ${\ms}$ is \emph{\textbf{uniform}} if each player  uses a
{\it uniform} probability distribution on its support; so, for each attacker ${\sf a}  \in {\cal A}$, for
each vertex $v \in {\sf Support}_{{\ms}}({\sf a} )$, $\sigma_{{\sf a} }(v) = \frac{\textstyle 1} {\textstyle
|{\sf Support}_{{\ms}}({\sf a} )|}$, and for each defender ${\sf d} \in {\cal D}$, for each edge $e \in {\sf
Support}_{\ms}({\sf d} )$, $\sigma_{{\sf d}}(e) = \frac{\textstyle 1}      {\textstyle |{\sf
Support}_{{\ms}}({\sf d})|}$.

A profile ${\ms}$ is \emph{\textbf{attacker-symmetric}} (resp., \emph{\textbf{defender-symmetric}}) if for
all pairs of attackers ${\sf a}_{i}$ and ${\sf a}_{k}$ (resp., all pairs of defenders ${\sf d}_{j}$ and ${\sf
d}_{k}$), for all vertices $v \in V$, (resp., all edges $e \in E$) $\sigma_{{\sf a}_{i}}(v) = \sigma_{{\sf
a}_{k}}(v)$ (resp., $\sigma_{{\sf d}_{j}}(e) = \sigma_{{\sf d}_{k}}(e)$). A profile is
\emph{\textbf{attacker-uniform}} (resp., \emph{\textbf{defender-uniform}}) if
 each attacker (resp., defender) uses a uniform probability
distribution on his support. Now,  \emph{\textbf{Attacker-Symmetric}}  (resp.,
\emph{\textbf{Defender-Symmetric}}) \emph{\textbf{Nash equilibria}} and \emph{\textbf{Attacker-Uniform}}
(resp., \emph{\textbf{Defender-Uniform}}) \emph{\textbf{Nash equilibria}} are defined in the natural way. A
\emph{\textbf{Symmetric Nash equilibrium}} is both Attacker-Symmetric and Defender-Symmetric. A
\emph{\textbf{Uniform Nash equilibrium}} is both Attacker-Uniform and Defender-Uniform.

A profile $\ms$ is \emph{\textbf{attacker-fullymixed}} (resp., \emph{\textbf{defender-fullymixed}}) if for
each attacker ${\sf a}$ (resp., for each defender ${\sf d}$), ${\sf Support}_{{\ms}}({\sf a}) = V$ (resp.,
${\sf Support}_{{\ms}}({\sf d})    =    E$). Now,  \emph{\textbf{Attacker-Fullymixed}}   (resp.,
\emph{\textbf{Defender-Fullymixed}}) \emph{\textbf{Nash equilibria}} are   defined in the natural way. A
\emph{\textbf{Fullymixed Nash equilibrium}} is both Attacker-Fullymixed and Defender-Fullymixed.

A profile $\ms$ is \emph{\textbf{defender-pure}} if   each defender chooses  a single strategy with
probability $1$ in $\ms$. Now    \emph{\textbf{Defender-Pure Nash equilibria}} are defined in the natural
way. Say that $G$ admits a  \emph{\textbf{Defender-Pure Nash equilibrium}}, or   $G$ is
\emph{\textbf{Defender-Pure}}, if there is a Defender-Pure Nash equilibrium for the strategic game ${\sf
AD}_{\alpha,\delta} (G)$.

Fix now a Perfect-Matching graph. Say that a profile is \emph{\textbf{perfect-matching}} if ${\sf
Supports}_{\ms}({\cal  D})$ is a Perfect Matching. Now,  \emph{\textbf{Perfect-Matching Nash equilibria}}
 are defined in the natural way.

\subsection{Notation}\label{Notation section}
 Fix a mixed profile $\ms$. For a vertex $v\in V$, set
\begin{eqnarray*}
{\sf Edges}_{{\ms}}(v)
&=&
 \left\{  e \in   {\sf Supports}_{{\ms}}({\cal D} ) \mid v\in e \right\};
\end{eqnarray*}
so, ${\sf Edges}_{{\ms}}(v)$  consists of all  edges incident to $v$ that are included in the union of
supports of the defenders.  For a vertex set $U \subseteq V$, set
\begin{eqnarray*}
{\sf Edges}_{{\ms}}( U )   &=& \bigcup_{v\in U} {\sf Edges}_{{\ms}} (v);
\end{eqnarray*}
 so, ${\sf Edges}_{{\ms}}(U)$ consists of all edges incident to a vertex in $U$ that are included in the union of supports
of the defenders.

For an edge $e\in E$, set
\begin{eqnarray*}
{\sf Vertices}_{{\ms}}( e )   &=& \left\{  v\in e
    \mid          v  \in {\sf Supports}_{{\ms}}({\cal A} ) \right\};
\end{eqnarray*}
so, $|{\sf Vertices}_{{\ms}}( e ) |\leq 2$.
 For an edge  set $F\subseteq E$, set
\begin{eqnarray*}
{\sf Vertices}_{{\ms}}(F) & =&   \bigcup_{e\in F} {\sf Vertices}_{{\ms}} (e);
\end{eqnarray*}
so, ${\sf Vertices}_{{\ms}}(F)$ consists of all vertices incident to an edge in  $U$ that are included in the
union of supports of the attackers.

\remove{ The \emph{\textbf{Price of Defense}} ${\sf PoD}_{G}$~\cite{MMPPS06} is the {\em worst-case}
Defense-Ratio over all Nash equilibria ${\ms}$; so, ${\sf PoD}_{G} = \max_{{\ms}}   {\sf DR}_{{\ms}}$. }

\section{The Structure of Nash Equilibria}
\label{structure of nash equilibria}
 We  provide an   analysis of the combinatorial structure of the Nash equilibria associated with the strategic
game ${\sf AD}_{\alpha, \delta}(G)$. Section \ref{Combinatorial Characterization section} presents a
combinatorial characterization of Nash equilibria. Some necessary conditions for Nash equilibria are derived
in Section \ref{Necessary Conditions}. Section \ref{Pure Nash Equilibria} treats the special case of Pure
Nash equilibria.

\subsection{Combinatorial Characterization}\label{Combinatorial Characterization section}

We show:
\begin{proposition}[Characterization of Nash Equilibria]
\label{characterization of nash equilibria} A profile $\ms$ is a Nash equilibrium if and only if the
 following conditions hold:
\begin{enumerate}

\item[{\sf (1)}] For each vertex $v \in {\sf Supports}_{\ms}( {\cal A} )$, $\mathbb{P}_{\ms}
   ({\sf Hit}(v)) = {\sf MinHit}_{\ms}$.

\item[{\sf (2)}] For each defender ${\sf d}\in {\cal D}$, for each edge $(u,v) \in {\sf Support}_{\ms}({\sf d})$,
{ \small
 \begin{eqnarray*}
 \lefteqn{   {\sf Prop}_{{\sf d}}({\ms_{-{\sf d}}} \diamond  u)   \cdot |{\sf A}|_{\ms}(u)   + {\sf
Prop}_{{\sf d}}({\ms_{-{\sf d}}} \diamond  v)\cdot  |{\sf A}|_{\ms}(v)  }&\\
   = &  \max_{(u',v') \in E}
    \Big\{
       {\sf Prop}_{{\sf d}}({\ms_{-{\sf d}}} \diamond  u')\cdot   |{\sf A}|_{\ms}(u') +
{\sf Prop}_{{\sf d}}({\ms_{-{\sf d}}} \diamond  v')    \cdot  |{\sf A}|_{\ms}(v')  \Big\}.
\end{eqnarray*}
}
\end{enumerate}
\end{proposition}

\begin{proof}
Assume first that $\ms$ is a Nash equilibrium. To establish Condition {\sf (1)}, consider any vertex $v \in
{\sf Supports}_{\ms}({\cal A})$; so, $v \in {\sf Support}_{\ms}({\sf a})$ for some attacker ${\sf a}$. Since
$\ms$ is a  Nash equilibrium,
 $\mathbb{P}_{\ms}({\sf Hit}(v'))$ is  constant
over all vertices $v' \in {\sf Support}_{\ms}({\sf a})$. We prove:
\begin{lemma}\label{characterization of nash equilibria_lemma1}
 Fix  any vertex $u \not\in {\sf Support}_{\ms}({\sf a})$. Then,
 \begin{eqnarray*}
\mathbb{P}_{\ms}({\sf Hit}(u))
&\geq &
 \mathbb{P}_{\ms}({\sf Hit}(v)) .
\end{eqnarray*}
\end{lemma}
\begin{proof}
Assume, by way of contradiction, that $\mathbb{P}_{\ms}({\sf Hit}(u)) < \mathbb{P}_{\ms}({\sf Hit}(v))$.
Define  ${\mt} = \ms_{-{\sf a}} \diamond \tau_{{\sf a}}$, where $\tau_{\sf a}$ is any mixed strategy of
attacker ${\sf a}$ such that   $u \in {\sf Support}_{\mt}({\sf a} )$. So, by construction,
$\mathbb{P}_{{\mt}}({\sf Hit}(u)) = \mathbb{P}_{\ms}({\sf Hit}(u))$.
 Then,
\begin{eqnarray*}
    \lefteqn{{\sf U}_{{\sf a}}(\ms_{-{\sf a}}\diamond \tau_{{\sf a}})}                                                      \\
= & 1 - \mathbb{P}_{{\mt}}({\sf Hit}(u))
  & \mbox{(since    $u \in {\sf Support}_{{\mt}}({\sf a})$)}                              \\
= & 1 - \mathbb{P}_{\ms}({\sf Hit}(u))
  &                                                                    \\
> & 1 - \mathbb{P}_{\ms}({\sf Hit}(v))
  & \mbox{(by assumption)}                                                                    \\
= & {\sf U}_{{\sf a}}(\ms) & \mbox{(since $v\in {\sf Support}_{\ms}({\sf a})$)},
\end{eqnarray*}
 a  contradiction.
\qed
\end{proof}

  We are now ready to prove Condition {\sf (1)}.
 Consider   any vertex $u \not\in {\sf Support}_{\ms}({\sf a})$ such that
$u \in {\sf Support}_{\ms}({\sf a}_{k})$ for some attacker ${\sf a}_k $.
 (If such a vertex does not exist, then
we are done). By Lemma \ref{characterization of nash equilibria_lemma1},  $\mathbb{P}_{\ms}({\sf Hit}(v))
\leq \mathbb{P}_{\ms}({\sf Hit}(u))$. Assume, by way of contradiction, that $\mathbb{P}_{\ms}({\sf Hit}(v)) <
\mathbb{P}_{\ms}({\sf Hit}(u))$.
 Since $\ms$ is a local maximizer of the Expected Utility of
  attacker ${\sf a}_{k}$, and ${\sf U}_{{\sf a}_{k}}(\ms )
=   1 - \mathbb{P}_{{\ms}}({\sf Hit}(u))$. Thus,
 $u \not\in {\sf
Support}_{\ms}({\sf a}_{k})$. A contradiction.

For Condition {\sf (2)}, fix a  defender ${\sf d}$ and consider an  edge $  (u, v) \in {\sf
Support}_{{\ms}}({\sf d})$.
 Since $\ms$ is a  Nash equilibrium, the quantity ${\sf Prop}_{\ms}({\sf d}, v') \cdot |{\sf A}|_{\ms}(v') + {\sf
Prop}_{\ms}({\sf d}, u') \cdot |{\sf A}|_{\ms}(u')$ is constant over all edges $(u', v') \in {\sf
Support}_{\ms}({\sf d})$. So, consider any edge $(u', v') \not\in {\sf Support}_{\ms}({\sf d})$. Assume, by
way of contradiction, that { \small
\begin{eqnarray*}
   & &    {\sf Prop}_{{\sf d}}({\ms_{-{\sf d}}} \diamond  u')    \cdot      |{\sf A}|_{\ms}(u')
    +       {\sf Prop}_{{\sf d}}({\ms_{-{\sf d}}} \diamond  v')      \cdot |{\sf A}|_{\ms}(v')\\
 & > &
 {\sf Prop}_{{\sf d}}({\ms_{-{\sf d}}} \diamond  u)   \cdot      |{\sf A}|_{\ms}(u)      +
     {\sf Prop}_{{\sf d}}({\ms_{-{\sf d}}} \diamond  v) \cdot      |{\sf A}|_{\ms}(v)  .
\end{eqnarray*}
} Denote ${\mt} = \ms_{-{\sf d}} \diamond \tau_{{\sf d}}$, where $\tau_{\sf d}$ is any mixed strategy of
defender ${\sf d}$ such that   $(u',v')  \in {\sf Support}_{\mt}({\sf d} )$. So, by construction, $|{\sf
A}|_{{\mt}}(u') = |{\sf A}|_{{\ms}}(u')$ and $|{\sf A}|_{{\mt}}(v') = |{\sf A}|_{{\ms}}(v')$. Then,
\begin{eqnarray*}
    \lefteqn{{\sf U}_{{\sf d}}({\ms}_{-{\sf d}}\diamond (u',v'))}                                                  \\
= &  {\sf Prop}_{{\sf d}}({\ms_{-{\sf d}}} \diamond  u')    \cdot    |{\sf A}|_{{\mt}}(u')    +
 {\sf Prop}_{{\sf d}}({\ms_{-{\sf d}}} \diamond  v')
\cdot    |{\sf A}|_{{\mt}}(v')
  & \mbox{(since $e' \in {\sf Support}_{{\mt}}({\sf d})$)}                               \\
= &  {\sf Prop}_{{\sf d}}({\ms_{-{\sf d}}} \diamond  u')    \cdot    |{\sf A}|_{\ms}(u') +     {\sf
Prop}_{{\sf d}}({\ms_{-{\sf d}}} \diamond  v')  \cdot |{\sf A}|_{\ms}(v')
  &                                        \\
> & {\sf Prop}_{{\sf d}}({\ms_{-{\sf d}}} \diamond  u)
    \cdot    |{\sf A}|_{\ms}(u)    +    {\sf Prop}_{{\sf d}}({\ms_{-{\sf d}}} \diamond  v)    \cdot
 |{\sf A}|_{\ms}(v)
  & \mbox{(by assumption)}                                                                       \\
= & {\sf U}_{{\sf d}}( {\ms} )  & \mbox{(since $(u,v) \in {\sf Support}_{{\ms}}({\sf d})$)}\, ,
\end{eqnarray*}
a contradiction.

Assume now that the mixed profile ${\ms}$ satisfies Conditions {\sf (1)} and {\sf (2)}. We will prove that
$\ms$ is a Nash equilibrium.
\begin{itemize}

\item Consider first an attacker ${\sf a}\in {\cal A}$. Then, for any  vertex $u \not\in {\sf Support}_{\ms}({\sf a})$,
\begin{eqnarray*}
       \lefteqn{{\sf U}_{{\sf a}}(\ms)}                                             \\
=    & 1 - \mathbb{P}_{\ms}({\sf Hit}(v))
     & \mbox{(where  $v \in {\sf Support}_{\ms}({\sf a})$)}                              \\
\geq & 1 - \mathbb{P}_{\ms}({\sf Hit}(u))   &
       \mbox{(by Condition {\sf (1)})}                                                           \\
=    & {\sf U}_{{\sf a}}(\ms_{-{\sf a} }\diamond  u)\, .
\end{eqnarray*}

\item Consider now a  defender ${\sf d}\in {\cal D}$. Then,
for any edge $  (u', v') \not\in {\sf Support}_{\ms}({\sf d})$,
\begin{eqnarray*}
    \lefteqn{{\sf U}_{{\sf d}}(\ms )}                                             \\
=  & {\sf Prop}_{{\sf d}}({\ms_{-{\sf d}}} \diamond  u)
    \cdot    |{\sf A}|_{\ms}(u)    +    {\sf Prop}_{{\sf d}}({\ms_{-{\sf d}}} \diamond  v)    \cdot    |{\sf A}|_{\ms}(v)
  &    \mbox{(where $(u,v)\in {\sf Support}_{\ms}({\sf d})$)}                                                                                       \\
\geq & {\sf Prop}_{{\sf d}}({\ms_{-{\sf d}}} \diamond  u')       \cdot       |{\sf A}|_{\ms}(u')       +
   {\sf Prop}_{{\sf d}}({\ms_{-{\sf d}}} \diamond  v')      \cdot |{\sf A}|_{\ms}(v')
  &    \mbox{(by Condition {\sf (2)})}\, .                                                     \\
\end{eqnarray*}
\end{itemize}
It follows that $\ms$ is a Nash equilibrium. The proof is now complete.
\qed\end{proof}

  \noindent We remark that Proposition~\ref{characterization of nash equilibria} generalizes a
corresponding characterization of Nash equilibria for ${\sf AD}_{\alpha, 1}(G)$ shown in~\cite[Theorem
3.1]{MPPS05a},  where Condition {\sf (2)} had the simpler counterpart {\sf (2')}:

\begin{proposition}
\label{characterization for one defender} A profile $\ms$ is a Nash equilibrium if and only if the following
conditions hold:
\begin{enumerate}
\item[{\sf (1)}] For each vertex $v \in {\sf Supports}_{\ms}({\cal A} )$, $\mathbb{P}_{\ms}
  ({\sf Hit}(v)) = {\sf MinHit}_{\ms}$.
\item[{\sf (2')}]  For each edge $e \in {\sf Supports}_{\ms}({\cal D} )$, $|{\sf A}|_{\ms}(e) =
\max_{e^{\prime} \in E}      \Big\{ |{\sf A}|_{\ms}(e')      \Big\}$.
\end{enumerate}
\end{proposition}

\subsection{Necessary Conditions}\label{Necessary Conditions}
 We now establish necessary conditions for Nash equilibria, which will follow from their
 characterization (Proposition \ref{characterization of nash equilibria}).
We first prove   a very simple expression for the {\em total} Expected Utility of the defenders:
\begin{proposition}
\label{sum of defender profits in a nash equilibrium} In a Nash equilibrium $\ms$,
\begin{eqnarray*}
\sum_{{\sf d} \in {\cal D}}   {\sf U}_{{\sf d}}(\ms)  & =& \alpha \cdot {\sf MinHit}_{\ms}.
\end{eqnarray*}
\end{proposition}
\begin{proof}
Clearly, { \small
\begin{eqnarray*}
   \lefteqn{\sum_{{\sf d}                   \in                   {\cal  D}}
        {\sf U}_{{\sf d}} ( \ms )}                                                        \\
= & \sum_{{\sf d} \in {\cal  D}}      \sum_{v \in V}        \mathbb{P}_{\ms} ({\sf Hit}({\sf d}, v)) \cdot
{\sf Prop}_{{\sf d}}({\ms_{-{\sf d}}} \diamond  v)        \cdot        |{\sf A}|_{\ms}(v)
  &  \mbox{(by Lemma \ref{expected utility of defender is})}                                                                                      \\
= & \sum_{v \in V}      \left( \sum_{{\sf d} \in {\cal  D}}               \mathbb{P}_{\ms} ({\sf Hit}({\sf
d}, v))               \cdot               {\sf Prop}_{{\sf d}}({\ms_{-{\sf d}}} \diamond  v)      \right)
\cdot |{\sf A}|_{\ms}(v)
  &                                                \\
= & \sum_{v \in V}      \mathbb{P}_{\ms}              ({\sf Hit}(v))      \cdot      |{\sf A}|_{\ms}(v)
  & \mbox{(by Lemma~\ref{tiny})}                                                            \\
= & \sum_{v \in {\sf Supports}_{\ms}({\cal A})}      \mathbb{P}_{\ms}              ({\sf Hit}(v))      \cdot
|{\sf A}|_{\ms}(v)
  &                                                                                         \\
= & \sum_{v \in {\sf Supports}_{\ms}({\cal A})}      {\sf MinHit}_{\ms}      \cdot      |{\sf A}|_{\ms}(v)
  & \mbox{(by Proposition~\ref{characterization of nash equilibria} (Condition {\sf (2)}))}
                                                                                            \\
= &  {\sf MinHit}_{\ms}   \cdot   \sum_{v \in {\sf Supports}_{\ms}({\cal A})}      |{\sf A}|_{\ms}(v)
  &                                                                                         \\
= & \alpha    \cdot    {\sf MinHit}_{\ms}\, & \mbox{(by Observation \ref{sum of VPs})} ,
\end{eqnarray*}
} as needed. \qed\end{proof}

 We continue to show:

\begin{proposition}
\label{edge cover necessary condition} In a Nash equilibrium $\ms$, ${\sf Supports}_{\ms}({\cal D} )$ is an Edge Cover.
\end{proposition}
\begin{proof}
Assume, by way of contradiction, that ${\sf Supports}_{\ms}({\cal D})$ is {\em not} an Edge Cover. Then,
choose a vertex $v \in V$ such that $v \not\in   {\sf Vertices}({\sf Supports}_{\ms}({\cal D}))$. So,
${\sf Edges}_{{\ms}}(v) = \emptyset$ and $\mathbb{P}_{\ms}({\sf Hit}(v)) = 0$.

Fix an attacker ${\sf a}\in {\cal A}$. Since $\ms$ is a local maximizer for the Expected Utility of     ${\sf a}$,
 which is at most $1$,
it follows that  $\sigma_{{\sf a}}(v) =1$. Hence, for each $  (u', v') \in {\sf Supports}_{\ms}({\cal D})$,
$|{\sf A}|_{\ms}((u',v')) = 0$, since   both $u'\neq u$ and $v'\neq v$ (by the choice  of vertex $v$). So,
  $|{\sf A}|_{\ms}(u') = |{\sf A}|_{\ms}(v') = 0$.  This implies that  for any defender ${\sf d}\in {\cal D}$,
\begin{eqnarray*}
      {\sf U}_{{\sf d}}({\ms}) & = & \sum_{  (u, v) \in
        {\sf Support}_{\ms}({\sf d})}        \sigma_{{\sf d}}((u,v)) \cdot \left( {\sf Prop}_{{\sf d}}({\ms_{-{\sf d}}} \diamond  u)
               \cdot               |{\sf A}|_{\ms}(u)               +
            {\sf Prop}_{{\sf d}}({\ms_{-{\sf d}}} \diamond  v)               \cdot               |{\sf A}|_{\ms}(v)
        \right)                                                                     \\
& = & 0\, .
\end{eqnarray*}
\noindent Since $\ms$ is a Nash equilibrium, ${\sf U}_{{\sf d}}(\ms)
 >
 0$. A contradiction.
\qed\end{proof}

\noindent Proposition \ref{edge cover necessary condition} immediately implies:
\begin{corollary}\label{unidefender nash is monodefender}
A unidefender Nash equilibrium is monodefender.
\end{corollary}

 We finally show:

\begin{proposition}
\label{vertex cover necessary condition} In a Nash equilibrium $\ms$, ${\sf Supports}_{\ms}({\cal  A} )$ is a
Vertex Cover of the graph $G({\sf Supports}_{\ms}({\cal D} ))$.
\end{proposition}
\begin{proof}
Assume, by way of contradiction, that ${\sf Supports}_{\ms}({\cal  A})$ is {\em not} a Vertex Cover of the
graph $G({\sf Supports}_{\ms}({\cal D}))$. Then, there is some edge $(u,v) \in {\sf Supports}_{\ms}({\cal
D})$ such that both  $ u \not\in {\sf Supports}_{\ms}({\cal A})$ and $ v \not\in {\sf Supports}_{\ms}({\cal
A})$. So,    $|{\sf A}|_{\ms}((u,v)) = 0$. Assume that $(u,v)\in {\sf Support}_{\ms}({\sf d})$ for some
defender ${\sf d}\in {\cal D}$. Since $\ms$ is a local maximizer for the Expected Utility of  defender ${\sf
d}$, it follows that $\sigma_{{\sf d}}((u,v)) = 0$. So, $(u,v) \not\in {\sf Support}_{\ms}({\sf  d})$. A
contradiction. \qed\end{proof}


 \subsection{Pure Nash Equilibria}\label{Pure Nash Equilibria}
 We observe that for the special case of Pure Nash equilibria,
Proposition~\ref{characterization of nash equilibria}
simplifies to:

\begin{proposition}[Characterization of Pure Nash Equilibria]
\label{characterization of pure nash equilibria} A pure profile ${\bf s}$ is a Pure Nash equilibrium if and
only if the following conditions hold:
\begin{enumerate}

\item[{\sf (1)}] ${\sf Supports}_{\ms}({\cal D} )$ is an Edge Cover.

\item[{\sf (2)}] For each attacker ${\sf d}\in {\cal D}$, for each edge $(u,v) \in {\sf Support}_{{\bf s}}({\sf d})$,
{ \small
\begin{eqnarray*}
      \frac{\textstyle |{\sf A}_{{\bf s}}(u)|}           {\textstyle |{\sf D}_{{\bf s}}(u)|}
 +
     \frac{\textstyle |{\sf A} _{{\bf s}}(v)|}           {\textstyle |{\sf D}_{{\bf s}}(v)|}
& = & \max_{(u',v') \in E}
     \left\{ \frac{\textstyle |{\sf A} _{{\bf s}}(u')|} {\textstyle |{\sf D}_{{\bf s}_{-  j}}(u')|
+ 1}                +
   \frac{\textstyle |{\sf A}_{{\bf s}}(v')|} {\textstyle |{\sf D}_{{\bf s}_{-j}}(v')| + 1}        \right\}\, .
\end{eqnarray*}
}
\end{enumerate}
\end{proposition}

  \noindent We now use Propositions~\ref{edge cover necessary condition} and~\ref{vertex cover necessary condition}
to show:

\begin{proposition}[Necessary Conditions for Pure Nash Equilibria]
\label{necessary condition for pure nash equilibria} Assume that $G$ is Pure. Then, {\sf (i)}
 $\delta \geq \beta'(G)$ and \myii  $\alpha \geq \min_{{\it EC} \in {\cal EC}(G)}   \beta (G(EC))$.
\end{proposition}
\begin{proof}
By contradiction. Consider a Pure Nash equilibrium ${\bf s}$. For Condition \myi,   assume   that $\delta < \beta'(G)$.
 Since
 $|{\sf Supports}_{{\bf s}}({\cal D})|
\leq \delta $, it follows that $|{\sf Supports}_{{\bf s}} ({\cal D})| <   \beta'(G)$. Hence, ${\sf
Supports}_{{\bf s}}({\cal D})$ is not an Edge Cover.
 A contradiction to   Proposition~\ref{edge
cover necessary condition}.

For Condition \myii, assume   that $\alpha < \min_{EC \in {\cal EC}(G)}   \beta (G(EC))$. Since   $|{\sf
Supports}_{{\bf s}}({\cal A})| \leq \alpha $,  it follows that
 $|{\sf Supports}_{{\bf
s}}({\cal  A})|    <  \min_{EC \in {\cal EC}(G)}   \beta (G(EC))$.   By Proposition~\ref{edge cover necessary
condition},  ${\sf Supports}_{{\bf s}}({\cal D})$ is an Edge Cover; so, $\beta(G({\sf
Supports}_{\ms}({\cal D}))) \geq \min_{EC\in {\cal EC}(G) } \beta(G(EC))$. It follows that $|{\sf
Supports}_{{\bf s}}({\cal A})| $ $ < \beta(G({\sf Supports}_{{\bf s}}({\cal D})))$. Thus, ${\sf
Supports}_{{\bf s}}({\cal A})$ is not a Vertex Cover of the graph  $G({\sf Supports}_{{\bf s}}({\cal D}))$. A
contradiction to Proposition~\ref{vertex cover necessary condition}. \qed\end{proof}

\begin{figure}[t]
\epsfclipon
  \centerline{\hbox{
 \epsfxsize=7.2cm
  \leavevmode
   \epsffile{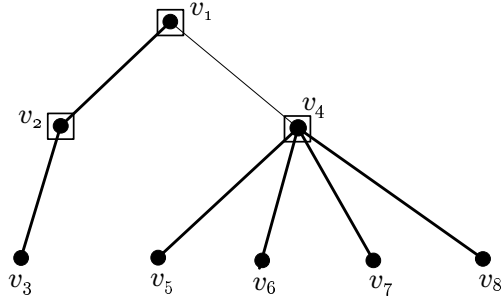}
  }
} \epsfclipoff \caption{The graph $G$ used in the proof of Proposition \ref{counterexample1}.
  Edges in  ${\sf Support
}_{\s}({\cal D})$ are drawn thick;  vertices in ${\sf Support }_{\s}({\cal A})$  are  squared.
 } \label{counterexample2figure}
\end{figure}

\noindent We remark that Condition \myi (resp., Condition \myii) in Proposition \ref{necessary condition for pure nash
equilibria} is necessary for Defender-Pure (resp., Attacker-Pure) Nash equilibria. We finally provide a counterexample to the converse of Proposition \ref{necessary condition for pure nash
equilibria}:

\begin{proposition}\label{counterexample2}
  There is a graph $G$ and integers $\alpha $ and  $\delta$ such that
{\sf (i)}  $\delta \geq \beta'(G)$ and \myii  $\alpha \geq \min_{{\it EC} \in {\cal EC}(G)}   \beta (G(EC))$
while  $ G $ is not   Pure.
\end{proposition}
\begin{proof}
Consider the graph  $G=(V, E)$ in Figure \ref{counterexample2figure}, and fix
  $\alpha =2$ and $\delta=6$. Clearly, $
\beta'(G)= 6$ and $\min_{{\it EC} \in {\cal EC}(G)} =2$. So, Conditions {\sf (i)} and {\sf (ii)} from the
claim hold. Towards a contradiction, assume that   $G$ is Pure; consider a Pure Nash equilibirum $\s$.
\begin{itemize}
\item By Proposition  \ref{edge cover necessary condition}, ${\sf Support }_{\s}({\cal D})$ is an Edge Cover. By the construction of $G$, this implies that ${\sf Supports
}_{\s}({\cal D}) = \{ (v_2, v_3) ,   (v_4, v_5), (v_4, v_6), (v_4, v_7),(v_4, v_8) , (v_1, v) \}$, where
$v\in\{ v_2, v_4  \}$. Since $\delta=6$, it follows that for each edge $e\in {\sf Supports
}_{\s}({\cal D})  $, there is a unique defender ${\sf d}$ such that $s_{{\sf d}} =e$.
\item By   Proposition    \ref{vertex cover necessary condition},   ${\sf
Supports }_{\s}({\cal A})$ is a Vertex Cover of the graph $G({\sf Supports }_{\s}({\cal D}))$. By the
construction of $G$, this implies that   ${\sf Supports }_{\s}({\cal A}) =\{ v_2,v_4  \}$. (Note that  $\{
v_2,v_4  \}$ is the {\em unique} Vertex Cover of the graph $G({\sf Supports }_{\s}({\cal D}))$ with size at
most $ 2$.) Since $\alpha = 2$, it follows that ${\sf Support }_{\s}({\sf a}_1) =v_2$ and ${\sf Support
}_{\s}({\sf a}_2) =v_4$.
\end{itemize}

\noindent Consider now the (unique) defender ${\sf d}\in {\cal D}$ such that $s_{{\sf d}} =(v_4,v_5)$.
Clearly,
   ${\sf U}_{{\sf d}}(\s) = \frac{\tx
1}{\tx 4}$, but
\begin{eqnarray*}
{\sf U}_{{\sf d}}(\s_{-{\sf d}} \diamond (v_2,v_3)) &=& \left\{  \begin{array}{ll}
 \frac{\tx 1}{\tx 3}, & \mbox{if $v=v_2$} \\
 \frac{\tx 1}{\tx 2}, & \mbox{if $v=v_4$}
\end{array}\right.\\
&\geq &  \frac{\tx 1}{\tx 3}.
\end{eqnarray*}
So,
${\sf U}_{{\sf d}}(\s_{-{\sf d}} \diamond (v_2,v_3)) > {\sf U}_{{\sf d}}(\s) $.
 A contradiction to the fact that   $\s$ is a
 Nash equilibrium.\qed\end{proof}

\section{Defense-Optimal Nash Equilibria} \label{Defense Optimal Graphs Section}
Section \ref{Defense-Optimal Nash Equilibria-Definition Section} introduces Defense-Optimal Nash equilibria
and Defense-Optimal graphs. Some sufficient conditions for Defense-Optimal graphs are presented in Section
\ref{Defense-Optimal Nash Equilibria-Sufficient Conditions Section}.

\subsection{Definitions}\label{Defense-Optimal Nash Equilibria-Definition Section}
  The \emph{\textbf{Defense-Ratio}} ${\sf DR}_{{\ms}}$ of a Nash equilibrium ${\ms}$
is the ratio of the {\em optimal} total Utility $\alpha$ of the defenders over their total Expected Utility
in ${\ms}$; so,
\begin{eqnarray*}
{\sf DR}_{{\ms}} &=& \frac{\textstyle \alpha}      {\textstyle \sum_{{\sf d}  \in {\cal  D}} {\sf U}_{{\sf d}
}( {\ms})} .
\end{eqnarray*}


  \noindent By the definition of Defense-Ratio, Proposition~\ref{sum of defender profits in a nash equilibrium}
immediately implies:

\begin{corollary}
\label{defense ratio} For  a Nash equilibrium $\ms$,
\begin{eqnarray*}
{\sf DR}_{\ms} &=& \frac{\textstyle 1}
    {\textstyle {\sf MinHit}_{\ms}}.
\end{eqnarray*}
\end{corollary}

  \noindent Clearly, ${\sf DR}_{\ms} \geq 1$.
  Furthermore, Lemma \ref{upper bound on minimum hitting probability}  implies a second lower bound on Defense-Ratio:

\begin{corollary}
\label{lower bound on defense ratio} For a Nash equilibrium $\ms$,
\begin{eqnarray*}
{\sf DR}_{\ms} &\geq& \frac{\textstyle |V|}
  {\textstyle 2\, \delta}.
\end{eqnarray*}
\end{corollary}

  \noindent Our next  major  definition  encompasses these two lower bounds
on Defense-Ratio.

\begin{definition}
\label{defense optimal nash equilibrium} A Nash equilibrium $\ms$ is {\em
\emph{\textbf{Defense-Optimal}}} if ${\sf DR}_{\ms} = \max \left\{ 1,
         \frac{\textstyle |V|}
    {\textstyle 2\, \delta} \right\}$.
\end{definition}

\noindent The justification for the definition of a Defense-Optimal Nash equilibrium will come later, when we
construct  Defense-Optimal Nash equilibria in two particular cases   (Proposition \ref{ridiculous one} and
Theorems~\ref{defender pure balanced equilibria} and~\ref{pure balanced equilibria});  these constructions
will establish that $\max \left\{ 1, \frac{\textstyle |V|} {\textstyle 2 \delta} \right\}$ is a {\em tight}
lower bound on Defense-Ratio.

 Say that that $G$ is \emph{\textbf{Defense-Optimal}}
  if $G$ admits  a Defense-Optimal Nash equilibrium.


\subsection{Sufficient Conditions}
\label{Defense-Optimal Nash Equilibria-Sufficient Conditions Section}

\noindent We  show:

\begin{theorem}
\label{lemma one} Assume that $G$ has a $\delta$-Partitionable Fractional Perfect Matching. Then,
 $G$ is Defense-Optimal.
\end{theorem}
 \begin{proof}
 Consider a $\delta$-Partitionable Fractional Perfect Matching $f$ and the corresponding (non-empty) partites  $E_1,
  \cdots , E_{\delta}$.  Recall that $E(f)$ is an Edge Cover.
Construct     $\ms$ as follows: 

 \begin{itemize}
  \item  For each attacker ${\sf a}  \in {\cal  A}$:
\begin{itemize}
\item For each vertex  $v\in V$,  set $\sigma_{\sf a}  (v) : = \frac{\textstyle 1}
            {\textstyle |V|}$; so,   ${\sf Support}_{\ms}({\sf a} ) =V$.
\end{itemize}
So, for each vertex $v\in V$, $|{\sf A}|_{\ms}(v)    = \sum_{{\sf a}\in {\cal A}} \frac{\textstyle 1}
            {\textstyle |V|} = \frac{\textstyle \alpha}
            {\textstyle |V|} $.
\item  For each defender ${\sf d}_j  \in {\cal D}$, with $j\in [\delta]$:
\begin{itemize}
\item  For each edge $e\in E$, set $\sigma_{{\sf d}_j }(e) :=
\frac {\textstyle 2\delta}      {\textstyle |V|}\cdot f(e)$ if $e \in E_j$, and $0$ otherwise; so, ${\sf
Support}_{\ms}({\sf d}_{j})  = E_j$ and all values of $\sigma_{{\sf d}_j}$ are non-negative.
\end{itemize}
 \end{itemize}

 \noindent Clearly, $\ms$ is  attacker-symmetric, attacker-uniform,   attacker-fullymixed and defender-symmetric; moreover,
$\ms$ is monodefender. Furthermore, for each vertex $v\in V$, ${\sf Edges}_{\ms}(v) =\{ e\in E(f) \mid v\in e
\}$.
 To   prove that $\ms$ is a (mixed) profile, we
prove that for each defender ${\sf d}_j \in {\cal D}$, $\sigma_{{\sf d}_j} $ is a probability distribution
(on $E$). Clearly,
\begin{eqnarray*}
\lefteqn{\sum_{e\in E}  \sigma_{{\sf d}_j}(e) }                         \\
 =& \sum_{e\in E_j}  \sigma_{{\sf d}_j}(e)
 &  \mbox{(since   ${\sf Support}_{\ms}({\sf d}_{j}) =E_j$)}      \\
  =& \sum_{e\in E_j}           \frac{\textstyle 2\delta }                {\textstyle |V|}           \cdot           f(e)
   & \mbox{(by construction)}                                        \\
  =& \frac{\textstyle 2\delta }          {\textstyle |V|}     \sum_{e\in E_j} f(e)
   &                                                                   \\
  =&  1
   & \mbox{(since $f$ is   $\delta$-Partitionable);}
\end{eqnarray*}
so,  $\sigma_{{\sf d}_j}$ is a probability distribution, which establishes that $\ms$ is a profile.

 \noindent We continue to prove that $\ms$ is a Nash equilibrium. We shall verify  Conditions
 {\sf (1)} and {\sf (2)} in the characterization of Nash equilibria
 ~(Proposition~\ref{characterization of nash equilibria}).


 For Condition {\sf (1)}, fix a  vertex $v\in V$. Since $E(f) $ is an Edge Cover,
 there is a  partite  $E_j\subseteq E(f) $  such that
$v\in {\sf Vertices}(E_j)$.  Since the partites  $E_1, \cdots , E_{\delta}$ are vertex-disjoint and $ {\sf
Support}_{\ms}({\sf d}_{j}) = E_j$, it follows that
  vertex $v$ is monodefender  in
$\ms$ with  ${\sf d}_{\ms}(v) = {\sf d}_j$.  We   prove:

\begin{claim}
\label{characterization of defense optimal graphs claim2}  $\mathbb{P}_{\ms}         ({\sf Hit}(v))  =
\frac{\textstyle 2\delta } {\textstyle |V|}$.
\end{claim}
\begin{proof}
Clearly,
\begin{eqnarray*}
\lefteqn{ \mathbb{P}_{\ms}
         ({\sf Hit}(v))}\\
 = &  \mathbb{P}_{\ms} ({\sf Hit}({\sf d}_j, v))
 &   \mbox{(since $v$ is monodefender in $\ms$)}     \\
 = & \sum_{e\in {\sf Support}_{\ms}({\sf d}_{j}) \mid v\in e} \sigma_{{\sf d}_{j}}(e)
   &                                                                                   \\
 = & \sum_{e\in {\sf Support}_{\ms}({\sf d}_{j})\mid v\in e }          \frac{\textstyle 2\delta }               {\textstyle |V|}          \cdot f(e)
   & \mbox{(by construction of $\ms$)}                                               \\
 = & \frac{\textstyle 2\delta }          {\textstyle |V|}     \sum_{e\in {\sf Support}_{\ms}({\sf d}_{j})\mid v\in e } f(e)
   &                                                                        \\
 = & \frac{\textstyle 2\delta }          {\textstyle |V|}     \sum_{e\in {\sf Edges}_{\ms}(v)} f(e)
   & \mbox{(since $v$ is monodefender in $\ms$)}                               \\
 = & \frac{\textstyle 2\delta }          {\textstyle |V|}     \sum_{e\in E(f) \mid  v \in e} f(e)
   &                              \\
 = & \frac{\textstyle 2\delta }      {\textstyle |V|}  \cdot  \sum_{e\in E \mid  v \in e} f(e) &
\mbox{(since $f(e) =0$ for $e\not\in E(f)$)}\\
 = & \frac{\textstyle 2\delta }      {\textstyle |V|}  \cdot 1 & \mbox{(since $f$ is a Fractional Perfect
Matching)},
 \end{eqnarray*}
as needed.
\qed\end{proof}

  \noindent By Claim \ref{characterization of defense optimal graphs claim2}, Condition {\sf (1)} holds trivially.

  For Condition {\sf (2)}, consider a  defender ${\sf d} \in {\cal  D}$. Fix an edge
$ (u,v) \in {\sf Support}_{\ms}({\sf d} )$. Since $\ms $ is monodefender,  ${\sf Prop}_{{\sf d}}(\ms_{-{\sf
d}} \diamond u) = {\sf Prop}_{{\sf d}}(\ms_{-{\sf d}} \diamond v) =1 $.
  Hence,
\begin{eqnarray*}
 {\sf Prop}_{{\sf d}}(\ms_{-{\sf
d}} \diamond u)\cdot  |{\sf A}|_{\ms}(u) +
  {\sf Prop}_{{\sf d}}(\ms_{-{\sf
d}} \diamond v) \cdot |{\sf A}|_{\ms}(v) & =&
 |{\sf A}|_{\ms}(u) + |{\sf A}|_{\ms}(v)  \\
& =&
 \frac{\textstyle 2\alpha}     {\textstyle |V|}\, .
\end{eqnarray*}
Fix now an  edge $  (u',v')\not\in  {\sf Support}_{\ms}({\sf d} )$. Since $E(f)$ is an Edge Cover, there are
edges $e_{u'}$ and  $e_{v'} \in E_f$ such that $u'\in e_{u'}$ and $v'\in e_{v'}$. By the construction of
$\ms$, this implies that there are defenders ${\sf d}_{u'}$ and ${\sf d}_{v'}$ such that $e_{u'} \in
 {\sf Support}_{\ms}(  {\sf d}_{u'})$ and  $e_{v'} \in  {\sf Support}_{\ms}({\sf d}_{v'})$.

There are two  cases for  ${\sf d}_{u'}$ (resp., ${\sf d}_{v'}$): either
 ${\sf d}_{u'} ={\sf d}$ or $ {\sf d}_{u'} \neq {\sf d}$ (resp., ${\sf d}_{v'} ={\sf d}$ or $ {\sf d}_{v'} \neq {\sf d}$).
We shall treat each of them separately.
\begin{itemize}
\item Assume first that ${\sf d}_{u'}={\sf d}$ (resp., ${\sf d}_{v'}={\sf d}$); since $u'$ is monodefender,
it follows that $ {\sf Prop}_{{\sf d}}(\ms_{-{\sf d}}\diamond u')  = 1$ (resp., $ {\sf Prop}_{{\sf
d}}(\ms_{-{\sf d}}\diamond v')    =1$).
\item Assume now that ${\sf d}_{u'}\neq {\sf d}  $ (resp., ${\sf d}_{v'}\neq {\sf d}$);
since $v'$ is monodefender, ${\sf Prop}_{{\sf d}}(\ms_{-{\sf d}}\diamond u') < 1$ (resp., ${\sf Prop}_{{\sf
d}}(\ms_{-{\sf d}}\diamond v') <1$).
\end{itemize}
So, in all cases, ${\sf Prop}_{{\sf d}}(\ms_{-{\sf d}}\diamond u') \leq 1$ and ${\sf Prop}_{{\sf
d}}(\ms_{-{\sf d}}\diamond v') \leq 1$. Thus,
\begin{eqnarray*}
{\sf Prop}_{{\sf d}}(\ms_{-{\sf d}}\diamond u') \cdot |{\sf A}|_{\ms}(u') + {\sf Prop}_{{\sf d}}(\ms_{-{\sf d}}\diamond v') \cdot |{\sf A}|_{\ms}(v') &\leq &  |{\sf A}|_{\ms}(u')       +       |{\sf A}|_{\ms}(v')\\
&=&
  \frac{\textstyle 2\alpha} {\textstyle |V|}\,.
\end{eqnarray*}
 Now, Condition {\sf (2)} follows.


  \noindent Hence, by Proposition~\ref{characterization of nash equilibria}, $\ms$ is a Nash equilibrium.
By Claim \ref{characterization of defense optimal graphs claim2}  and    Condition {\sf (1)} of Proposition
~\ref{characterization of nash equilibria}, it follows that  ${\sf MinHit}_{\ms} = \frac{\textstyle 2\delta}
{\textstyle |V|}$. By Corollary~\ref{defense ratio}, it follows that ${\sf DR}_{\ms} = \frac{\textstyle |V|}
{\textstyle 2\delta}$. Since $G$ has a $\delta$-Partitionable
  Fractional  Perfect Matching,   Corollary~\ref{Partitionable Fractional Perfect Matchings divides lemma} implies that
$\delta\leq \frac{\tx |V|}{\tx 2}$, so that
 $\max \left\{ 1,
         \frac{\textstyle |V|}
    {\textstyle 2\, \delta} \right\} = \frac{\tx |V|}{\tx 2\, \delta }$.
  This implies that   $ {\sf DR}_{\ms}  = \max \left\{ 1,
         \frac{\textstyle |V|}
    {\textstyle 2\, \delta} \right\}$. Hence,    $\ms$ is Defense-Optimal, as needed.
\qed\end{proof}


  We continue with another sufficient condition:
\begin{theorem}
\label{defender pure is defense optimal} Assume that $G$ is Defender-Pure. Then,     $G$ is
Defense-Optimal.
\end{theorem}

\begin{proof}
Fix an arbitrary Defender-Pure Nash equilibrium $\ms$; so ${\sf DR}_{\ms} =1$. For each defender ${\sf d} \in
{\cal D}$, denote $s_{{\sf d} } = (u_{\sf d}, v_{\sf d})\in E$. Since $\ms$ is a Nash equilibrium,  $ {\sf
U}_{{\sf d} }(\ms) =
   {\sf U}_{{\sf d} }(\ms_{-{\sf d} } \diamond (u_{\sf d}, v_{\sf d}) )$.  So,
{\small \begin{eqnarray*}
    \lefteqn{{\sf DR}_{\ms}}                                                               \\
= & \frac{\textstyle \alpha}         {\textstyle \sum_{{\sf d}  \in {\cal  D}} {\sf U}_{{\sf d} }(\ms)}
  &                                                                                            \\
= & \frac{\textstyle \alpha}         {\textstyle \sum_{{\sf d}  \in {\cal  D}}
   {\sf U}_{{\sf d} }(\ms_{-{\sf d} } \diamond  (u_{\sf d}, v_{\sf d}) )}
  & \\
&&   \\
= & \frac{\textstyle \alpha}         {\textstyle \sum_{{\sf d}  \in {\cal  D}} \left( {\sf Prop}_{{\sf
d}}(\ms_{-{\sf d}} \diamond  u_{{\sf d}})                       \cdot                       |{\sf
A}|_{\ms}(u_{{\sf d}}) + {\sf Prop}_{{\sf d}}(\ms_{-{\sf d}} \diamond  v_{{\sf d}})
                 \cdot |{\sf A}|_{\ms}(v_{{\sf d}}) \right)}  & \\
= & \frac{\textstyle \alpha}         {\textstyle \sum_{{\sf d}  \in {\cal  D}} \left( \frac{\textstyle |{\sf
A}|_{\ms}(u_{{\sf d}})}                            {\textstyle |{\sf D}_{\ms}(u_{{\sf d}})|} +
\frac{\textstyle |{\sf A}|_{\ms}(v_{{\sf d}})}                            {\textstyle |{\sf D}_{\ms}(v_{{\sf
d}})|} \right)}
  & \ \mbox{(since $\ms$ is Defender-Pure)} \\
= & \frac{\textstyle \alpha}         {\textstyle \sum_{v\in V} \sum_{{\sf d}  \in {\cal  D}\mid v \in {\sf
Vertices}_{\ms}({\sf Support}_{\ms}({\sf d} )) }
 \frac{\textstyle |{\sf A}|_{\ms}(v)} {\textstyle |{\sf D}_{\ms}(v)|}}   & \\
  = &
 \frac{\textstyle \alpha}         {\textstyle \sum_{v \in {\sf Supports}_{\ms}({\cal A})} |{\sf D} _{\ms}(v)|
\cdot \frac{\textstyle |{\sf A}|_{\ms}(v)} {\textstyle |{\sf D}_{\ms}(v)|}}
 & \\
 = &   \frac{\tx \alpha}         {\tx \sum_{v \in {\sf Supports}_{\ms}({\cal A})}            |{\sf A}|_{\ms}(v)}
          &    \\
 = & 1\, .   &
\end{eqnarray*}
}
By   Corollary~\ref{defense ratio}, it follows that     ${\sf MinHit}_{\ms} =1$. Hence,   Lemma \ref{upper
bound on minimum hitting probability} implies that   $\delta \geq \frac{\textstyle |V|}{\textstyle 2 } $.
Since ${\sf DR}_{\ms} = 1$,  it follows that    ${\sf DR}_{\ms}=
 \max \left\{ 1,
         \frac{\textstyle |V|}
    {\textstyle 2\, \delta} \right\}$,   and Condition   {\sf (i)} follows.
\qed\end{proof}

\noindent We finally compare the sufficient conditions for a Defense-Optimal graph from Theorems \ref{lemma
one} and \ref{defender pure is defense optimal}:

\begin{figure}[t]
\epsfclipon
  \centerline{\hbox{
  \epsfxsize=6.9cm
  \leavevmode
   \epsffile{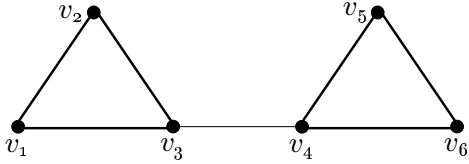}
  }
} \epsfclipoff \caption{The graph $G$ used in the proof of Proposition \ref{counterexample1}.
Edges in $E(f)$ for the $2$-Partitionable Fractional Perfect Matching $f$ are drawn thick.
 } \label{counterexample1figure}
\end{figure}

\begin{proposition}\label{counterexample1}
  There is a graph $G$ and an integer  $\delta$ such that
 $G$ has a  $\delta$-Partitionable Fractional Perfect Matching  while  $G$ is not Defender-Pure.
\end{proposition}
\begin{proof}
Consider the graph  $G=(V, E)$  in Figure \ref{counterexample1figure}, and fix   $\delta=2$. Consider the function
   $f : E \rightarrow [0,1]$ with $f(e) =\frac{\tx 1}{\tx 2}$  for each edge $e\in E \setminus \{(v_3,v_4)\} $
   and $f(e)=0$ for $e= (v_3, v_4)$. Clearly,     $f$ is a
2-Partitionable Fractional Perfect Matching with
$E_1= \{ (v_1, v_2), (v_2, v_3), (v_1,v_3)\}$ and $E_2= \{ (v_4, v_5), (v_5, v_6), (v_4,v_6)\}$. Since
$\delta =2$ and $\beta'(G)=3$, it follows by Proposition \ref{necessary condition for pure nash equilibria}
 (Condition {\sf (i)}) that $G$  is not Defender-Pure.
 \qed\end{proof}

\section{Few Defenders}
\label{few defenders}

We consider the case of \emph{\textbf{few defenders}}   where $\delta \leq \frac{\textstyle |V|} {\textstyle
2}$; there, a Defense-Optimal Nash equilibrium $\ms$ has Defense-Ratio ${\sf DR}_{\ms} = \max \left\{ 1,
\frac{\textstyle |V|} {\textstyle 2\, \delta}      \right\} = \frac{\textstyle |V|} {\textstyle 2 \delta}$,
so that by Corollary \ref{defense ratio}, $  {\sf MinHit}_{\ms} = \frac{\textstyle 2\delta}{\tx |V|}$. This
implies that $\sum_{v\in V}  \mathbb{P}_{\ms}({\sf Hit}(v))   \geq 2\delta$.  By Lemma \ref{upper bound on
sum of hit probabilities}, it follows that $\sum_{v\in V}  \mathbb{P}_{\ms}({\sf Hit}(v))   =2\delta$, so
that $\ms$ is unidefender. By Corollary \ref{unidefender nash is monodefender}, $\ms$ is monodefender.

   Section \ref{Necessary Condition for Defense-Optimal Nash Equilibria}
provides some necessary conditions for Defense-Optimal Nash equilibria and Defense-Optimal graphs. A
combinatorial characterization  of Defense-Optimal graphs is presented in
Section \ref{characterization of defense optimal graphs section}, with an implication on the associated complexity. Section \ref{Perfect Matching Graphs}
considers the special case of Perfect-Matching graphs.

\subsection{Necessary Conditions}\label{Necessary Condition for Defense-Optimal Nash Equilibria}

We   show a necessary condition for Defense-Optimal graphs:
\begin{proposition}
\label{lemma two} Assume that $\delta \leq \frac{\textstyle |V|}               {\textstyle 2}$. Then, a
Defense-Optimal graph      has a $\delta$-Partitionable Fractional Perfect Matching.
\end{proposition}
\begin{proof}
Consider a Defense-Optimal  Nash equilibrium $\ms$. Recall that $\ms$ is monodefender.
 Since  $ {\sf MinHit}_{\ms}(v) = \frac{\textstyle
2\delta} {\textstyle |V|}$ and $\sum_{v\in V} \mathbb{P}_{\ms}({\sf Hit}(v)) =2 \delta $, it follows that
  for each vertex $v\in V$, $\mathbb{P}_{\ms} ({\sf Hit}(v))  =\frac{\textstyle 2\delta}
{\textstyle |V|}$.

We  now define a function $f:E\rightarrow {\mathbb R}$; we will then prove that $f$ is a
$\delta$-Partitionable Fractional Perfect Matching. For each edge $e\in E$, set
\begin{eqnarray*}
f(e) &:=& \left\{
\begin{array}{ll}
\frac{\textstyle |V|}             {\textstyle 2\delta}        \cdot
  \sigma_{{\sf d}_{\ms}(e)} (e),   & \mbox{if }  e\in {\sf Supports}_{{\ms}}({\cal D})  \\
 0 , & \mbox{otherwise}
\end{array} \right. .
\end{eqnarray*}

\remove{
\begin{claim}
\label{characterization of defense optimal graphs claim3} $\sum_{e\in E} f(e) =   \frac{\textstyle |V|}
{\textstyle 2}$.
\end{claim}
\begin{proof}
By Corollary \ref{Defense-Optimal Nash equilibrium is monodefender}, $\ms$ is monodefender. Hence, by the
construction of $f$,
\begin{eqnarray*}
 \sum_{e\in E} f(e) &= & \sum_{e\in {\sf Supports}_{\ms}({\cal D})}f(e)\\
&= & \sum_{{\sf d}\in {\cal D}}\sum_{ e \in {\sf Support}_{\ms}({\sf d})} f(e)\\
&= & \sum_{{\sf d} \in {\cal  D}}        \sum_{e\in {\sf Support}_{\ms}({\sf d} )} \frac{\textstyle |V|}
{\textstyle 2\delta}              \cdot              \sigma_{{\sf d} }(e)\\
&= & \frac{\textstyle |V|} {\textstyle 2\delta}     \cdot      \sum_{{\sf d}  \in {\cal  D}} \sum_{e\in {\sf
Support}_{\ms}({\sf d} )} \sigma_{{\sf d} }(e)\\                                                        \\
&= & \frac{\textstyle |V|}         {\textstyle 2\delta}    \cdot    \delta \\
& = & \frac{\textstyle |V|}         {\textstyle 2 },
\end{eqnarray*}
as needed. \qed\end{proof}

}

By construction,    $E(f) = {\sf Supports}_{\ms}({\cal D})$; so, for each vertex $v\in V$, $ \{e\in E(f)\mid
v\in e \} = {\sf Edges}_{\ms}(v)$.
   Since $\ms$ is monodefender,  it follows that for each vertex $v\in V$,
$ \mathbb{P}_{\ms}            ({\sf Hit}(v))             = \mathbb{P}_{\ms}({\sf Hit}({\sf d}_{\ms}(v) ,
v))$.
 We prove:

\begin{claim}
\label{characterization of defense optimal graphs claim4} For each vertex $v\in V$,
\begin{eqnarray*}
\sum_{e\in {\sf Edges}_{\ms}(v) } f(e) &=& 1.
\end{eqnarray*}
\end{claim}
\begin{proof}
By the construction of $f$,
 \begin{eqnarray*}
  \sum_{e\in {\sf Edges}_\ms(v) } f(e) & = & \sum_{e\in {\sf Support}_{\ms}({\sf d}_{\ms}(v)) \mid v\in e} f(e) \\                                                                                      \\
& = & \sum_{e\in {\sf Support}_{\ms}({\sf d}_{\ms}(v))\mid v\in e }          \frac{\textstyle |V|}               {\textstyle 2\delta }          \cdot \sigma_{{\sf d}_{\ms}(v)}(e)\\
& = & \frac{\textstyle |V| }          {\textstyle 2\delta}     \sum_{e\in {\sf Support}_{\ms}({\sf
d}_{\ms}(v))\mid v\in e }  \sigma_{{\sf d}_{\ms}(v)}(e)
\\
& = & \frac{\textstyle |V| }          {\textstyle 2\delta}     \cdot     \mathbb{P}_{\ms}({\sf Hit}( {\sf
d}_{\ms}(v), v))
\\
& = & \frac{\textstyle |V| }          {\textstyle 2\delta}     \cdot     \mathbb{P}_{\ms}({\sf Hit}(   v))
\\
& = & \frac{\textstyle |V| }          {\textstyle 2\delta}     \cdot    \frac{\textstyle 2\delta} {\textstyle
|V| }
 \\
& =  & 1     ,
 \end{eqnarray*}
 as needed.
\qed\end{proof}

  \noindent Since ${\sf Edges}_{\ms}(v) =\{ e\in E(f)\mid v\in e \}$,  Claim \ref{characterization of defense optimal graphs claim4}
implies that  $f$ is a Fractional Perfect Matching. To prove that $f$ is $\delta$-Partitionable, define the
(non-empty) sets $E_1, \cdots, E_{\delta}$ where for each $j \in [\delta]$,  $E_j := {\sf Support}_{\ms}({\sf
d}_{j})$. Clearly,
\begin{eqnarray*}
\bigcup_{j\in[\delta] } E_j &=& \bigcup _{j \in [\delta]}  {\sf Support}_{\ms}({\sf d}_{j})\\
& =& {\sf Supports}_{\ms}( {\cal D} ) \\
&=& E(f) \,.
\end{eqnarray*}
Since $\ms$ is monodefender, it follows that for all pairs of distinct defenders ${\sf d}_k$ and ${\sf d}_l$, ${\sf
Support}_{\ms}({\sf d}_{k}) \cap {\sf Support}_{\ms}({\sf d}_{l}) =\emptyset$. Hence,  it follows that  the
sets $E_1, \cdots, E_{\delta}$, partition the set $E(f)$;  so, we shall call them partites. We observe:

\begin{claim}
\label{characterization of defense optimal graphs claim5} For each index $j\in [\delta]$,
 \begin{eqnarray*}
\sum_{e \in E_{j}} f(e) &=& \frac{\textstyle |V| } {\textstyle 2\delta}.
 \end{eqnarray*}
\end{claim}
\begin{proof}
By the construction of $f$ and the partites $E_1, \cdots, E_{\delta}$,
\begin{eqnarray*}
   \sum_{e\in E_j } f(e)
& = & \sum_{e\in {\sf Support}_{\ms}({\sf d}_{j})} f(e) \\
& = & \sum_{e\in {\sf Support}_{\ms}({\sf d}_{j})}          \frac{\textstyle |V|}               {\textstyle
2\delta }          \cdot          \sigma_{{\sf d}_j}(e)   \\
& = & \frac{\textstyle |V| }          {\textstyle 2\delta}     \sum_{e\in {\sf Support}_{\ms}({\sf d}_{j})}
\sigma_{{\sf d}_j}(e) \\
 & = & \frac{\textstyle |V| }          {\textstyle 2\delta},
\end{eqnarray*}
as needed. \qed\end{proof}

  \noindent Claim \ref{characterization of defense optimal graphs claim5} implies that $f$ is   $\delta$-Partitionable,
and the claim follows.   \qed
\end{proof}

\noindent Proposition \ref{lemma two} establishes that the sufficient condition for  a Defense-Optimal graph from
Theorem \ref{lemma one} is also necessary when  $\delta \leq \frac{\tx |V|}{\tx 2}$.

\subsection{Characterization and Complexity of Defense-Optimal Graphs}\label{characterization of defense optimal graphs section}

\noindent We now state a combinatorial characterization of Defense-Optimal graphs $\Big($for
$\delta\leq\frac{\tx |V| }{\tx 2}\Big)$; sufficiency and necessity follow from Theorem \ref{lemma one} and
Proposition \ref{lemma two}, respectively.

\begin{theorem}
\label{characterization of defense optimal graphs} Assume that $\delta \leq \frac{\textstyle |V|} {\textstyle
2}$. Then,  $G$  is Defense-Optimal if and only if $G$  has a $\delta$-Partitionable Fractional
Perfect Matching.
\end{theorem}

  \noindent  We observe three implications of Theorem \ref{characterization of defense optimal graphs}. The first one is an
immediate consequence of Theorem \ref{characterization of defense optimal graphs} and Corollary
\ref{Partitionable Fractional Perfect Matchings divides lemma}.

\begin{corollary}
\label{defense optimal graphs corollary} Assume that   $\delta \leq \frac{\textstyle |V|} {\textstyle 2}$ and
  $G$ is   Defense-Optimal. Then,   $\delta$ divides $|V|$.
\end{corollary}
\remove{
\begin{proof} Assume that   $G$ is Defense-Optimal.  By Theorem \ref{characterization of defense
optimal graphs}, $G$ has a $\delta$-Partitionable Fractional Perfect Matching. Corollary~\ref{Partitionable
Fractional Perfect Matchings divides lemma} implies now the claim. \qed\end{proof} }

  \noindent The second   implication is an immediate consequence
of Theorem \ref{characterization of defense optimal graphs} and Proposition  \ref{m-fractional perfect
matching for perfect graphs}.
\begin{corollary}
\label{impossibility result for many defenders no contradiction} Assume that $\delta =\frac{\textstyle |V|}
{\textstyle 2} $. Then,    $G$ is Defense-Optimal if and only if   it is Perfect-Matching.
\end{corollary}
\remove{
\begin{proof}
 Assume first that $G$ is Defense-Optimal. Since $\delta =\frac{\textstyle |V|}      {\textstyle 2} $,
 Theorem \ref{characterization of defense optimal graphs} implies that
 $G$  has a $\frac{\textstyle |V|}      {\textstyle 2}$-Partitionable Fractional Perfect
Matching.  So,
 Proposition \ref{m-fractional perfect matching for perfect graphs} implies that
$ \beta'(G) = \frac{\textstyle |V|}      {\textstyle 2}    $.
  Assume now that   $\beta'(G) =\frac{\textstyle |V|}      {\textstyle 2} $.
 Proposition \ref{m-fractional perfect matching for perfect graphs} implies that $G$ has
 a $ \frac{\textstyle |V|}      {\textstyle 2}    $-Partitionable Fractional Perfect
Matching. Hence, Theorem \ref{characterization of defense optimal graphs} implies that $G$ is Defense-Optimal.
\qed
\end{proof}
}

\noindent Corollary \ref{impossibility result for many defenders no contradiction} identifies a particular
value of $\delta$ $\Big($namely, $\delta =\frac{\tx |V|}{\tx 2}\Big)$ for which the recognition problem for
Defense-Optimal graphs is tractable. For the third implication,     Theorem~\ref{characterization of defense optimal graphs}
implies that   the complexity of recognizing Defense-Optimal
 graphs is that of      {\sf $\delta$-PARTITIONABLE FPM}.
Hence, Proposition~\ref{Fractional Perfect Matching theo NPcompleteness1}   immediately implies:

\begin{corollary}\label{Fractional Perfect Matching theo NPcompleteness}
 Assume that $\delta \leq \frac{\textstyle |V|}{\textstyle 2}$. Then, the recognition problem for
Defense-Optimal graphs    is ${\cal NP}$-complete.
\end{corollary}

\subsection{Perfect-Matching Graphs}\label{Perfect Matching Graphs}
  We    show:

\begin{theorem}
\label{perfect matching few defenders characterization} Assume that $\delta \leq \frac{\textstyle |V|}
{\textstyle 2}$ for  a Perfect-Matching graph $G$. Then, $G$ admits a Defense-Optimal, Perfect-Matching Nash
equilibrium  if and only if $2\, \delta$ divides $|V|$.
\end{theorem}
\begin{proof}
\noindent The claim   will follow from Propositions~\ref{ridiculous one} and~\ref{ridiculous two}.
\begin{proposition}
\label{ridiculous one}  Assume that  $\delta \leq \frac{\textstyle |V|} {\textstyle 2}$ for a
Perfect-Matching graph $G$, where     $2\, \delta$ divides $|V|$. Then,  $G$ admits a Defense-Optimal,
Perfect-Matching Nash equilibrium.
\end{proposition}
\begin{proof}
Consider a Perfect Matching $M$. Construct a profile $\ms$ as follows:
\begin{itemize}

\item For each attacker
${\sf a}  \in {\cal  A}$ and for each vertex $v\in V$, set
\begin{eqnarray*}
\sigma_{{\sf a} }(v) &: =& \frac{\textstyle 1 } {\textstyle |V|}  .
\end{eqnarray*}
So, $\ms$  is attacker-symmetric, attacker-uniform and attacker-fullymixed. Clearly,   for each vertex $v\in
V$,    $|{\sf A}|_{\ms}(v) = \frac{\textstyle \alpha} {\textstyle |V|}$.

\item Partition $M$ into $\delta $  sets $M_1, \cdots, M_{\delta}$, each with $\frac{\textstyle |V|}      {\textstyle 2\delta}$
edges; each defender ${\sf d}_j$ with $j \in[\delta] $ uses a uniform probability distribution  over  the set
$M_j$.  So, for each edge $e\in M_j$, set
\begin{eqnarray*}
\ms_{{\sf d}_{\ms}(e)}(e) &:=& \frac{\tx 2\delta }{\tx |V|}.
\end{eqnarray*}
 Thus,  ${\sf Support}_{\ms}({\sf d}_j ) = M_j$
for each $j \in [\delta]$, so that  $ {\sf Supports}_{\ms}({\cal D}) = M $. Clearly, each vertex $v\in V$ is
monodefender in $\ms$ with $\mathbb{P}_{\ms}         ({\sf Hit}(v)) = \mathbb{P}_{\ms} ({\sf Hit}({\sf
d}_{\ms}(v),v)) = \frac{\textstyle 2\delta }      {\textstyle |V|}$.
\end{itemize}

  \noindent We shall verify Conditions  {\sf (1)} and {\sf (2)} in the characterization of Nash
equilibria (Proposition~\ref{characterization of nash equilibria}).
 For Condition {\sf (1)}, fix a vertex $v\in V$. Since
 $\mathbb{P}_{\ms}   ({\sf Hit}(v))  =
\frac{\textstyle 2\delta }      {\textstyle |V|}$,   Condition {\sf (1)} follows trivially.
   For Condition {\sf (2)}, consider any defender ${\sf d} \in {\cal  D}$.
\begin{itemize}
\item
Fix an edge $(u,v) \in {\sf Support}_{\ms}({\sf d} )$. Since each edge  is monodefender in $\ms$, it follows
that ${\sf Prop}_{{\sf d}}( {\ms}_{-{\sf d}} \diamond u) = {\sf Prop}_{{\sf d}}( {\ms}_{-{\sf d}} \diamond v)
= 1$. Hence,
\begin{eqnarray*}
  {\sf Prop}_{{\sf d}}(\ms_{-{\sf d}} \diamond  u) \cdot |{\sf A}|_{\ms}(u)
+ {\sf Prop}_{ {\sf d}}(\ms_{-{\sf d}} \diamond  v) \cdot |{\sf A}|_{\ms}(v)
 &= & \frac{\textstyle 2\alpha}     {\textstyle |V|}.
\end{eqnarray*}
\item
Fix now an  edge $ (u',v')\not\in  {\sf Support}_{\ms}({\sf d} )$. Since $M$ is an Edge Cover,  there are
edges $e_{u'}, e_{v'}\in M$  such that $u' \in e_{u'}$ and $v' \in e_{v'}$. By the construction of $\ms$,
this implies that there are defenders ${\sf d}_{u'}$ and ${\sf d}_{v'}$ such that $ e_{u'} \in  {\sf
Support}_{\ms}({\sf d}_{ u'})$ and   $ e_{v'} \in  {\sf Support}_{\ms}({\sf d}_{ v'})$. Since each vertex is
monodefender in $\ms$, it follows that ${\sf d} \neq {\sf d}_{u'} $ and ${\sf d} \neq {\sf d}_{v'} $.
 Hence,
${\sf Prop}_{{\sf d}}( {\ms}_{-{\sf d}} \diamond u')  \leq \frac{\tx 1}{\tx 2}   $ and  ${\sf Prop}_{{\sf
d}}( {\ms}_{-{\sf d}} \diamond v) \leq \frac{\tx 1}{\tx 2}$, so that
\begin{eqnarray*}
{\sf Prop}_{{\sf d}}( {\ms}_{-{\sf d}} \diamond u')       \cdot      |{\sf A}|_{\ms}(u') +
 {\sf Prop}_{{\sf d}}( {\ms}_{-{\sf d}} \diamond v')
\cdot      |{\sf A}|_{\ms}(v')  & \leq &
\frac{\tx 1}{\tx 2} \cdot \left( |{\sf A}|_{\ms}(u') + |{\sf A}|_{\ms}(v') \right)\\
&= & \frac{\textstyle  \alpha} {\textstyle |V|}\,.
\end{eqnarray*}
\end{itemize}
Now, Condition {\sf (2)} follows.
 Hence, by Proposition~\ref{characterization of nash equilibria}, $\ms$ is a Nash equilibrium.

 To prove  that $\ms$ is Defense-Optimal, recall that for each vertex $v\in V$,  $\mathbb{P}_{\ms}
     ({\sf Hit}(v))  =\frac{\textstyle 2\delta}        {\textstyle |V|}$. Hence, ${\sf MinHit}_{\ms} =
\frac{\textstyle 2\delta}     {\textstyle |V|}$. By Corollary~\ref{defense ratio}, it follows that  ${\sf
DR}_{\ms} = \frac{\textstyle  |V|}      {\textstyle 2\delta}$. Since $\delta\leq \frac{\tx |V|}{\tx 2}$,
 it follows that ${\sf DR}_{\ms}=
 \max \left\{ 1,
         \frac{\textstyle |V|}
    {\textstyle 2\, \delta} \right\} $.
Hence,      $\ms$ is Defense-Optimal.
\qed\end{proof}

\noindent We continue to prove:
\begin{proposition}
\label{ridiculous two} Assume that   $\delta \leq \frac{\textstyle |V|} {\textstyle 2}$ for a
Perfect-Matching graph $G$, which   admits a Defense-Optimal,  Perfect-Matching  Nash equilibrium. Then, $ 2\,
\delta$ divides $|V|$.
\end{proposition}
\begin{proof}
Consider such a Nash equilibrium $\ms$, and  recall that   ${\sf MinHit}_{\ms}   = \frac{\textstyle 2\delta}
{\textstyle |V|}$. Consider an edge $(u,v)\in {\sf Supports}_{\ms}(  {\cal D} )$; so, $e\in {\sf
Support}_{\ms}({\sf d})$ for some defender ${\sf d}\in {\cal  D}$.
Proposition \ref{vertex cover necessary condition} implies that ${\sf Supports}_{\ms}({\cal A} )$ is a Vertex
Cover of the graph $G( {\sf Supports}_{\ms}( {\cal D} ))$. Hence, either    $u\in {\sf Supports}_{\ms}(
{\cal A} )$ or $v \in {\sf Supports}_{\ms}( {\cal A} )$ (or both). Assume without loss of generality, that
$u\in {\sf Supports}_{\ms}( {\cal A} )$.
 Since $\ms$ is  monodefender,
  there is a single  defender ${\sf d}_k$ such that $u\in {\sf Vertices}  ( {\sf
Support}_{\ms}({\sf d}))$.
Hence,   ${\sf d}_k$ is identified with ${\sf d}$. Since $\ms$ is Perfect-Matching, $ {\sf
Support}_{\ms}( {\sf d} )$  is a Perfect Matching; this implies that $ \mathbb{P}_{\ms} ({\sf Hit}(v)) = s
_{{\sf d}}(e)$. We prove:

\begin{claim}
\label{perfect matching few defenders characterization claim1} $| {\sf Support}_{\ms}({\sf d})  | =
\frac{\textstyle |V|}     {\textstyle 2\delta}$
\end{claim}
\begin{proof} Clearly,
 \begin{eqnarray*}
  \sum_{e\in {\sf Support}_{\ms}({\sf d})} \sigma_{{\sf d}} (e)
  &
 =&  \sum_{e\in {\sf Support}_{\ms}({\sf d})}
                          \mathbb{P}_{\ms}({\sf Hit}(v))\\
& =&  |{\sf Support}_{\ms}({\sf d})|     \cdot     \frac{\textstyle 2\delta }          {\textstyle |V|}
 \end{eqnarray*}
Since $\ms$ is a profile,    $\sum_{e\in {\sf Support}_{\ms}({\sf d})}      \sigma_{{\sf d}} (e) = 1$.
Hence,  $|{\sf Support}_{\ms}({\sf d})| = \frac{\textstyle |V|}     {\textstyle 2\delta } $, as needed.
\qed
\end{proof}

  \noindent
Claim~\ref{perfect matching few defenders characterization claim1} immediately implies that $2 \delta$
divides $|V|$, as needed. \qed\end{proof}

 \noindent The claim follows now from Propositions \ref{ridiculous one} and \ref{ridiculous two}.
  \qed\end{proof}

 \noindent Note that while Corollary~\ref{defense optimal graphs corollary} applies to {\em all} graphs,
Proposition~\ref{ridiculous two} applies only to  Perfect-Matching  graphs. However,   the restriction of
Corollary~\ref{defense optimal graphs corollary} to  Perfect-Matching graphs  does {\em not} imply
Proposition~\ref{ridiculous two} {\em unless} $\delta$ is odd. (This is because $2$  divides $|V|$ and
$\delta$ divides $|V|$ imply together that $2 \delta$ divides $|V|$ exactly when $\delta$ is odd.) Hence,
Proposition \ref{ridiculous two} strictly strengthens Corollary~\ref{defense optimal graphs corollary} for
the case where $\delta$ is even.


\section{Many Defenders}
\label{moderate defenders}

  We now consider the case of \emph{\textbf{many defenders}}, where $\frac{\textstyle |V|}
  {\textstyle 2} < \delta < \beta'(G)$. In this case, a Defense-Optimal Nash equilibrium $\ms$
has Defense-Ratio ${\sf DR}_{\ms} = \max \left\{ 1, \frac{\textstyle |V|} {\textstyle 2\, \delta}  \right\} =
1$. By Corollary \ref{defense ratio}, this implies that ${\sf MinHit}_{{\ms}} =1$. It follows that for each
vertex $v\in V$,  $\mathbb{P}_{\ms}({\sf Hit}(v))  =1$, so that the number of maxhit vertices in $\ms$ is
$|V|$.  We show:

 \begin{theorem}
\label{impossibility result for many defenders} Assume that $\frac{\textstyle |V|}      {\textstyle 2} <
\delta < \beta'(G)$. Then, $G$ is not   Defense-Optimal.
\end{theorem}
\begin{proof}
Towards   a  contradiction, consider a Defense-Optimal Nash equilibrium $\ms$. Consider any   (maxhit) vertex
$v \in V$. By Lemma   \ref{PsHit equals one case},   there is a    maxhitter ${\sf d}  \in {\cal D}$ in $\ms$
with  $\mathbb{P}_{\ms}({\sf Hit}({\sf d} , v))    =    1$.
 Use $\ms$ to construct a defender-pure profile ${\mt}$ as follows:
\begin{itemize}
\item Fix a defender   ${\sf d}\in {\cal D} $. If ${\sf d}$ is maxhitter in $\ms$, then
${\mt}_{{\sf d}}$ is any edge  $(u,v) \in {\sf Support}_{\ms} ({\sf d} )$ such that   ${\sf d} $ is maxhitter in $\ms$ for the
vertex $v\in V$; else, $\mt_{\sf d}$ is any arbitrary edge $(u,v)\in {\sf Support}_{\ms} ({\sf d})$.
\end{itemize}

 By construction of ${\mt}$, the following conditions hold:
  \begin{itemize}
 \item[{\sf (1)}] $|{\sf Supports}_{{\mt}}({\cal D}) |\leq \delta$.
 \item[{\sf (2)}] Each maxhit vertex in $\ms$ remains a maxhit vertex in $\mt$; so, the number of maxhit vertices in $\mt$
is $|V|$.
 \end{itemize}

 Since $\delta < \beta^{'}(G)$, Condition {\sf (1)} implies that $|{\sf Supports}_{{\mt}}({\cal  D})| < \beta'(G)$.  Hence,
 ${\sf Supports}_{{\mt}}({\cal D})$ is not an Edge Cover. So, there is some vertex $v \in V$ with
$\mathbb{P}_{{\mt}} ({\sf Hit}(v)) = 0$. It follows that the number
 of maxhit vertices in ${\mt}$ is less than $|V| $.   A contradiction.
\qed\end{proof}

\section{Too Many Defenders}
\label{too many defenders}

  \noindent We finally  turn to the case of \emph{\textbf{too many defenders}}   where $\delta \geq \beta'(G)$.
In this case, $\frac{\textstyle |V|} {\textstyle 2\, \delta} \leq \frac{\textstyle |V|} {\textstyle 2
\beta'(G)}\leq 1$; so, a   Defense-Optimal Nash equilibrium $\ms$ has Defense-Ratio
 ${\sf DR}_{\ms} =   1  $. By Corollary \ref{defense ratio}, this implies that ${\sf MinHit}_{{\ms}} =1$;
so,  that for each vertex $v\in V$, $\mathbb{P}_{\ms}({\sf Hit}(v))  =1$.

 \noindent Section \ref{Balanced Profiles Section} introduces     vertex-balanced profiles. These profiles give rise to Defender-Pure,  Vertex-Balanced Nash
equilibria and Pure, Vertex-Balanced Nash equilibria, which will be treated in Sections \ref{defender-pure
Balanced Nash Equilibria Section} and \ref{Pure Balanced Nash Equilibria Section}, respectively.

\subsection{(Defender-Pure and Pure,) Vertex-Balanced Profiles}\label{Balanced Profiles Section}

We start with a significant definition:

\begin{definition}\label{Defender-pure Vertex-balanced Profile definition}
A mixed profile  $\ms $ is {\textbf{\em\emph{vertex-$\!\!$ balanced}}} if   there is a constant $c
> 0$ such that for each vertex $v \in V$,
\begin{eqnarray*}
\frac{\textstyle |{\sf A}|_{\ms}(v)}      {\textstyle |{\sf D}_{\ms}(v)|} &=& c.
\end{eqnarray*}
\end{definition}

The following properties follow trivially for a   vertex-balanced profile $\ms$:

\begin{itemize}
\item[{\sf 1}] The set ${\sf Supports}_{{\ms}}({\cal D})$ is an Edge Cover. This matches the necessary condition
for an arbitrary Nash equilibrium from Proposition \ref{edge cover necessary condition}.

\item[{\sf 2}] The set ${\sf Supports}_{{\ms}}({\cal A})$ is $V$. Note that this property is strictly weaker
than the   condition defining an  attacker-fullymixed profile $\ms$, which requires that for each attacker
${\sf a}\in {\cal A}$, ${\sf Support}_{{\ms}}( {\sf  a} ) = V$.
\end{itemize}

 We shall consider defender-pure,  vertex-balanced profiles and pure, vertex-balanced
profiles. We prove  a nice property of  defender-pure, vertex-balanced profiles.

\begin{proposition}
\label{defender pure balanced is defender nash} A defender-pure,  vertex-balanced  profile is a local
maximizer for the Expected Utility of each defender.
\end{proposition}
\begin{proof}
Consider  such  a   profile $\ms$ and a defender ${\sf d}  \in {\cal D}$ with
   $ \sigma_{\sf d} = (u, v)   $. Clearly,
\begin{eqnarray*}
 {\sf U}_{{\sf d} }(\ms) & =& \frac{\textstyle |{\sf A}|_{\ms}(u)}
{\textstyle |{\sf D}_{\ms}(u)|} + \frac{\textstyle |{\sf A}|_{\ms}(v)}      {\textstyle |{\sf D}_{\ms} (v)|}\\
& = & 2c.
\end{eqnarray*}
  Fix now    an edge $ (u', v') \not\in {\sf Support}_{\ms}({\sf d} )$. Clearly,
\begin{eqnarray*}
     {\sf U}_{{\sf d} }(\ms_{-{\sf d} } \diamond (u', v'))
 & =& \frac{\textstyle |{\sf A}|_{\ms}(u')} {\textstyle |{\sf D}_{\ms}(u')| + 1} + \frac{\textstyle |{\sf A}|_{\ms}(v')} {\textstyle |{\sf D}_{\ms}(v')| + 1}   \\
  & < & \frac{\textstyle |{\sf A}|_{\ms}(u')} {\textstyle |{\sf D}_{\ms}(u')|  } + \frac{\textstyle |{\sf A}|_{\ms}(v')} {\textstyle |{\sf D}_{\ms}(v')| }   \\
 &=  & 2c ,
\end{eqnarray*}
 and the claim follows.
 \qed
 \end{proof}

\noindent Proposition \ref{defender pure balanced is defender nash} implies that a defender-pure,
vertex-balanced profile, which is a local maximizer for the Expected Utility of each attacker, is a Nash
equilibrium.
 We shall present polynomial time algorithms to compute Defender-Pure, Vertex-Balanced Nash equilibria
 and Pure, Vertex-Balanced Nash equilibria which are Defense-Optimal for the case where $\delta \geq \beta'(G)$; the second algorithm will require an additional assumption.

\subsection{Defense-Optimal, Defender-Pure, Vertex-Balanced Nash Equilibria}
\label{defender-pure Balanced Nash Equilibria Section}
\begin{figure}[h]
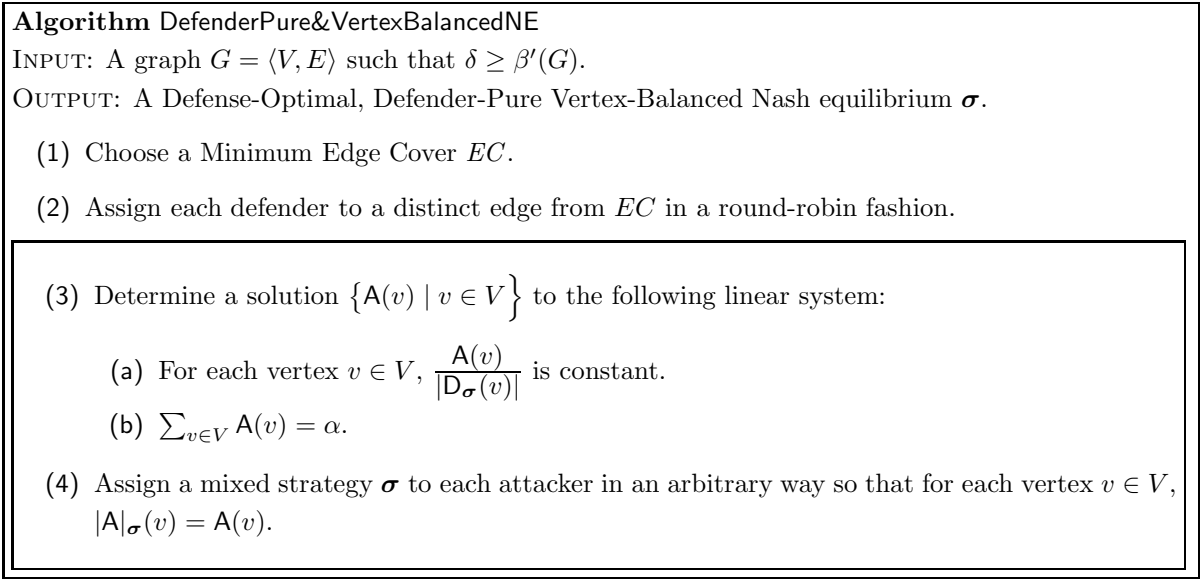

\fbox{\parbox{15.5cm}{ { \small

{\bf Algorithm} {\sf DefenderPure\&VertexBalancedNE}

{\sc Input}:  A graph  $G=\langle V,E\rangle$        such that $\delta \geq  \beta'(G) $.

{\sc Output}: A Defense-Optimal, Defender-Pure Vertex-Balanced Nash equilibrium $\ms$.
\begin{enumerate}

\item[{\sf (1)}] Choose  a Minimum Edge Cover ${\it EC}  $.

\item[{\sf (2)}] Assign each defender  to a distinct edge  from $EC$ in a round-robin fashion.
\end{enumerate}
\fbox{\parbox{15.2cm}{
\begin{enumerate}
\item[{\sf (3)}] Determine a solution $\big\{  {\sf A} (v)    \mid    v \in V \Big\}$ to the
following linear system:
\begin{itemize}
\item[{\sf (a)}] For each  vertex $v \in V$,
$\frac{\textstyle  {\sf A} (v)}      {\textstyle |{\sf D}_{\ms}(v)|}  $ is  constant.
\item[{\sf (b)}] $\sum_{v \in V}    {\sf A} (v) = \alpha$.
\end{itemize}
\item[{\sf (4)}] Assign a mixed strategy $\ms$ to each attacker in an arbitrary
way so that  for each vertex
  $v \in V$,\\ $|{\sf A}|_{\ms}      (v ) =  {\sf A} (v )$.
\end{enumerate}
}}
 }}} \caption{The algorithm {\sf DefenderPure\&VertexBalancedNE}. By Step {\sf (2)}, $\ms$ is
defender-pure; note that the assignment exchausts all edges from the Minimum Edge Cover $EC$ due to the assumption that
$\delta\geq \beta'(G)$.
Step {\sf (3)} provisions for  $\ms$ to be vertex-balanced; towards this end, it provides for
the ratio $\frac{\textstyle {\sf A} (v)}      {\textstyle |{\sf D}_{\ms}(v)|}$ to be constant over all
vertices $v\in V$. Finally, Step {(\sf 4)} provides mixed strategies to the attackers that induce $| {\sf A}
|_{\ms}(v) = {\sf A} (v)$ for each vertex $v\in V$; by Step {\sf (3/a)} this implies that   $\ms$ is
vertex-balanced. Since a Minimum Edge Cover is computable in polynomial time, the algorithm
{\sf DefenderPure\&VertexBalancedNE} is polynomial time.    }\label{DefenderPure_VertexBalancedNE_alg}
\end{figure}

 We show:

\begin{theorem}
\label{defender pure balanced equilibria} Assume that $\delta \geq \beta'(G)$. Then, $G$ admits a
   Defender-Pure, Vertex-Balanced      Nash equilibrium, which is computable in polynomial time.
\end{theorem}
 \noindent
 To prove the claim, we   present
 the    algorithm {\sf DefenderPure\&VertexBalancedNE} in  Figure \ref{DefenderPure_VertexBalancedNE_alg}.

\begin{proof}
\remove{ \noindent Since a Minimum Edge Cover can be computed in polynomial time, it follows that algorithm
{\sf defender-pure\&Balanced} runs in polynomial time. We continue to establish correctness of the algorithm.
}  By construction (Steps {\sf (1)} and {\sf (2)}) and the assumption that $\delta \geq \beta'(G)$, it
follows that ${\sf Supports}_{\ms}({\cal D})$ is a Minimum Edge Cover. Since $\ms$ is defender-pure, this
implies that for each vertex $v \in V$, $\mathbb{P}_{\ms}({\sf Hit}(v)) = 1$; hence, for each attacker ${\sf
a}\in {\cal A}$,
\begin{eqnarray*}
{\sf U}_{{\sf a} }(\ms_{-{\sf a} } \diamond v) &=& 1 - \mathbb{P}_{\ms}({\sf Hit}(v)) \\
&= & 0.
\end{eqnarray*}
This implies that   $\ms$ is (vacuously) a local maximizer for the Expected Utility of each attacker. By
Proposition \ref{defender pure balanced is defender nash}, it follows that  $\ms$ is a Nash equilibrium.
  \qed\end{proof}

   By Theorem \ref{defender pure is defense optimal} and Theorem \ref{defender pure balanced
equilibria}, it immediately follows:

\begin{corollary}\label{d more than beta' means Defense Optimal G}
Assume that $\delta\geq \beta'(G)$. Then, $G$ is Defense Optimal.
 \end{corollary}

\noindent By Theorem \ref{necessary condition for pure nash equilibria} (Condition {\sf (i)}) and Theorem
\ref{defender pure balanced equilibria}, it finally follows:

\begin{corollary}\label{defender pure if and only if condition}
$G$ is Defender-Pure if and only if    $\delta \geq \beta'(G)  $.
\end{corollary}

\noindent Since a Minimum Edge Cover is computable in polynomial time, Corollary \ref{defender pure if and
only if condition} implies that the class of Defender-Pure graphs is recognizable in polynomial time (for an
arbitrary value of $\delta$).

  \subsection{Defense-Optimal, Pure, Vertex-Balanced   Nash Equilibria}\label{Pure Balanced Nash Equilibria Section}
  We show:

\begin{theorem}
\label{pure balanced equilibria} Assume that $\delta \geq \beta'(G)$ and $2\, \delta$ divides $\alpha$. Then,
$G$ admits a Defense-Optimal, Pure, Vertex-Balanced   Nash equilibrium, which is computable in polynomial
time.
\end{theorem}

 \noindent To prove the claim, we present  the
 algorithm {\sf Pure\&VertexBalanced} in  Figure \ref{Pure_VertexBalancedNE_alg}.
The proof of Theorem \ref{pure balanced equilibria} is identical to the proof of Theorem
\ref{defender pure balanced equilibria}.

\begin{figure}[h]
{\small

\fbox{\parbox{15.5cm}{

{\bf Algorithm} {\sf Pure\&VertexBalancedNE}

{\sc Input}:  A graph  $G=\langle V,E\rangle$               such that $ \delta \geq \beta'(G)  $  with  $
 2\delta$ divides $\alpha$.

{\sc Output}: A Defense-Optimal, Pure, Vertex-Balanced Nash equilibrium ${\bf s}$.
\begin{enumerate}

\item[{\sf (1)}] Choose a Minimum Edge Cover ${\it EC}$.

\item[{\sf (2)}] Assign each defender  to a distinct edge  from $EC$ in a round-robin fashion.

\end{enumerate}
\fbox{\parbox{15.2cm}{
\begin{enumerate}

\item[{\sf (3)}] For each vertex $v\in V$, set
${\sf A}(v) := |{\sf D}_{\ms}(v)| \cdot  \frac{ \tx \alpha } {  \tx 2\, \delta}$.

\item[{\sf (4)}] Assign   each attacker to a vertex from $V$ in an arbitrary way so that
for each vertex $v\in V$,  $|{\sf A}|_{\ms}(v)= {\sf A}(v)$.

\end{enumerate}}}
 }}} \caption{The algorithm {\sf Pure\&VertexBalancedNE}.
The algorithm {\sf Pure\&VertexBalancedNE} differs from the the algorithm {\sf
DefenderPure\&VertexBalancedNE} only in Steps {\sf (3)} and {\sf (4)}. The additional assumption that
$2\delta$ divides $\alpha$ suffices for Steps {\sf (3)} and {\sf (4)} to construct an attacker-pure profile.
By Step {\sf (2)}, $\ms$ is defender-pure. Step {\sf (3)} provisions for  $\ms$ to be vertex-balanced;
towards this end, it sets the ratio $\frac{\textstyle  {\sf A} (v)}      {\textstyle |{\sf D}_{\ms}(v)|} $ to
the fixed (integer) value $\frac{\textstyle  \alpha}      {\textstyle  2 \delta} $ for each vertex $v\in V$.
Hence, for each vertex $v\in V$, ${\sf A}(v) $  is integer. Finally,  Step {\sf (4)}   assigns pure
strategies to the attackers that induce the (integer) value $|{\sf A}|_{\ms}(v) = {\sf A}(v) $ for each
vertex $v \in V$; hence, $\ms$ is vertex-balanced by construction. Since a Minimum Edge Cover is computable in polynomial time, the algortihm {\sf Pure\&VertexBalancedNE} is polynomial time.
 }\label{Pure_VertexBalancedNE_alg}
\end{figure}

\section{Epilogue}
\label{conclusion}

  We proposed  and analyzed a new combinatorial model for a distributed system like the Internet with
selfish, malicious attacks and selfish, non-malicious,  interdependent defenses. Through an extensive
combinatorial analysis of Nash equilibria for this model, we derived a comprehensive collection of (in some
cases surprising) trade-off results between the number of defenders and the best possible Defense-Ratio of
associated Nash equilibria.

Our work leaves numerous open  problems relating to ({\it i}) the {\em worst-case} Nash equilibria for this
model, ({\it ii}) the investigation of alternative reward-sharing schemes for the defenders  and ({\it iii}) the complexity  of
computing and verifying (Defense-Optimal) Nash equilibria (especially for the case of too many defenders) in
this model.

\paragraph{Acknowledgments.}  We  thank      Martin Gairing, Loizos Michael, Florian Schoppmann,
Karsten Tiemann and the anonymous {\it HICSS 2008} reviewers for many helpful comments and suggestions on
earlier versions of this  paper.

\end{document}